\documentclass[12pt]{article}
\usepackage{amsmath}
\usepackage{epsfig}
\usepackage{amssymb}
\input{epsf}

\def \be  {\begin{equation}}
\def \ee  {\end{equation}}
\def \ba  {\begin{eqnarray}}
\def \ea  {\end{eqnarray}}
\def \baa {\begin{eqnarray*}}
\def \eaa {\end{eqnarray*}}
\def \bb  {\begin{thebibliography}}  
\def \eb  {\end{thebibliography}}

\def \lab #1 {\label{#1}}

\def \P {{\cal P}}

\def \fracs #1#2 {\mbox{\small $\frac{#1}{#2}$}}

\def \bin #1#2 {{\left({#1}\atop{#2}\right)}}
\def \as {\relax\ifmmode\alpha_s\else{$\alpha_s${ }}\fi}

\def \al #1 {\frac {\as({#1})}{\pi} }
\def \ds #1 {\ooalign{$\hfil/\hfil$\crcr$#1$}}

\newcommand \bea{\begin{eqnarray}}
\newcommand \eea{\end{eqnarray}}

\newcommand \ep {\epsilon}
\newcommand \vep {\varepsilon}

\def\hepph  #1 {{\tt hep-ph/#1}}

\begin{document}


\begin{flushright}
YITP-SB-05-26\\
Revised November 18, 2005
\end{flushright}

\vbox{\vskip .5 in}

\begin{center}
{\Large \bf Fragmentation, NRQCD and NNLO Factorization \\
\medskip

Analysis in  Heavy Quarkonium Production}

\vbox{\vskip 0.25 in}

{\large Gouranga C.\ Nayak$^a$, Jian-Wei Qiu$^b$ and George Sterman$^a$}

\vbox{\vskip 0.25 in}

{\it {}$^a$C.N.\ Yang Institute for Theoretical Physics,
Stony Brook University, SUNY\\
Stony Brook, New York 11794-3840, U.S.A.}

\bigskip

{\it {}$^b$Department of Physics and Astronomy,
Iowa State University\\
Ames, Iowa 50011-3160, U.S.A.}

\end{center}

\begin{abstract}
We discuss heavy quarkonium production
through parton fragmentation,
including a review of arguments for the factorization of
high-$p_T$ particles into fragmentation functions
for hadronic initial states.  We
investigate the further
factorization of fragmentation
functions in the NRQCD formalism, and 
argue that this requires a modification of
NRQCD octet production matrix elements to
include nonabelian phases, which
makes them gauge invariant.  We
describe the calculation of uncanceled infrared divergences
in fragmentation functions
that must be factorized at NNLO, and verify
that they are absorbed into the new,
gauge invariant matrix elements.
\end{abstract}

\bigskip


\section{Introduction}

The  production of bound states of 
heavy quark pairs played an historic role in the 
development of the Standard Model \cite{der04}.  This subject
has retained a continuing fascination, in part
because it offers unique perspectives into
the formation of QCD bound states.  
The first step in quarkonium production, the inclusive creation of a pair of
charm or bottom quarks, is an essentially perturbative
process.  In particular, at high transverse momentum in
hard-scattering processes, the dominant
mechanism for the production of heavy quarkonium  is the perturbative
fragmentation of lighter partons, especially the gluon \cite{braatenhipt}.

A basic result of perturbative QCD for the production of
hadrons $H$ at high transverse momentum 
from the scattering of initial-state particles $A$ and $B$ is the factorization
of universal fragmentation functions, 
\cite{mueller78,Col81a,CSSrv,1pIfact} 
\begin{equation}
d\sigma_{A+B\to H+X}(p_T) = 
\sum_i\; d\tilde\sigma_{A+B\to i+X}(p_T/z,\mu) \otimes
D_{H/i}(z,m_c,\mu) + {\mathcal O}(m_H^2/p_T^2)\, .
\label{cofact}
\end{equation}
Here, $\otimes$ represents a convolution in the momentum fraction $z$.
 The cross section $d\hat\sigma_{A+B\to i+X}$ includes all 
 information on the incoming state, including convolutions with
 parton distributions when $A$ and $B$ are hadrons,
 at factorization scale $\mu$.  We exhibit the $m_c$-dependence
 of the fragmentation function $D_{H/i}$ in anticipation of
 our interest in $H$ as a charm-anticharm bound state,
 for which the boundary condition for evolution is naturally taken at $\mu={\cal O}(m_c)$.
 
The actual transformation of a heavy quark pair
into a heavy quarkonium
$H$ with momentum fraction $z$ from parton $i$ 
requires the introduction of fragmentation functions $D_{H/i}(z,\mu)$.
This reasoning applies in principle to both charm and bottom quarks.
For definiteness, we
will generally refer to the heavy quark mass by $m_c$, and  
to the produced hadrons as $H=J/\psi$ etc.,
of mass $m_H \sim 2m_c$.

At moderate $z$
only the evolution, that is, the $\mu$-dependence, of
the fragmentation function is computable
perturbatively, with the remaining information
encoded in some initial function $D_{H/g}(z,\mu_0)$,
where we may take $\mu_0 \sim m_H$.
The effective field theory 
nonrelativistic QCD (NRQCD), however, 
has been invoked to simplify the
nonperturbative content of the fragmentation
functions $D_{H/i}(z,\mu_0)$, in terms of 
a few (or anyway finite number  of)
nonperturbative matrix elements.   

The logic behind the application of
NRQCD to fragmentation functions is readily summarized.
One applies the NRQCD expansion in the
relative velocity of the  produced quark pair, assuming
that a bound state will form only if this
relative velocity is small to begin with.  
One then argues that the formation of the
bound state is not affected by the
exchange of soft gluons with other hard
partons, only by exchanges between the quark and
antiquark, and with the vacuum \cite{bodwin94}.
NRQCD then specifies a complementary 
factorization theorem, often written as
\begin{equation}
d\sigma_{A+B\to H+X}(p_T) = \sum_n\; d\hat\sigma_{A+B\to 
c\bar{c}[n]+X}(p_T)\,
\langle {\mathcal  O}^H_n\rangle\, ,
\label{nrfact}
\end{equation}
where the ${\mathcal O}^{H}_n$ are universal NRQCD 
production operators, 
organized according to 
powers of the relative velocity of the $c\bar{c}$ state $[n]$, 
and their rotational and color
quantum numbers.  We will encounter explicit forms below.

Next, accepting that both (\ref{cofact}) and (\ref{nrfact}) hold
at high-$p_T$,
the fragmentation function and 
NRQCD matrix elements are related by \cite{braaten96}
\begin{equation}
D_{H/i}(z,m_c,\mu) = \sum_n\; d_{i\to c\bar{c}[n]}(z,\mu,m_c) \, \langle 
{\mathcal O}^H_n\rangle\, ,
\label{combofact}
\end{equation}
where $d_{i\to c\bar{c}[n]}(z,\mu,m_c)$ describes the evolution of an 
off-shell parton into a quark pair in state $[n]$, including logarithms of 
$\mu/m_c$.   
This formalism has been extensively
applied to heavy quarkonium production
\cite{QWGBrambilla,CDFoctet,CDFpolarization,HERAoctet,LEPoctet,RHICoctet,fixedtargetoctet,bodwin03}.
At the same time, it has been observed that
the applicability of NRQCD to production processes has
not been fully demonstrated \cite{bodwin03,nayak05}.

 In this paper we
revisit the formation of heavy quarkonia 
in fragmentation, with the aim of testing
the relation (\ref{combofact}).
We  will, however, first outline the proof of
 Eq.\ (\ref{cofact}) in leptonic annihilation and hadronic
 scattering, emphasizing that the arguments that apply
to light quark bound states apply as
well to heavy quarks, with corrections
suppressed by powers of the mass of
the quark divided by the transverse momentum.   
 While our arguments for the factorization
 of fragmentation functions will not actually
 cover new ground, we are aware of no other unified
 presentation of the steps leading to (\ref{cofact}) 
 for hadronic scattering in nonabelian gauge theories.

Once we have established (\ref{cofact}), we can
test NRQCD factorization in its more specific form,
Eq.\ (\ref{combofact}), which will simplify our analysis
considerably.
We study the factorization of the fragmentation
functions into perturbative coefficient
functions times NRQCD matrix elements,
with evolution included in the former.  We
shed new light on the necessary cancellation
of infrared divergences in the perturbative
calculation of coefficient functions.
In particular, we will show that to carry out such a
factorization, it is useful to redefine
conventional NRQCD production matrix
elements, with the addition of extra gauge
links, or Wilson lines, a process that we termed ``gauge
completion" in Ref.\ \cite{nayak05}.  
This modification renders the 
matrix elements gauge invariant.

Gauge completion is also necessary in order to absorb 
certain infrared divergences, beginning
at next-to-next-to leading order (NNLO), that
were not covered by the
original arguments for NRQCD
factorization.   We should note that our NNLO
infrared effects will appear only at order $\as^3$
in inclusive heavy quark pair
production cross sections of the type calculated
to order $\as^2$ in \cite{czar97} (see also the corresponding two loop
decay cross sections in \cite{ben97}).
Indeed we will encounter two loop
corrections with a quark pair and an
additional hard gluon in the final state.
The calculation is only possible, of
course, because we restrict ourselves
to the infrared sector.

In any case, however, we are not yet able to prove Eq.\ (\ref{combofact})
to all orders in perturbation theory. 
The basic results of this paper were outlined in
\cite{nayak05}.  Here, in addition to the
arguments on factorization,
we will provide rather
complete details on the methods
used to identify the relevant infrared
behavior, and on the necessary
two-loop calculations.

\section{Factorization of Fragmentation Functions}

\subsection{Long-distance dynamics in high-$p_T$ production.}

   The analysis of hard scattering cross sections begins with
   the identification of leading regions in the momentum space integrals
   of perturbative amplitudes and phase space \cite{CSSrv,Ste78,Stbook}.  Regions in
   this multidimensional space are conveniently classified
in cut diagram notation, in which graphical contributions to the
amplitude are
represented to the left of the final state, and contributions to the
complex conjugate amplitude to the right.  In the complex conjugate
graphs
the roles of final and initial states are reversed.


Because loop integrals are defined by contours
in complex momentum space, it is only at momentum
configurations where some subset of loop  momenta
are pinched that the contours are forced to or near
mass-shell poles that correspond to long-distance behavior.
These ``pinch surfaces", or subspaces, in turn can be classified
   according to their reduced diagrams, found by contracting
off-shell lines to points.  

The basic result is this.  Any reduced
diagram corresponding to a pinch surface can be interpreted
as a physical process, in which each vertex can be assigned
positions $x_i^\mu$ in space-time in such a way that if
$x_i$ and $x_j$ are connected by line $h$ carrying 
nonzero momentum
$p_h$, then
\bea
\Delta x_{ij}^\mu \equiv
\left( x_i - x_j\right)^\mu = (x_i-x_j)^0\, \beta_h^\mu\, ,
\label{deltaxij}
\eea
where  $\beta_h^\mu = p_h^\mu/p_h^0$ is the four-velocity of
   particle $h$.   At a pinch surface, this relation can be imposed for every line
   and vertex of the reduced diagram.    Consistency
   then requires that the sum of $\Delta x_{ij}$'s
   around any loop vanishes.  This is
   enough to ensure 
  that the reduced diagram does
  correspond to a physical picture, in which
   on-shell lines describe free, classical motion between the vertices.   
   To this physical picture for finite-energy lines,
   lines with vanishing momenta may be attached
   in an arbitrary manner \cite{Ste78,Stbook}.

   The proof of the relationship between pinches in
   loop momentum space and physical
   pictures described by (\ref{deltaxij}) is not difficult \cite{Ste78,Col65,Edenetal},
   but we shall not review it here.  Its consequences,
   however, are important and easily drawn.

\begin{figure}[h]
\begin{center}
\epsfig{figure=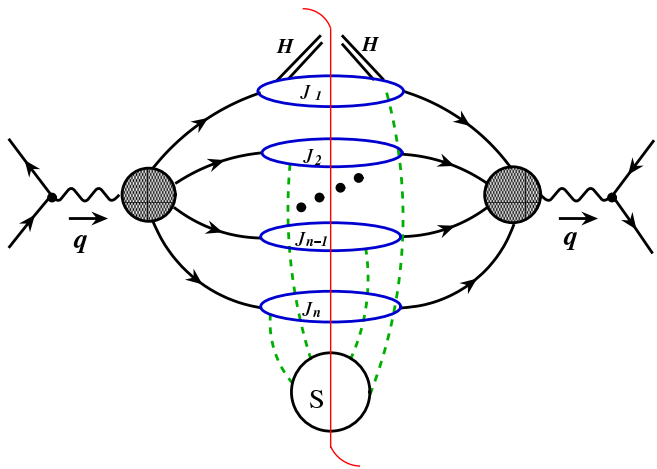,width=0.5\textwidth}
\caption{Reduced diagrams for high-$p_T$ particle production
in leptonic annihilation. \label{reducedlep}}
\end{center}
\end{figure}

\begin{figure}[h]
\begin{center}
\epsfig{figure=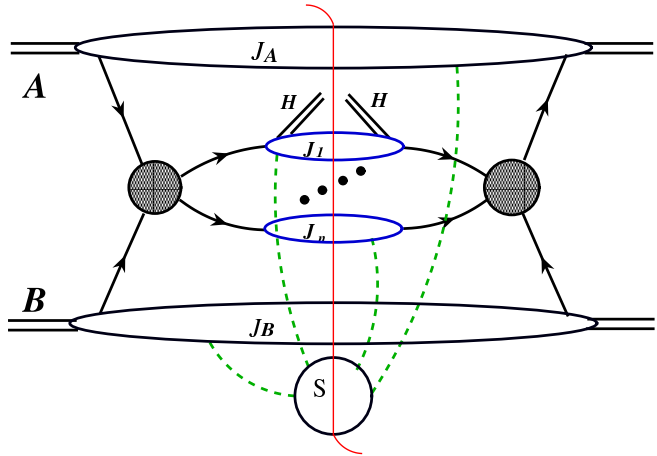,width=0.5\textwidth}
\caption{Reduced diagrams for high-$p_T$ particle production
in hadronic collisions. \label{reducedhad}}
\end{center}
\end{figure}

   Let us apply the above considerations to the production of
   a hadron $H$, with momentum $P$, in the scattering
   of particles $A$ and $B$,
   \bea
   A(p_A)+B(p_B) \rightarrow H(\vec P) +X\, .
   \eea
   We will assume that $P_T$ is a large scale,
   of the order of the center-of-mass energy, and 
   far above the strong coupling scale $\Lambda_{\mathrm QCD}$.
The relevant reduced diagrams for this
process are illustrated in cut diagram notation  by Fig.\ \ref{reducedlep}
when $A$ and $B$ are a leptonic pair,
and by Fig.\ \ref{reducedhad} when $A$ and $B$ are
strongly interacting (partons or physical hadrons).


At the  pinch surfaces, there is a single hard-scattering, labelled by
a shaded circle, in the amplitude and its complex conjugate.
For dilepton annihilation, Fig.\ \ref{reducedlep}, the hard scattering is the
result of the decay of the (real or virtual) electroweak boson formed
in the annihilation process.  For the hadronic process, the hard
scattering results from the collisions of a single parton from each of
the colliding hadrons
\footnote{There are, in fact, physical pictures corresponding to
collisions involving more than one parton from each hadron.
These, however, are suppressed by powers after summing
over gauge-invariant sets of diagrams \cite{Lab84}.}. 
Finite-energy partons emerge from the hard scatterings and connect to
subdiagrams of on-shell collinear lines, the jets, $J_i$.
All finite-energy particles of the 
final states are in one of these jets.
In addition zero-momentum lines may be exchanged 
 between the lines of the jets, and interact arbitrarily
 in a cut ``soft subdiagram" $S$, consisting 
 entirely of such lines.  In
effect, there are no final-state interactions
involving finite momentum transfer in
the reduced diagram for these processes.
Thus, the total momentum of each jet is
determined at the hard scattering, and
the distributions of jet energies are calculable
in perturbation theory.
The essential observation to obtain this result is that once the parent
partons of the jets emerge from the hard scattering, they separate
at the speed of light, and subsequently cannot interact locally in any
physical picture defined as above.  
 The observed hadron $H$ appears in one
of the jets \footnote{Here, we treat $H$ as massless, on
the scale of $P_T$.}.

We will not attempt a full review of power-counting
analysis in the neighborhood an arbitrary
reduced diagram of Figs.\ \ref{reducedlep} and \ref{reducedhad}.
It is worthwhile to recall, however, that we may characterize these
regions of momentum space by introducing scaling
variables, conventionally denoted $\lambda$, which control
the relative rates at which components of loop momenta vanish
near the pinch surfaces.
In the terminology of \cite{Ste78}, a leading region
is one for which a vanishing region of loop momentum
space near a pinch surface produces leading-power
behavior.  Such behavior is naturally associated with logarithmic
singularities at the corresponding pinch surface.

For the pinch surfaces at hand, we assign to
the $j$th jet a light-like vector in the jet direction,
$\bar{n}_j^\mu$, $\bar{n}_j^2=0$,
and an opposite-moving vector $n_j^\mu$,
$ n_j^2=0$, normalized such that $\bar{n}_j\cdot  n_j=1$.
For each jet, the  combination of $\bar n_j$ and $ n_j$
define a two-dimensional transverse space, which
we will denote as ${\perp^{(j)}}$.
The fundamental
leading regions are characterized by a familiar
scaling behavior for the loop momenta
internal to jet $j$,
\be
{\rm loop}\ l\ {\rm in\ jet}\ j: \quad
\left(n_j\cdot l \sim E_j,\ \bar n_j\cdot l \sim \lambda E_j,\
l_{\perp^{(j)}} \sim \lambda^{1/2}E_j\right)\, ,
\label{scalingjet}
\ee
where $E_j$ is the energy characteristic
of jet $j$, which we take to be of
the order of the overall center-of-mass energy, denoted $Q$.
In a similar notation, the scaling for soft
loop momenta, internal to the soft
subdiagram $S$ or flowing between $S$ and
any of the jets, is
\bea
{\rm soft\ loop}:\ k^\mu \sim \lambda\, Q\, .
\label{scalingsoft}
\eea
In principle, the two scales $\lambda$ in
Eq.\ (\ref{scalingjet}) and (\ref{scalingsoft}) need not be
the same.  A complete discussion includes
power counting for subdivergences, as
some lines approach the mass shell faster
than others \cite{Ste78,Stbook}.  

   \subsection{Jet-soft factorization}

Arguments for the factorization of soft 
gluons from jets
were given in some detail in Ref.\ \cite{CSSrv}
for $\rm e^+e^-$  annihilation.  
We review these arguments here, and discuss
their extension to hadronic scattering.
Jet-soft factorization is made
possible by the singularity structure
of loop integrals near pinch surfaces \cite{Col81a,Col81b}.
Many of these results have been rederived in
the language of soft-collinear effective theory \cite{bau01,bau02}.
In graphical terms, the factorization
 is most clearly illustrated for leptonic
annihilation, as in Fig.\ \ref{reducedlep}.   

\subsubsection{The soft approximation in leptonic annihilation}

In Fig.\ 1 let us consider a
loop momentum $k_i$, flowing from the soft subdiagram
into jet $J_j$, through the hard
scattering to another jet, and finally back to the  soft subdiagram.
We will examine poles in the integral near $k_i^\mu=0$
due to the denominators of $J_j$.
For definiteness, we assume the $k_i$ loop
is in the amplitude.  

The soft momentum appears in propagator
denominators of a set of lines in the jet function.
Let $r_j$ be the momentum of any such line at
$k_i=0$.  
When momenta are scaled as in Eqs.\ (\ref{scalingjet}) and (\ref{scalingsoft}),
any denominator in the jet function
is of the general form
\bea
\left( r_j \pm k_i\right)^2 +i\epsilon &=&
r_j^2 
\pm \left [\,  2n_j\cdot r_j\, \bar{n}_j\cdot k_i + 2\bar{n}_j\cdot r_j\, n_j\cdot k_i
- 2 r_{j\perp} \cdot  k_{i\perp}\, \right] + k_i^2 +i\epsilon
\nonumber\\
&=& {\mathcal O}(\lambda) \pm \left[\, {\mathcal O}(\lambda) + {\mathcal
O}(\lambda^2)
+ {\mathcal O}(\lambda^{3/2})\, \right] + {\mathcal O}(\lambda^2) +i\epsilon
\nonumber\\
&=& r_j^2 
\pm 2n_j\cdot r_j\, \bar{n}_j\cdot k_i + {\mathcal O}(\lambda^{3/2})+i\epsilon\, ,
\label{jsdenom}
\eea
where the second  equality gives the scaling
behavior of each of the terms in order,
according to (\ref{scalingjet}) and (\ref{scalingsoft}).
In the third equality we exhibit the leading behavior,
which depends only on the single component
$\bar n_j\cdot k_i$ of the soft momentum $k_i$.
This approximation will hold so long as the
contour of momentum component $\bar{n}_j\cdot k_i$ 
is not pinched in such a way as to violate the
 scaling of Eq.\ (\ref{scalingsoft}).

Within jet $J_j$, the soft loop momentum $k_i$ can always be re-routed
by shifts of the jet's loop momenta. 
In this way we may choose $k_i$ to flow 
on each jet line in a sense
opposite to the direction of the jet momentum $P_j\propto \bar{n}_j$
(minus sign in Eq.\ (\ref{jsdenom})).
With this choice 
all singularities in the variable $\bar{n}_j\cdot k_i$ are
in the same (here upper) half-plane.
Thus, although the poles in 
$\bar{n}_j\cdot k_i$ due to jet lines are generally
quite close to the origin,
they do not pinch the integration contour.  As a result,
the presence of poles in the jet subdiagrams does
not  take
us outside the scaling region of Eq.\ (\ref{scalingsoft}).

We now consider other possible sources of 
singularities in the variable $\bar{n}_j\cdot k_i$.
Every
virtual soft loop momentum that attaches at least
one jet to the soft subdiagram may be routed so
that it flows into no more than two jets.
This is because once it reaches a second jet, it
can be routed back to the first through the hard scattering subdiagram
(where it is neglected in the off-shell propagators
and vertices).   The two jets determine  distinct
contour deformations for the soft loop momentum.
These deformations are guaranteed to be compatible, however,
because we can always specify them
in a frame where the two jets in question, say $j$ and $j'$, are back-to-back.
In this frame, we may identify $\bar{n}_j\cdot k_i \equiv k_i^-$
and $\bar{n}_{j'}\cdot k_i=k_i^+$, for example. The consistency
of the two contour deformations is then clear. 
Momentum $k_i$ must also flow through the soft
subdiagram.  Here, singularities are 
generically a distance ${\cal O}(\lambda)$ from
the origin, except at lower-dimensional 
spaces.   If such a subspace corresponds to a
leading region, it may be treated separately, by
the same arguments \cite{Stbook}.
Finally, we note that if $k_i$ is the momentum
of an on-shell gluon (or decays into
a set of on-shell gluons), Eq.\ (\ref{jsdenom}) always
holds, unless $k_i$ is itself collinear to the jet
momentum.  In this case, the line carrying $k_i$ should be treated as part of
the jet.  

In summary, we have learned that the leading power behavior
of the cross section from any leading region $R$
may  be found by keeping only the $\bar{n}_j\cdot k_i$ component
of soft momenta $k_i$ within the jet subdiagram,
and setting the remaining components  to zero.
Similarly, in region $R$
the soft gluons couple to the jet subdiagram
only through the polarization component proportional to the
jet direction, because at the pinch surface all other 
components of the jet tensor vanish as a power of $\lambda$.

Now we consider
the subdiagram consisting of all  lines in 
jet $J_j$, connected to the
hard subdiagram by parton $j$  in
leading region $R$, not including the propagators
of its external soft lines \footnote{In a covariant gauge, parton $j$
is accompanied by a set of collinear vector
lines with scalar polarizations at the coupling of jet $J^{(R,C)}_j$ to the
hard scattering \cite{CSSrv}.  These gluons may be
factored from the hard scattering and are included in the jet.}.   
To be specific, we assume there are $m$ soft
gluons connected to the jet in the amplitude, and
$n-m$ in the complex conjugate amplitude.
We introduce a jet function $J^{(R,C)}_j$
for leading region $R$, where $C$ labels the particular
cut of the jet subdiagram.  With a given cut $C$, of course, the 
assignment of soft lines to the amplitude and its
complex conjugate is specified.

Our considerations lead
us to a ``soft approximation" for the function $J^{(R,C)}_j$ \cite{CSSrv}.
Within leading region $R$ we may make a replacement
that isolates the leading soft gluon momentum and polarization components.
In these terms, the soft approximation may be defined by
\bea
J^{(R,C)}_j{}_{ML,a_1\dots a_n}^{\mu_1\dots\mu_n} \left(k_1  \dots k_n,P_j\right) =
J^{(R,C)}_j{}_{ML,a_1\dots a_n}^{\nu_1\dots\nu_n} 
\left(\tilde{k}_1 \dots  \tilde{k}_n,P\right)n_{\nu_1}\dots n_{\nu_n}
\, \bar{n}^{\mu_1}\dots \bar{n}^{\mu_n}\, ,
\label{softapprox}
\eea
where we define, for any momentum $k_i$,
\bea
\tilde{k_i}^\sigma =k_i\cdot \bar n_j\, n_j^\sigma
\label{tildedef}
\eea
as a vector with only the ``opposite moving"
component of momentum.  (Of course, the
definition of $\tilde k_i$ varies from jet to jet.)
The indices $a_i$ ($\mu_i$) are the color (vector) indices of the
external soft gluons of momentum $k_i$, while $M,\, L$ are
the color indices of parent parton $j$, in the
appropriate  color representation.
Corrections to the soft approximation are suppressed by powers
of the scaling variable $\lambda$, and hence by the overall hard scale $Q$.

From the soft approximation Eq.\ (\ref{softapprox}),
the coupling of soft gluons to jet $J_j$ is identical
to the coupling of a set of unphysical gluons to the jet,
whose polarizations are proportional to their momenta.
Once we have made the soft approximation, 
it becomes straightforward
to apply the nonabelian Ward identities of QCD
to the connections of soft gluons to the jet \cite{CSSrv}.
This has a simple classical analogy.    In
the rest frames of particles within jet $J_j$,
the classical fields due to particles in other jets, all of
which are separating with relative velocities $\beta_{\mathrm rel}\sim c$,
reduce to pure gauge fields, up to corrections of
order $(\beta_{\mathrm rel}-c)$ \cite{bas84}.

\subsubsection{Factorization and the residual jet factor}

Once we have used the soft approximation
and the Ward identities,
the entire effect of the soft gluons external to the jets
  is to produce, order by order,
a product of eikonal factors, 
\bea
J^{(R,C)}_j{}_{ML,a_1\dots a_n}^{\nu_1\dots\nu_n}
\left(\tilde{k}_1 \dots \tilde{k}_n,P_j\right)n_{\nu_1}\dots n_{\nu_n}
\, \bar{n}^{\mu_1}\dots \bar{n}^{\mu_n} && \nonumber\\
&& \hspace{-80mm}
=J^{(R,C)}_j{}(P_j)\; 
E^{(j)}_{MK,}{}_{a_{m+1}\dots a_n}^{\mu_{m+1}\dots\mu_n}\,  {}^\dagger{}
\left(- \tilde{k}_{m+1}\dots - \tilde{k}_n\right)
E_{KL,}^{(j)}{}_{a_1\dots a_m}^{\mu_1\dots\mu_m}\left(\tilde{k}_1\dots \tilde{k}_m\right)\, .
\label{jetsoftfact}
\eea
Here, $m$ is the number of soft gluons that couple to the jet subdiagram in
the  amplidue, and  $n-m$ the number  in the complex conjugate
amplitude.  
The eikonal factors $E$ and $E^\dagger$ 
reproduce all momentum and color dependence, but are insensitive
to the internal dynamics of the jet, and depend only
on the 4-velocity $\bar{n}_j$, in the jet direction, and the
color representation of parton $j$.  Specifically, they 
are given by
\bea
E_{KI,}^{(j)}{}_{a_1\dots a_m}^{\mu_1\dots\mu_m}\left(\tilde{k}_1\dots \tilde{k}_m\right) = 
\sum_{\mathrm perms}\, \left[\; P\, \prod_{i=1}^m\ 
 \frac{g\, \bar{n}_j^{\mu_i}\, T^{[j]}_{a_i}}
      {- \bar{n}_j\cdot (k_1 + \cdots + k_i)+i\epsilon}\; \right]_{KI}
\, ,
\label{eikonalfactors}
\eea
where $P$ implies ordering of the
color matrices $T^{[j]}_a$ according to the  
permutation of soft gluon connections.  (As above, soft momenta flow into
the jet.)  
The function $J^{(R,C)}_j(P_j)$ in (\ref{jetsoftfact})
represents what we will refer to as the residual
jet factor in region $R$ with cut $C$.  It is given by the normalized color trace of the jet
function with no external soft gluons,
\bea
J^{(R,C)}_j(P_j) =  \frac{1}{d(j)}\; \sum_L 
J^{(R,C)}_j{}_{LL}
\left(P_j\right)\, ,
\eea
where $d(j)$ is the
dimension of the color representation of parton $j$.

To apply the Ward identities
that lead to Eq.\ (\ref{jetsoftfact}) we need only
integrate over the opposite-moving components,
$\bar n_j\cdot l$ of the jet loop momenta $l$.
This is because the Ward identities only require shifts
in loop momentum equal to the momenta
that flow into the jet from the external lines.

In general, the residual jet function includes
  contributions from soft gluons
 for which the soft approximation fails,  but
 which remain internal to the jet.  It is not necessary
 that the soft approximation apply to every soft gluon.
 Rather, for this analysis to hold it is only necessary
 that in every leading region we can find 
a set of soft lines for which it holds, and for which we
may apply Eq.\ (\ref{jetsoftfact}).  
Equation (\ref{jetsoftfact}) is a general result
for final-state jets in arbitrary leading regions.
We will see below
how function $J^{(R,C)}_j$ can be
identified as a contribution to a fragmentation
function.  

At each  (here $n$th) order, the factorized
$\tilde{k}_i$ dependence in the eikonal factor $E$ in (\ref{eikonalfactors})
is identical to the corresponding dependence in the
expansion of the ordered exponential 
\bea
\Phi^{(j)}_{\bar{n}_j}(0,\infty) = \P \exp \left[ -ig\int_0^\infty d\lambda\, \bar{n}_j\cdot A^{(j)}(\bar{n}\lambda)\right]\, ,
\label{oexp}
\eea
where now $P$ denotes path ordering, and where $A^{(j)}$ is the
gauge field in the matrix 
color representation of the parent parton of the jet (quark, antiquark or gluon)
\footnote{This is the factorization effected in soft-collinear effective theory
by a redefinition for collinear fields. \cite{bau02}}.   At leading power 
in $\lambda$, and hence in the large momentum scale $Q$,
soft gluons couple to jets only through the operators $\bar{n}_j\cdot A$,
restricted to the light cone along the jet directions.  We will
see below how other operators arise at nonleading powers.

All of the reasoning above may be applied to the 
particular jet ($J_1$ in Fig.\ 1) from which the
observed hadron $H$ arises.  
The entire leading-power dependence on the masses 
and relative momentum of the quarks, as well
as on the momentum fraction ($z$) of the pair  is in 
the functions $J^{(R,C)}_j(P)$ at each leading region.
The influence of
soft gluon emission on $z$ can be neglected,
precisely because of the soft approximation (\ref{softapprox}).
Thus, in each leading region, the jet dynamics
that produces an observed particle decouples from
soft gluons that could link it to the other jets in the final state.
In this way, fragmentation is
seen to be universal, depending only on
the parent parton, the  produced hadron, its momentum fraction $z$,
and eventually a
factorization scale.

\subsubsection{Hadronic scattering}

The  arguments for jet-soft factorization in hadronic
scattering are similar to those
for leptonic annihilation\footnote{They are also essentially
identical to those for lepton-hadron scattering.}, but special care must be
taken because of the ``initial-state" jet subdiagrams $J_A$ and $J_B$
consisting of lines collinear to particles $A$ and $B$ in Fig.\ \ref{reducedhad}.
As we shall see, however, the factorization of fragmentation within a final-state jet
holds in this case as well, and
is actually somewhat more general than collinear factorization
in terms of parton distributions
\cite{CSSrv}.    

The essential difference between leptonic annihilation
and hadronic scattering may be seen by comparing
Figs.\ \ref{reducedlep} and \ref{reducedhad}.
In the former, although the poles from jet subdiagram $J_j$
in the soft momentum components $\bar{n}_j\cdot k_i$
are closer to the origin than ${\mathcal O}(\lambda)$ in
general, they are
all in the same half-plane.
As a result, these momentum
components may be deformed away from
the poles into a region where the soft approximation
holds. 

For hadronic scattering, precisely the same reasoning
applies for soft momenta that flow
only between final-state jets,
and/or through the hard scattering.
 It also applies for soft loops that
connect to an initial-state jet only via lines whose large
momenta flow directly from the initial state into the
hard interaction.  In these connections, to what are
sometimes called ``active" jet lines, all poles are  again
in the same half-plane, and the same reasoning
allows us to deform contours as above to justify
the soft approximation.

A difference arises, however,
when soft lines connect to the initial state jets
by  ``spectator" jet lines,
whose momenta flow into the final state without
passing through the hard scattering.
In this case, to complete the soft
loop through the hard interaction,
the momentum must flow ``back" to
a vertex at which spectator lines and active lines connect,
and then flow once again forward into the
hard scattering.   Suppose this occurs for
initial-state jet $A$.
Both spectator and active lines in $A$
produce poles close to the origin
for a soft component
$\bar{n}_A\cdot k_i$, and these poles are in opposite
half-planes.  The resulting pinch forces us into a leading
region where $\bar{n}_A\cdot k_i\ll k_{i\perp}\sim \lambda$,
which is generally referred to as a ``Glauber region" \cite{Col81a,bod81}. 
In this region, the scaling (\ref{scalingsoft}) 
does not hold, and the soft approximation fails for this jet.
Because the soft approximation fails, 
soft gluons ``resolve" the internal structure of
the jet, and the factorization arguments given
above may not apply.

When the soft loop flows between an initial
state jet and a final state jet, however, only a single
light-cone component is pinched, associated with the
initial-state jet.  The soft approximation
may still be applied to the final-state jet,
giving eikonal factors as in Eqs.\ (\ref{jetsoftfact}) and 
(\ref{eikonalfactors}).  The eikonal factors associated with
the outgoing jet then cancel in a single-particle inclusive
cross section, in same way that soft divergences
cancel in  jet cross sections.\footnote{The argument
for this cancellation in the case of hadronic
scattering is given in the first part of Sec.\ V
of Ref.\ \cite{lib78}.}
    This decoupling and cancellation of soft gluons enables us
    to identify universal fragmentation functions, in terms of
    universal matrix elements, in hadronic scattering as
    well as leptonic annihilation,
    independent
    of the jet structure of the particular hard scattering.

Finally,  consider those
soft loops that flow between the two initial-state jets and/or
through the hard scattering.  In general such loops encounter Glauber
pinches in two light-cone components.  
For cross sections that are inclusive in soft gluon emission
and in the fragmentation of the forward jet remnants,
these pinches nevertheless cancel in the sum over final states.
We then have collinear factorization into independent parton
distributions  for incoming
hadrons $A$ and $B$ and a fragmentation function
for hadron $H$ \cite{CSSrv}.
It is worthwhile noting, however, that even when
these criteria are not satisfied, and the overall
cross section does not factorize into incoming
parton distributions (as, for example, in diffractive
scattering in hadron-hadron collisions \cite{diffact}), the 
final-state jets still factorize from the incoming jets
and their soft exchanges, and the single-particle cross section
at high $p_T$ is still governed by a universal fragmentation function.

\subsection{Power corrections}

Once we have determined that the leading-power
contributions factorize for the leading regions
associated with Fig.\ \ref{reducedhad}, we
naturally turn our attention to power corrections \cite{bau03}.
These may be classified by an expansion
in nonleading contributions to the integrand
near the pinch surfaces.  It is therefore
an expansion in terms of ratios such as
$f(k)/{q}_j^2(\tilde k)$, where the numerator
$f(k)$ represents any of the terms involving soft momenta
$k$ that can be neglected at leading power.  These are
the terms in Eq.\ (\ref{jsdenom}) that scale as $\lambda^{3/2}$ or
higher, as  well as non-leading terms from numerator momenta.
The denominator represents the squared momentum
of the  jet line $q_j$, after the
soft momenta  $k$ flowing on jet line $q_j$ is
replaced by $\tilde k=n_j(\bar n_j\cdot k) \sim {\cal O}(\lambda)$, 
Eq.\ (\ref{tildedef}).    Although $q_j^2(\tilde k)$ is not large,
the ratio  $f(k)/q_j^2$ is small in the leading region.
(As we have seen, this may require
contour deformations.)

Keeping only $\bar n_j\cdot k$ terms in the denominators,
the numerator terms $f(k)$ are polynomials in the $n_j\cdot k$
and $\perp^{(j)}$ components of soft gluon momenta.
In position space,
these vertices, connected to jet lines,
correspond to operators that are
local with respect to the $n_j^\mu$ and $\perp^{(j)}$
directions, but are
relatively on the $\bar n_j^\mu$ light-cone.
The gauge invariance of the theory requires that these
vertices, representing the interactions
of soft gluons with the jet functions, combine to
form gauge covariant operators.
We may think of these vertices as supplementing
the leading-power $\bar{n}_j\cdot A$
vertices of the ``soft approximation", identified above
with the Wilson lines of Eq.\ (\ref{oexp}).  As above, the
application of Ward identities, or equivalently
a redefinition of collinear fields as in soft-collinear
effective theory, organizes all leading vertices
into nonabelian phase operators, but now
acting as color rotations on the nonleading vertices
as well as on the ``parent" parton lines of the
jet functions.
For the
purposes of factorization at leading power
in $Q$, however, we need not enumerate these nonleading operators
or vertices.

\subsection{Fragmentation functions}

So far, we have identified the leading regions in
cut diagrams that are associated with infrared
dynamics in single-particle inclusive cross sections.
We have seen that at each leading region
the cross section breaks up into a factor
associated with the production of a parton ($j$ above),
times a jet function $J_j^{(R,C)}(P)$ that
describes a contribution to the formation of hadron $H$.
In this section, we will show that the jet
functions identified above are in one-to-one
correspondence with 
leading regions for the standard fragmentation functions, $D_{H/j}$.

Fragmentation functions may be defined 
in terms of expectation values \cite{col82}.  For example,
consider the production of hadron $H$
from a parent gluon at momentum fraction $z$,
taken, for definiteness along the 3-direction.
The relevant matrix element is then, in $D=4-2\vep$ dimensions,
\bea
D_{H/g}(z,\mu)
&=&
 {-z^{2-2\vep} \over 16(2-2\vep)\pi P^+}\ {\rm Tr_{color}}\;
 \int dx^- \, {\rm e}^{-i(P^+/z)x^-}\
 \langle 0|\, F^{+\lambda}(0)\, \left[\, \Phi_n^{(A)}(0,\infty)\, \right]^\dagger
\nonumber\\
&\ & \hspace{10mm} \times 
a_H(P^+,0_\perp)a_H^\dagger(P^+,0_\perp)
 \Phi_n^{(A)}(0,\infty)\; F^+{}_\lambda(0^+,x^-,0_\perp)\, |0\rangle\, ,
 \label{fragfn1}
 \eea
 where $a^\dagger_H$ is the creation operator for particle $H$ at momentum $P$
  and $F^{+\lambda}$ is the
  gluon field-strength.  The operator $\Phi^{(A)}$ is defined as in Eq.\ (\ref{oexp}),
 but in the direction $n^\mu=\delta_{\mu -}$, opposite to the 
 direction of hadron $H$.  Its fields are  in the adjoint matrix
 representation of color.
  The product of operators on
  the light cone requires renormalization and the introduction
  of a scale $\mu$, as described in \cite{col82}.

  The expectation value in Eq.\ (\ref{fragfn1})
may also be expressed as  a sum over all states including
 hadron $H$,
 \bea
 D_{H/g}(z,\mu)
&=&
 {- z^{2-2\vep} \over 16(2-2\vep)\pi P^+}\ {\rm Tr_{color}}\; \sum_N
 \int dx^- \, {\rm e}^{-i(P+/z)x^-}\
 \langle 0|\, F^{+\lambda}(0)\, \left[\, \Phi_n^{(A)}(0,\infty)\, \right]^\dagger
\nonumber\\
&\ & \hspace{10mm} \times 
|\, H(P^+,0_\perp)\, N\rangle\langle N\, H(P^+,0_\perp)\,  |
 \Phi_n^{(A)}(0,\infty)\; F^+{}_\lambda(0^+,x^-,0_\perp)\, |0\rangle\, .
 \label{fragfn2}
 \eea
 This form shows its close correspondence to a cross section.

The leading regions of the expectation values (\ref{fragfn1}) and  (\ref{fragfn2})
are, in fact, very similar to those of leptonic annihilation cross sections
discussed above.  Every leading region
includes in its reduced diagram
 a jet $J(P)$ that provides the particle of momentum $P$ in the final state,
 in addition to a jet in the opposite-moving direction $n^\mu$,
 and possibly other jets and arbitrary soft radiation  
 (subject to the effective phase space limitations imposed by renormalization 
 at scale $\mu$).  
 At each such leading region $R$, the same arguments as
 for leptonic annihilation lead to the precise analog of
 Eq.\ (\ref{jetsoftfact}), with exactly the same residual
 jet functions $J_j^{(R,C)}(P)$.
 Because the cross section is otherwise inclusive,
 the sum over final states  results in the cancellation of
 all soft and collinear singularities except for those
 associated with $J_j^{(R,C)}(P)$.
Therefore, we recognize a
one-to-one matching of every leading region
in the fragmentation function with a corresponding region
 in the total cross section.  This is the case
 for both leptonic and hadronic initial states,
 because the residual jet functions are the same in each case.

Strictly speaking, of course, the above discussion 
applies only to perturbation theory, which requires that
we impose an infrared regulation, presumably
dimensional regularization.    Because our
arguments extend to all orders in perturbation theory, however,
we may in principle introduce an interpolating field 
with the quantum numbers of hadron $H$, sum to
all orders in that channel, and isolate the S-matrix
elements for $H$ in the regulated theory.   
In this sense our arguments demonstrate factorization for
bound state $H$ in the regulated
theory.  We assume that the continuation back to
physical QCD in four dimensions respects this result.
This assumption is shared with essentially all demonstrations
of infrared safety and factorization.

   \section{NRQCD Factorization and Gauge Completion}

 Having reviewed
 arguments for the factorization of fragmentation functions, Eq.\ (\ref{cofact}),
   up to corrections in powers of $m_H/P_T$, we 
  are ready to rephrase the question of NRQCD factorization
   in terms of the fragmentation functions  themselves,
   as in Eq.\ (\ref{combofact}).
    We begin with a further examination of the leading regions of the 
   fragmentation functions,
 and we discuss evolution to
  the mass scale of the heavy quarkonium $m_H$.  We then analyze
 the refactorization, Eq.\ (\ref{combofact}) of the gluon fragmentation function
 in terms of NRQCD production operators, and propose a gauge-invariant 
 extension of the conventional operators.

   \subsection{Refactorization at the heavy quark mass}

   Our first goal is to separate logarithms associated with
   evolution from dynamics at the scale of the heavy
   quark mass.  This can be done by invoking the evolution
   equations for the gluon fragmentation functions in Eqs. (\ref{fragfn1}) and (\ref{fragfn2}),
   \bea
   \mu\frac{d}{d\mu}\,  D_{H/g}(z,\mu) 
   =
   \sum_i\; 
   \int_z^1 {d\xi\over \xi} \ P_{ig}\left({z\over \xi},\alpha_s(\mu)\right)\,
    D_{H/i}(\xi,\mu)\, , 
    \label{fragevol}
   \eea
   with a sum over
   partons $i$, and similarly when the
   gluon is replaced by a quark or antiquark.\footnote{The evolution kernels 
   for heavy quarks may be chosen identical to those
   for massless quarks in the case of parton distibutions \cite{col86}.  
   A similar relation should hold here, although we will not attempt a formal
   proof.}
    The solution to (\ref{fragevol}) enables us
   to relate fragmentation at the conventional scale
   $P_T$ with the mass scale of the produced hadron, $H$,
   \bea
   D_{H/g}(z,P_T)
=
  \sum_i\; 
  \int_z^1 \frac{d\xi}{\xi}\ {\cal C}_{gi}\left( {z\over \xi}, P_T,m_H\right)\;  D_{H/i}(\xi,m_H)\, ,
  \label{evolvedd}
  \eea
  where ${\cal C}_{gi}$  is a perturbative factor.  

  We will want to
  study the expansion in relative velocity of the heavy quarks in
  the fragmentation function evaluated at a scale on the
  order of the heavy quarkonium masss.  It is natural, of course, to
  carry out this expansion in the rest frame of the heavy
  quark pair.  Since this is not the usual frame in which to
  discuss fragmentation or the evolution (\ref{fragevol})
  associated with it,  we will briefly discuss how
  evolution appears in this frame.  Specifically, we need to show that
evolution logarithms factorize from the decay of an off-shell
gluon, with mass of order $m_H$, as seen in the rest
frame of hadron $H$.  

 The transverse momentum of the observed
  heavy quarkonium in the fragmentation function (\ref{fragfn1}) is by definition zero.
  Thus,  the transformation to its rest frame is a boost in the
  direction of its momentum as seen in the lab.  
  For convenience we take this momentum
  in the ``plus" direction, as in (\ref{fragfn1}).
  In both the lab frame and
  the quarkonium rest frame, evolution then
  results from the strongly ordered transverse momenta
  of partonic radiation.   

  To confirm Eq.\ (\ref{evolvedd}), we
  should verify that we can factorize soft gluons that
  connect partons with transverse momenta $k_\perp \gg m_H$
  from those of lower transverse momentum.  The former will
  appear in the evolution functions ${\cal C}_{gi}$, the latter in
  the fragmentation function at the scale of $m_H$.
  This separation  of low- from high-$k_\perp$ gluons 
  as seen in the $H$ rest frame follows
  exactly the same pattern as the factorization of gluons
  from the jets in Sec.\ 2 above.  

  Consider a parton $d$ of  
transverse momentum $k_{d\perp}\gg m_H$ and longitudinal 
momentum $k_d^+=z_dQ$,
  as seen in the lab frame (or the center of mass frame of the overall collision), with $Q$
  the  energy of the jet in that frame.   In the same frame and notation, the 
  heavy quarkonium $H$ has transverse momentum $k_{\perp,H}=0$, and 
energy $E_H=zQ\gg m_H$.
  A boost to the rest frame, where the energy of $H$ is $m_H$,
  leaves the transverse momentum  $k_{d\perp}$ unchanged,
  while transforming the plus and minus components of 
  $k_d$ according to
  \bea
  k_d^+ = z_dQ \qquad \Rightarrow && z_dQ\; \frac{m_H}{\sqrt{2}zQ}
  \nonumber\\
  k_d^- = \frac{k^2_{d\perp}}{2z_dQ} \qquad \Rightarrow && 
   \frac{k^2_{d\perp}}{2z_dQ}\; \frac{\sqrt{2}zQ}{m_H}\, .
  \label{pdtransform}
  \eea
  Equivalently, the rapidity of parton $d$ transforms according to
  \bea
  \eta_d = \frac{1}{2}\ \ln \left(\, \frac{2(z_dQ)^2}{k^2_{d\perp}}\, \right)
  \qquad
  \Rightarrow \qquad
  \frac{1}{2}\ \ln \left(\, \frac{z_d^2}{z^2}\, \frac{m^2_H}{k^2_{d\perp}} \, \right)\, .
  \eea
  By assumption, $k_{d\perp}\gg m_H$.  Therefore, as long as $z$ is not itself
  small, that is, assuming that $H$ is one of the ``leading" hadrons
  in the jet, the rapidity of parton $d$, which is large and positive in the
   center of mass frame, is large and negative in the rest frame of
   hadron $H$.   In this frame, all strongly ordered (in transverse momentum)
   partons are moving in the direction opposite to the original jet direction.
     The soft approximation can now be applied to soft gluons connecting
   the heavy quark pair that forms the quarkonium to the strongly ordered gluons.
   The interactions of these soft gluons may then be approximated
   by an eikonal line in the direction $n^\mu$, opposite to the
   jet's direction.  The only difference from soft-jet factorization
   in a cross section is that now the soft gluons' transverse 
   momenta are smaller than $m_H$.  The result is exactly a
   fragmentation function with upper limit $m_H$ on gluon transverse momentum
    in convolution with a perturbative function, 
   as in Eq.\ (\ref{evolvedd}), which is what we set out to show.   

   \subsection{Long and short distance dependence at the scale $m_H$}

To make contact with NRQCD applied to a fragmentation
function, we explore further the sources of its long-
and short-distance behavior.  This can be done as in
the discussion of  cross sections and fragmentation
functions above, in Sec.\ 2, although now we will carry out
our analysis in the rest frame of the
heavy quarkonium.  We begin, as above, with the
physical pictures associated with pinch surfaces.

\begin{figure}[h]
\begin{center}
\epsfig{figure=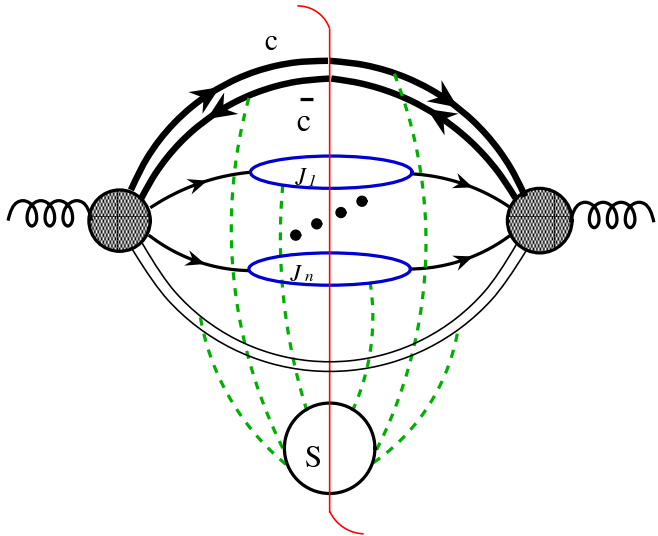,width=0.5\textwidth}
\caption{Physical pictures for heavy quarkonium
  production. \label{fragphyspict}} 
\end{center}
\end{figure}

The relevant physical pictures for fragmentation into
hadron $H$ are shown in Fig.\ \ref{fragphyspict}.
Since we are working in infrared regularized perturbation theory, the
heavy quarks appear in the final states.  We recall our discussion
above, however, in which we argued that in principle
the reduction of the bound-state pole does not modify
factorization.  We will continue with this assumption.

A related point is that at the bound state pole the relative momenta of the
quark pairs on either side of the cut need not be the same.
In principle, then, we should take the relative momentum of
$c\bar{c}$ pair in the amplitude,
$q$ below, to be independent of the relative momentum, $q'$
of the $c\bar{c}'$ pair to the right.  This is the 
method employed in the explicit calculations of
Refs.\ \cite{ma94,braaten97,petrelli98,ma05,braaten00,lee05}, for   
example.  Powers of $q$ and 
$q'$, however, are employed to identify
operators in NRQCD, terms linear in $q$
corresponding to the lowest order of the 
covariant derivative.  Since 
we are interested primarily in separating infrared
poles from coefficient functions, we will not
distinguish between $q$ and $q'$ below,
and simply calculate the fixed-order 
eikonal cross section for a quark pair.

As noted in the discussion of Sec.\ 2.4, the physical pictures for
fragmentation 
are similar to those for the hadronic final state interactions
in leptonic annihilation.  The process begins with
a short-distance subdiagram, represented by a shaded
circle in Fig.\ \ref{fragphyspict}.  In this case, short-distance
refers to virtualities at the order of $m_H$.
 Lightlike jets, $J_j$, develop from 
energetic ($E\sim m_H$) ``semihard" quarks, antiquarks or gluons, which emerge from
the hard scattering.  
These jets may be connected
by a diagram consisting entirely of soft quanta, $S$,
to each other, and to the Wilson line that is part of the
construction of the fragmentation function.
In the case at hand, the
heavy quark pair also emerges from the hard scattering, and soft quanta may connect
to the jets and/or the Wilson line.   To prove NRQCD factorization, Eq.\ (\ref{combofact}),
it will be necessary to show that all of this long-distance behavior
either cancels or matches entirely to NRQCD matrix elements.

Considered abstractly, the connection to NRQCD is made by ``integrating out"
degrees of freedom at the mass scale $m_H$ in the calculation
of the fragmentation function.   
In practice, that is in perturbation theory, 
the NRQCD operators ${\mathcal O}_n$ can be identified once we 
consistently separate long and short distance contributions. 
As the figure shows, a generic pinch surface in phase space
involves not only a truly short distance part, but also a variety of
semi-hard jets.  The question we must ask is to what extent hadronization
is affected by the presence of these jets.  In the original discussion
of NRQCD factorization given
in Ref.\ \cite{bodwin94}, it was argued that in the inclusive
sum over cuts in $H$ production, all infrared
divergences due to soft exchanges between the heavy quarks and
the extra jets cancel in the inclusive sum, even while we
fix the final state of the quark pair to be a gauge singlet.  
Notice that even in the absence of semi-hard gluons,
soft gluons may be exchanged with the Wilson line
that is part of the definition of the fragmentation function.
Indeed, this Wilson line is what remains of all
exchanges of soft gluons between the heavy quarks and
partons at relative momenta greater than $m_H$.

\begin{figure}[h]
\begin{center}
\epsfig{figure=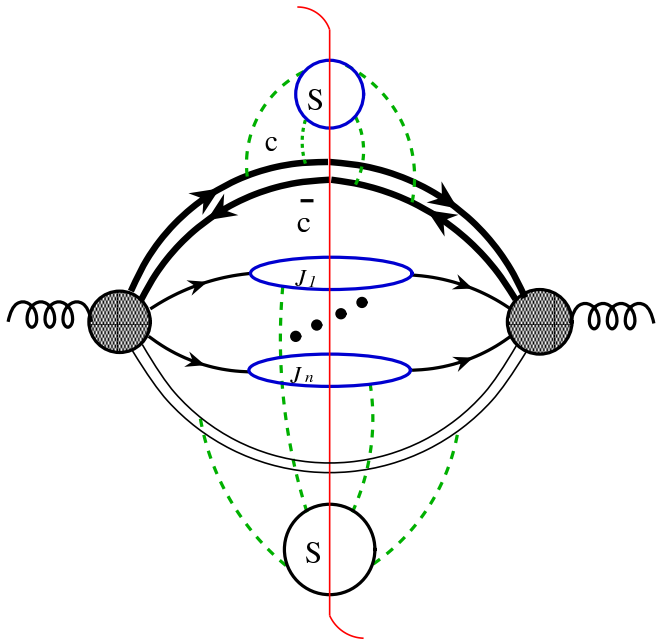,width=0.5\textwidth}
\caption{Topologically factored physical pictures for heavy quarkonium
  production. 
\label{physaftercancel}}
\end{center}
\end{figure}

In the absence of the soft gluon connections
between the heavy quarks and semi-hard gluons, the
remaining physical pictures can be
 represented as in Fig.\ \ref{physaftercancel}. 
 In this case, all collinear and soft divergences associated
 with the jets, whose final states are summed over inclusively,
 cancel, just as in leptonic annihilation.  Soft singularities may,
 and in general do, remain in the transition of the heavy quarks
 from short distances to hadronization, but such soft divergences
 are said to be ``topologically factorized" \cite{braaten97b}, and are readily
 factorized from the hard scattering function by a standard expansion
 in relative velocity, as we now sketch.

In a topologically factorized diagram like Fig.\ \ref{physaftercancel},
we can expand the short-distance function
around vanishing relative momentum, or equivalently relative
velocity $v$ of the heavy quarks.   
Similarly, we may decompose each diagram according
to the color state (singlet or octet) of the heavy quark pair,
and may also expand in the momenta of any light
quanta (gluons or quark pairs) that also emerge from
the short distance subdiagram.
This leads precisely to an expansion in terms of
local operators, creating heavy quark
pairs in states $n$, labelled $c\bar{c}[n]$, where
in general $[n]$ also labels the term in the 
expansion in relative velocity and light parton quanta,
so that the corresponding operator, $\psi^\dagger(0){\mathcal \kappa}'_n\chi(0)$,
which always includes a quark pair, may also 
create light quanta.  
Each such operator will be accompanied by the
sum of all hard subdiagrams, evaluated at
zero relative velocity and at zero light parton
momentum.  We will refer to this sum as the
hard scattering,
or coefficient, function for operator $\psi^\dagger(0){\mathcal \kappa}'_n\chi(0)$.

 Combining the expansions from the amplitude and
 its complex conjugate,
 we derive Eq.\ (\ref{combofact}), with operators that 
 describe the creation of a heavy quark pair from the
 vacuum, summing over all final states that include hadron $H$.
 The general form  of these operators
  \cite{bodwin94} is
  \ba
{\mathcal O}^H_n(0)
&=&
\sum_N\ \chi^\dagger(0){\mathcal \kappa}_n\psi(0)\, \left | N,H\right\rangle\,
\left\langle N,H \right|\, \psi^\dagger(0){\mathcal \kappa}'_n\chi(0)
\nonumber\\
&=&
\chi^\dagger(0){\mathcal \kappa}_n\psi(0)\, \left(a^\dagger_Ha_H\right)\,
\psi^\dagger(0){\mathcal \kappa}'_n\chi(0)
\, ,
\label{Ondef1}
\ea
where the insertion of the creation operator $a^\dagger_H$, which is understood
to act on out states to produce hadron $H$,
and its conjugate enable us to sum over the complete set of out
states between the creation and annihilation operators.
The first form defines the sum over
final states appropriate to quarkonium production, while
the second form is a convenient shorthand.

The following discussion is an attempt to analyze the
basic assumption that enables us to expand in $v$
in this manner.  That is, we will begin with the general
momentum region illustrated by
Fig.\ \ref{fragphyspict} and explore the reduction to 
the simpler ``topologically factorized" picture of Fig.\ \ref{physaftercancel},
by testing the cancellation of soft exchanges between the
heavy quarks and semi-hard gluons, or equivalently,
the Wilson line.

\subsection{Operators and gauge completion}

 Our first observation, already described in \cite{nayak05}, is that matrix
 elements of the form (\ref{Ondef1}) are not invariant under operator-valued
 gauge transformations.   In general, the onium creation operators
 $a_H$ and $a^\dagger_H$, which act on out states,  
 need not commute with gauge transformations carried out
 at the origin, even though they are themselves color singlets.
As a result, it seems most natural to us to modify the
operators (\ref{Ondef1}) to provide a form precisely analogous to 
the gauge-invariant definitions of fragmentation functions
in Eq.\ (\ref{fragfn1}) above,
        \ba
{\mathcal O}^H_n(0)
\to
\chi^\dagger(0){\mathcal \kappa}_{n,c}\psi(0)\, \Phi_l^{(A)}{}^\dagger(0){}_{cb}\, 
\left(a^\dagger_Ha_H\right)\,
\Phi_l^{(A)} (0)_{ba}\, \chi^\dagger(0) {\mathcal \kappa}'_{n,a}\psi(0)\, ,
\label{replace}
\ea
in terms of ordered exponentials, defined as in
Eq.\ (\ref{oexp}).  In the (complex conjugate) amplitudes, (anti)time-ordering
is understood.
We emphasize that such a redefinition is not
required for self-consistency.  If one can demonstrate
NRQCD factorization in terms of operators in any specific gauge, 
a gauge-dependent definition of the operator matrix elements
is admissible, as long as the gauge-dependence
is not infrared sensitive.   Indeed, this is the case
for fragmentation functions, because of the 
cancellation of infrared divergences in final-state interactions
at high $P_T$, as observed above.  In the absence of 
a similar demonstration of infrared finiteness for
the refactorization (\ref{combofact}) of fragmentation
functions in terms of NRQCD operators, however,
it seems natural to entertain (\ref{replace}) 
as a plausible replacement.   We now turn to the
expansion in relative velocity, which will enable us
to test our suggestion.

           \section{Velocity expansion}

      \subsection{Requirements for NRQCD factorization}

\begin{figure}[h]
\begin{center}
\epsfig{figure=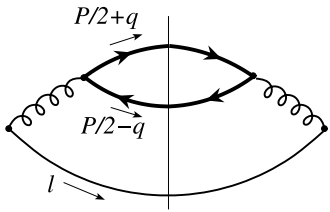,width=0.25\textwidth}
\caption{ Lowest-order fragmentation function for $g\to c\bar{c}$.  
There are no interactions on the eikonal quark pair or the Wilson line 
that corresponds to
an eikonal gluon of four-velocity $l$.  
\label{lofig}}
\end{center}
\end{figure}

  To study the role of soft gluon emission in heavy quarkonium 
  production, we will analyze
  infrared divergences in the production amplitude 
  for two heavy quarks, of total momentum $P$ and 
 relative momentum, $q$: 
 \ba
 P_{1}= \frac{P}{2} + q \equiv p+q \qquad P_{2}= \frac{P}{2} - q \equiv p-q\, .
 \label{p12def}
 \ea
 That is, we study the process $g\to c\bar{c}[{n_0}]+X$ with
$c\bar{c}[{n_0}]= c(p+q)\bar c(p-q)$.
   The lowest-order diagram for this fragmentation function is
  shown in Fig.\ \ref{lofig}.  It consists of a single gluon  
  splitting into the quark-anti-quark pair.  

In the following, we will study infrared divergences in soft gluon
corrections to 
this process, when the quark-antiquark pair is created
as a color octet, but is restricted to a singlet
in the final state.   Otherwise, we sum over all perturbative final states.

To form a heavy quarkonium, of course, these quarks cannot be truly on-shell.
Rather, they are off-shell by an energy of order $q^2/m_c$, characteristic
of a Coulomb bound state \cite{Luke:1999kz}.   
These are nonperturbative effects, however,
while coefficient functions are calculated in perturbation theory.   The
cancellation of divergences, and/or their matching to matrix elements
 in soft-gluon corrections to Fig.\ \ref{lofig}
is a necessary condition for NRQCD factorization.
Any remaining  divergences would be a violation
of factorization.  In our calculation below, we will  
   find uncanceled divergences at NNLO for conventional operators, which, however, may
   be absorbed into gauge-completed NRQCD operators.

 \subsection{Expansion in the eikonal approximation}  

 Because our calculation will be carried out with on-shell quarks,
 we can use the eikonal approximation for the coupling
 of soft gluons to the quarks in order to identify infrared divergences in the cross section.  
 Equivalently, we may
 treat the quarks in heavy-quark effective theory to leading order in their mass.
  Yet another equivalent, and for us particularly convenient, approach is to
  replace the quarks by path-ordered exponentials, similar
  to Eq.\ (\ref{oexp}) above, but now with time-like velocities representing the quark and antiquark.

  The dimensions of the velocity in ordered exponential (\ref{oexp}) can be
 shifted by a change of variables in the parameter $\lambda$.
For this reason, 
we are free to identify the quark velocities directly with their momenta
   $P_{1,2}=(P/2\pm q)$. 
   At fixed (and unequal) values of $P_1$ and $P_2$, 
  all infrared divergences can be found by the eikonal approximation.  
    The eikonal approximation and hence infrared divergences
   are completely independent of any spin
  projections that we may make on the state of the quark pair.
  As a result,  soft gluon emission separates from all dynamical
  factors that involve the spins of the quarks, and enters
  as a multiplicative factor.   We will come back to the limit $P_1=P_2$ 
  below.

  We can classify the eikonal infrared-sensitive factors by the
  color of the $c\bar c$ pair at creation (the origin of the ordered exponentials)
  and  by their color in the final state.  Gluon emission, of course, will mix these states.
  For an NRQCD-like factorization to hold, if we fix
  the color of the $c\bar c$ pair in the final state, infrared divergences
  either cancel or can be matched with matrix elements \cite{bodwin94}.   Finite remainders
  will be associated with coefficient functions, as in 
  the NLO calculations of Refs.\ \cite{ma94}-\cite{lee05}

  In summary, we will study the infrared factor
  associated with the creation of a $c\bar c$ pair in an octet configuration,
  and its evolution into a singlet in the final state.
  This infrared factor may be written in the notation
  of Eq.\ (\ref{oexp}) as
  \ba
{\cal I}^{(8\to 1)}(P_1,P_2)
&=& 
\sum_N
<0|\, \left [\Phi_{P_2}^{(\bar{q})}{}^\dagger (0)\right]_{IJ} \left[T_d\right]_{JK}\ 
\left[\Phi_{P_1}^{(q)}{}^\dagger(0) \right]_{KI}\, 
\Phi_l^{(A)}{}^\dagger (0)_{db}\, \left| N\right\rangle
\nonumber\\
&\, & \hspace{20mm} \times
\left< N \right| 
\Phi_l^{(A)} (0)_{bc}\, 
\left[\Phi_{P_1}^{(q)}(0)\right]_{LM} \left[T_c\right]_{MN} 
\left[ \Phi_{P_2}^{(\bar{q})}(0)\right]_{NL} 
\, |0\rangle \, ,
\label{sigma81}
\ea
where we have exhibited all color indices: those in
adjoint representation by $a,\, b \dots$, and those in
the fundamental representation by $I,\, J\dots $, to indicate
the trace structure, which imposes a color singlet configuration
in the final state.  

The operator $\Phi^{(\bar q)}$ is the
ordered exponential that represents the antiquark.
It has the opposite sign on the coupling compared to the quark operator, and
has color matrices ordered in the reverse sense
to time ordering.  In the notation of the standard definition, Eq.\ (\ref{oexp}),  we represent
this matrix ordering  by $\bar \P$, and define 
\bea
\Phi_{P_2}^{(\bar{q})}(0)
= \bar \P \exp \left[ ig\int_0^\infty d\lambda\,  P_2\cdot A^{(q)}(P_2\lambda)\right]\, .
\label{Phiqbar}
\ea
Here  $A^{(q)}_\nu \equiv \sum_a T_a\, A_{\nu,\, a}$ is the
matrix-valued field in the quark fundamental representation.
For classical fields, $\Phi_n^{(\bar q)}(0)$ is the  hermitian conjugate
of $\Phi_n^{(q)}(0)$.
In Eq.\ (\ref{sigma81}) and below,
overall time-ordering of the field operators is understood in the amplitude,
and anti-time ordering in its complex conjugate.
For explicit computations,
we restrict the sum over final states $N$ in Eq.\ (\ref{sigma81}) to 
soft gluon emission only.

\begin{figure}[h]
\begin{center}
\epsfig{figure=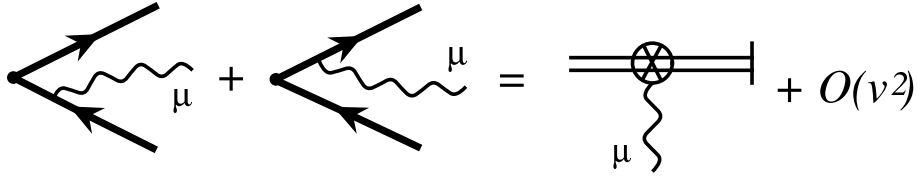,width=0.7\textwidth}
\caption{ Velocity expansion.  The heavy solid lines represent heavy
quark propagators in the eikonal approximation.
\label{velexp}}
\end{center}
\end{figure}

  The graphical rules for the interactions of gluons with
  the ordered exponentials are exactly the same as the
  eikonal approximation, and
  propagators and vertices are given by 
  \bea
  \frac{i}{(\beta\cdot k+i\epsilon)} \quad  ,
  \quad
  \pm ig_sT_a\beta^\mu\, , 
 \label{eikrules}
 \eea
with the plus for antiquarks and the minus for quarks
on the vertex  and with
$\beta^\mu$ the time-like quark four-velocity.   
The quark and antiquark eikonal propagators are represented as
heavy lines on the left-hand side of Fig.\ \ref{velexp}. 
In this notation, Eq.\ (\ref{sigma81}) describes
a product of color traces in the fundamental representation.
Our ability to use the same notation for velocities
as for momenta is manifest since the combination of 
each eikonal vertex and propagator is
scale-invariant.  In the pair rest frame
the {\it relative velocity} of the members of the pair is
proportional to the ratio $q^2/m_c^2 = 4q^2/P^2$.  

 In the spirit of NRQCD analysis, and because it leads
 to some simplification, we will study corrections to Fig.\ \ref{lofig}
 to order $q^2$, which is the first nontrivial order.
 At zeroth order in $q^2$, the quark and antiquark  
    never separate, and all infrared divergences cancel,
    since there are no color multipoles to which they can couple.
    We can see this in Eq.\ (\ref{sigma81}), in which both
    the amplitude and complex conjugate amplitude reduce to
    unity in the limit $P_1= P_2\to P/2$.  
    This is easily proved by considering the $A(x)$-field with
    the largest time in the amplitude.  This field may come
    either from the quark exponential, $\Phi_{P_1}^{(q)}(0)$,
    or the antiquark exponential $\Phi_{P_2}^{(\bar{q})}(0)$.
    When $P_1=P_2$, the only difference between these
    two terms is the relative minus sign between the quark
    and antiquark vertices.   Every such pair of terms cancels pairwise.
    An identical argument applies to the complex conjugate amplitude, and
        there is therefore no overall $q$ term, and $q^2$ can be
    reached only by expanding both the amplitude and its
    complex conjugate to order $q$ independently.

    The expansion to order $q$ is straightforward, and has
    a nice interpretation in terms of fields.  We start with the
    expansion for the individual ordered exponentials,
    \bea
    q^\nu\frac{\partial}{\partial p^\nu}\ \Phi_{p}^{(q)}(0)
    && =
    -ig\int_0^\infty d\lambda'\,  \lambda' \,
   \P \exp\left[ -ig\int_{\lambda'}^\infty d\lambda\; p\cdot  A^{(q)}(\lambda p)\right]
    \nonumber\\
    && \hspace{20mm} \times\; \left[\, p^\mu q^\nu F_{\nu\mu,a}(\lambda' p)T_a\, \right]
    \; \P\exp\left[ -ig\int_0^{\lambda'}d\lambda\; p\cdot  A^{(q)}(\lambda p)\right]
    \nonumber\\
    - q^\nu\frac{\partial}{\partial p^\nu}\ \Phi_{p}^{({\bar q})}(0)
    && =
   -  ig\int_0^\infty d\lambda'\,  \lambda' \, T\, \Bigg \{\,
    \bar \P\exp\left[ ig\int_0^{\lambda'}d\lambda\; p\cdot  A^{(q)}(\lambda p)\right]
    \nonumber\\
    && \hspace{20mm} \times\; \left[\, p^\mu q^\nu F_{\nu\mu,a}(\lambda' p)T_a\, \right]
    \;  \bar \P\exp\left[ ig\int_{\lambda'}^\infty d\lambda\; p\cdot  A^{(q)}(\lambda p)\right]
    \Bigg\}
    \, ,
    \label{derivoexp}
    \ea
    where the explicit minus sign on the left in the second expression anticipates that 
    we will be expanding in the momentum of the anti-quark, $P_2 = p-q$.
    In both of these  expressions, the time-ordering is from
   right (earlier) to left (later), with an (opposite)
    identical ordering of color matrices for the (anti) quark exponential.
    We have inserted an explicit $T$ in the antiquark expression, to
    remind ourselves that the operators and color matrices have the opposite
    ordering in this case.
    The operator $F_{\mu\nu,a}$ is the gluon field strength with
    tensor and color indices.  Note the overall
    factor of $\lambda'$, which reflects the increasing separation
    of the quark and antiquark paths with increasing distance from
    the origin when $q$ is changed by a constant amount.

We now apply Eq.\ (\ref{derivoexp}) to the amplitudes in Eq.\ (\ref{sigma81}).
Expanding $P_1$ and $P_2$ about $p\equiv P/2$ we find
    \bea
   q^\nu\frac{\partial}{\partial q^\nu}\  
    \left< N \right|  \Phi_l^{(A)} (0)_{bc}\,
\left[\Phi_{P_1}^{(q)}(0)\right]_{LM} \left[T_c\right]_{MN} 
\left[ \Phi_{P_2}^{(\bar{q})}(0)\right]_{NL} \ \, |0\rangle\bigg|_{q=0}
&=&
\nonumber\\
&\ & \hspace{-85mm}
-ig \left< N \right| \,  \Phi_l^{(A)} (0)_{bc}\,
  \left[ \P\exp\left[ -ig\int_0^\infty d\lambda\, p\cdot  A^{({q})}(\lambda p)\right]\, \right]_{LM}
  \left[T_c\right]_{MN}
  \nonumber\\
  &\ &  \hspace{-70mm} \times\, 
  \int_0^\infty d\lambda' \,  \lambda' \,
  \Bigg \{\,
    \bar \P\exp\left[ ig\int_0^{\lambda'}d\lambda\; p\cdot  A^{(q)}(\lambda p)\right]
    \nonumber\\
    && \hspace{-70mm} \times\; \left[\, p^\mu q^\nu F_{\nu\mu,a}(\lambda' p)T_a\, \right]
    \;  \bar \P\exp\left[ ig\int_{\lambda'}^\infty d\lambda\; p\cdot  A^{(q)}(\lambda p)\right]
    \Bigg\}
_{NL}\, \, |0\rangle
\nonumber\\
&\ &
\hspace{-85mm} 
-ig \left< N \right| \,  \Phi_l^{(A)} (0)_{bc}\,   \int_0^\infty d\lambda' \,  \lambda' \,
 \Bigg\{   \P\exp\ \left[ -ig\int_{\lambda'}^\infty d\lambda\, p\cdot  A^{(q)}(\lambda p)\right] 
    \nonumber\\
    && \hspace{-70mm} \times\; \left[\, p^\mu q^\nu F_{\nu\mu,a}(\lambda' p)T_a\, \right]
    \;  \P\exp\left[ -ig \int_0^{\lambda'}d\lambda\, p\cdot  A^{(q)}(\lambda p)\right]\Bigg\}_{LM}
 \left[T_c\right]_{MN}
 \nonumber\\
&\ &  \hspace{-70mm} \times
  \left[ \bar \P \exp\left[ ig\int_0^\infty d\lambda\, p\cdot  A^{(q)}(\lambda p)\right]\, \right]_{NL}\, 
  |0\rangle\, .
  \label{expandprod1}
\eea
Here again, time ordering is understood for all field
operators.  
The lowest order of the expansions of the
left- and right-hand sides of Eq.\ (\ref{expandprod1}) are shown
graphically in Fig.\ \ref{velexp}.
On the right, the vertex associated with the field strength 
in the final form is represented by $\otimes$.

    We next apply reasoning similar to that which led to the cancellation of the ordered
    exponentials at $q=0$.  Again, consider the $A$-field with largest variable $\lambda$,
    assuming that there is at least one such field with $\lambda>\lambda'$, that is,
    at least one field at a larger time than the field strength $p^\mu q^\nu F_{\nu\mu}$.
    We recognize that whenever we find such a field, there is a cancellation between 
    the cases when that field is associated with the quark and
    antiquark ordered exponentials.   
    All fields at times  greater than that of the field strength cancel, and we have
    \bea
   q^\nu\frac{\partial}{\partial q^\nu}\  
    \left< N \right|  \Phi_l^{(A)} (0)_{bc}\,
\left[\Phi_{P_1}^{(q)}(0)\right]_{LM} \left[T_c\right]_{MN} 
\left[ \Phi_{P_2}^{(\bar{q})}(0)\right]_{NL} \ \, |0\rangle\bigg|_{q=0}
&=&
\nonumber\\
&\ & \hspace{-95mm}
-ig \left< N \right| \,  \Phi_l^{(A)} (0)_{bc}\,  \int_0^\infty d\lambda' \,  \lambda' \,
  \left[ \P\exp\left[ -ig\int_0^{\lambda'} d\lambda\, p\cdot  A^{({q})}(\lambda p)\right]\, \right]_{LM}
  \left[T_c\right]_{MN}
  \nonumber\\
  &\ &  \hspace{-70mm} \times\, 
  \Bigg \{\,
    \bar \P\exp\left[ ig\int_0^{\lambda'}d\lambda\; p\cdot  A^{(q)}(\lambda p)\right]
    \; \left[\, p^\mu q^\nu F_{\nu\mu,a}(\lambda' p)T_a\, \right]
    \;  
    \Bigg\}
_{NL}\, \, |0\rangle
\nonumber\\
&\ &
\hspace{-115mm} 
-ig \left< N \right| \,  \Phi_l^{(A)} (0)_{bc}\,   \int_0^\infty d\lambda' \,  \lambda' \,
 \Bigg\{ \; \left[\, p^\mu q^\nu F_{\nu\mu,a}(\lambda' p)T_a\, \right]
    \;  \P\exp\left[ -ig \int_0^{\lambda'}d\lambda\, p\cdot  A^{(q)}(\lambda p)\right]\Bigg\}_{LM}
 \left[T_c\right]_{MN}
 \nonumber\\
&\ &  \hspace{-70mm} \times
  \left[ \bar \P \exp\left[ ig\int_0^{\lambda'} d\lambda\, p\cdot  A^{(q)}(\lambda p)\right]\, \right]_{NL}\, 
  |0\rangle
  \nonumber
  \eea
  \bea
  && \hspace{5mm} =
  -2ig \left< N \right| \,  \Phi_l^{(A)} (0)_{bc}\,   \int_0^\infty d\lambda' \,  \lambda' \, {\rm Tr}\,
 \Bigg\{ \; \left[\, p^\mu q^\nu F_{\nu\mu,a}(\lambda' p)T_a\, \right]
  \nonumber\\
&&  \hspace{10mm} \times
    \;  \P\exp\left[ -ig \int_0^{\lambda'}d\lambda\, p\cdot  A^{(q)}(\lambda p)\right]\;
 T_c\;
   \bar \P \exp\left[ ig\int_0^{\lambda'} d\lambda\, p\cdot  A^{(q)}(\lambda p)\right]\, \Bigg\}\,
  |0\rangle\, .
  \label{expandprod2}
\eea
In the last line we have used the cyclic nature of the color trace
and the anti-path ordering of the antiquark exponential
to show that the two terms above are equal.
Next, we note that were it not for the generator $T_c$, we could
use the same reasoning as above to show that the $A$-field of lowest
$\lambda'$ cancels between the quark and antiquark  exponentials.
The nonvanishing remainder, therefore, is a commutator,
\bea
-ig A_d\left(\lambda_{\mathrm min}p\right)\left[\, T_d, T_c\,  \right] 
&=&
g A_d\left(\lambda_{\mathrm min}p\right)\, f_{dce} T_e
\nonumber\\
&=& -ig A_d\left(\lambda_{\mathrm min}p\right)\, [T_d^{(A)}]_{ec}\, T_e\, ,
\nonumber\\
&=& T_e\, \left(-ig A^{(A)}\left(\lambda_{\mathrm min}p\right)\, \right)_{ec} \, ,
\label{convertcolor}
\eea
where $[T^{(A)}_d]_{ec}= -if_{dec}$ is a generator in the adjoint representation.
In effect, the gluon field is converted from the fundamental representation to
the adjoint.

At any order in $g$, this procedure may be repeated
until all $A$-fields from the remaining ordered exponentials have
been converted from fundamental to
adjoint representation in  Eq.\ (\ref{expandprod2}).
The final color trace in fundamental representation is trivial
(and gives $1/2$),
and we derive the relatively simple  form
   \bea
      q^\nu\frac{\partial}{\partial q^\nu}\  
    \left< N \right|  \Phi_l^{(A)} (0)_{bc}\, 
\left[\Phi_{P_1}^{(q)}(0)\right]_{LM} \left[T_c\right]_{MN} 
\left[ \Phi_{P_2}^{(\bar{q})}(0)\right]_{NL} 
\, |0\rangle\Bigg |_{q=0}
&=&
\nonumber\\
&\ & \hspace{-115mm}
- ig   \int_0^\infty d\lambda' \,  \lambda' \, \left< N \right| \,  \Phi_l^{(A)} (0)_{bc}\,
   \left[\, p^\mu q^\nu F_{\nu\mu,a}(\lambda' p)\, \right]
     \Bigg\{   \;
       \P\exp\left[ -ig \int_0^{\lambda'}d\lambda\, p\cdot  A^{(A)}(\lambda\, p)\right]\Bigg\}_{ac}\, \, |0\rangle
    \nonumber\\
    &\ & \hspace{-95mm}
= - ig   \int_0^\infty d\lambda' \,  \lambda' \, \left< N \right| \,  \Phi_l^{(A)} (0)_{bc}\,
   \left[\, p^\mu q^\nu F_{\nu\mu,a}(\lambda' p)\, \right]
    \Phi_p^{(A)}(\lambda')_{ac}\, \, |0\rangle\, ,
    \label{expandprod3}
    \eea
    in which all fields are in adjoint representation.  There is only
    a single ordered exponential, linking the gauge index at the origin ($c$) 
    with the field strength at the variable  point $\lambda'$.  
    Note that the index $c$ of $\Phi_p^{(A)}$ is itself linked to the final state by
    the auxiliary ordered exponential that we have added in
    the $l$ direction through gauge completion, as described above and in Ref.\ \cite{nayak05}.

    To derive contributions of order $v^2$, or equivalently $q^2/m_c^2=4q^2/P^2$, we will study the
    nonlocal matrix element
    \bea
    {\cal I}_2(p,q) &\equiv&
    \nonumber\\
    &\ & \hspace{-20mm}
    \sum_N
     \int_0^\infty d\lambda' \,  \lambda' \, \left< 0 \right| \,  \Phi_l^{(A)}{}^\dagger (0)_{bc'}\,
     \Phi_p^{(A)}(\lambda'){}^\dagger_{a'c'}\,
        \left[\, p^\mu q^\nu F_{\nu\mu,a'}(\lambda' p)\, \right]\,
       \left|  N\right\rangle
       \nonumber\\
       &\ & \hspace{-15mm} \times  \left\langle N\right|
    \int_0^\infty d\lambda' \,  \lambda' \,  \Phi_l^{(A)} (0)_{bc}\,
   \left[\, p^\mu q^\nu F_{\nu\mu,a}(\lambda' p)\, \right]
     \Phi_p^{(A)}(\lambda')_{ac}\, \, |0\rangle\, .
            \label{M2def}
       \eea
       As above, (anti-) time ordering is implicit in the (complex conjugate) amplitudes.
    This is the complete  ${\cal O}(v^2)$ result for ${\cal I}^{(8\to 1)}$, Eq.\ (\ref{sigma81}),
    because, as we have seen
    above, at  order  ${\cal O}(v^0)$, the quark and anitquark eikonal lines in both 
    the amplitude or its complex conjugate cancel completely,  and thus decouple
    for soft radiation.  

\begin{figure}[h]
\begin{center}
\epsfig{figure=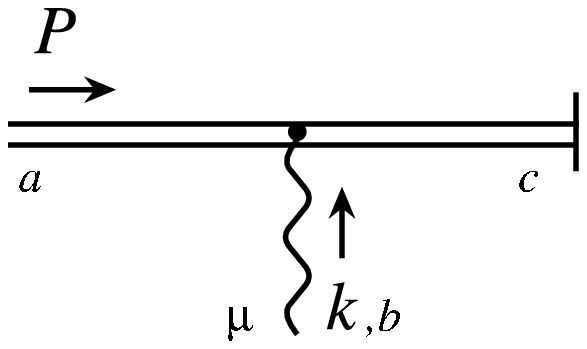,width=0.2\textwidth}
\hfil
\epsfig{figure=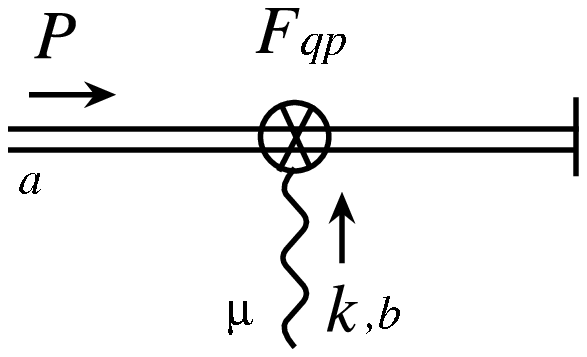,width=0.2\textwidth}
\hfil
\epsfig{figure=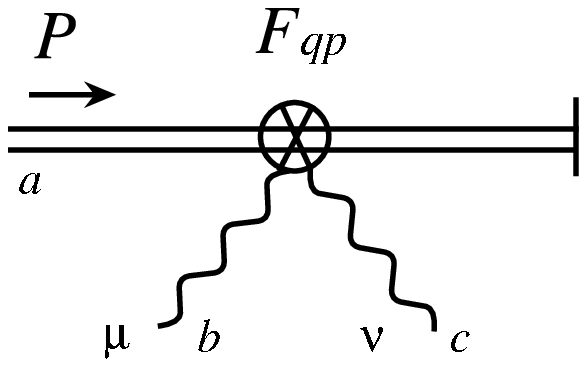,width=0.2\textwidth}
\hfil
\epsfig{figure=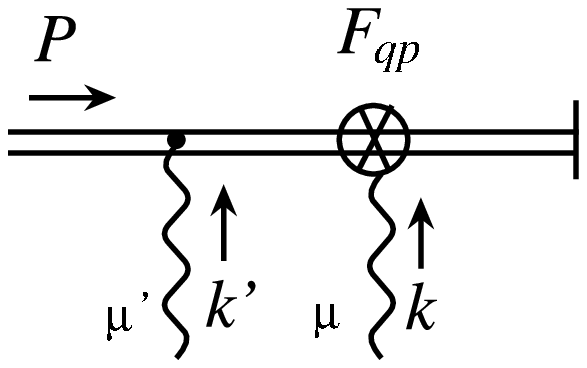,width=0.2\textwidth}

(a) \hskip 0.21\textwidth
(b) \hskip 0.21\textwidth
(c) \hskip 0.21\textwidth
(d)
\caption{a) Eikonal vertex; (b,c) vertices for the field strength; (d) line
with eikonal and field strength interactions.
\label{Fvertices}} 
\end{center}
\end{figure}

    In the following, we will study the expansion of Eq.\ (\ref{M2def}) to
    NNLO.   
   The explicit factor of $\lambda'$ in Eq.\ (\ref{expandprod3}) modifies the
eikonal propagators.  To see this, we can formally evaluate the $\lambda'$ integral in
Eq.\ (\ref{expandprod3}) in terms of the Fourier  transform of the field strength,
$\tilde F_{\nu\mu,a}(k) \equiv \int d^4x F_{\nu\mu,a}(x) \exp [ - ik\cdot  x]$.  With
this convention, momentum $k$ flows into the field strength (and hence
out of the eikonal lines).   For a given  order
in the expansion of the adjoint ordered exponential in Eq.\ (\ref{expandprod3}),
the lower limit of the $\lambda'$ integral is
some value $\lambda_m$, the maximum value of $\lambda$ 
in the ordered fields $p\cdot A^{(A)}(\lambda p)$
from $\Phi_p^{(A)}$.  The relevant integral is then
\bea
-ig  \int_{\lambda_m}^\infty d\lambda' \,\lambda'
\int {d^4k \over (2\pi)^4} \,  {\rm  e}^{i\lambda'(n\cdot k+i\epsilon)}\, \tilde F_{\nu\mu,a}(k)
&=& \nonumber\\
&\ & \hspace{-50mm}
-ig \int {d^4k \over (2\pi)^4} \,  {\rm  e}^{i\lambda_m(n\cdot k+i\epsilon)}\, \tilde F_{\nu\mu,a}(k)\,
\left[\, \lambda_m\, \frac{i}{n\cdot k + i\epsilon} - \frac{1}{(n\cdot k+i\epsilon)^2}
\, \right]\, ,
\eea
where we  have integrated by parts.
The second term gives
a squared eikonal propagator. 
The first (boundary) term  in brackets on the right-hand side gives the  standard
eikonal propagator of Eq.\ (\ref{eikrules}), times a factor of $\lambda_m$,
producing a similar pattern in the next integral.    The next integral will again give a    
squared propagator plus a boundary term, until the final $\lambda$ integral,
for which the lower limit is zero and the boundary term vanishes.
 The  result for a specific diagram is to replace the standard product
of eikonal propagators by a sum of terms, in each of which one of the propagators 
is squared.    Vertices for the operators $p\cdot A$ are
unchanged.  The relevant
 graphical notations for vertices  are shown in Fig.\ \ref{Fvertices}. 
  The three-point field strength vertex may
    be represented as 
    \bea
    U^\mu_{F,ac}(p,q,k) = - g\, \delta_{ac}\, \left( (p\cdot k) q^\mu - (q\cdot k) p^\mu\right)\, ,
    \label{Udef}
    \eea
    and the four-point vertex as
    \bea
    W^{\mu\nu}_{F,abc}(p,q) =
  ig^2  f_{abc}\, \left(\, p^\mu q^\nu -  q^\mu  p^\nu\, \right)\, .
    \ea
    In both cases, $c$ represents the color factor of the field strength
    tensor of Eq.\ (\ref{expandprod3}), while $a$ and/or $b$ are
    the color indices of the gluon(s) that couple to the field strength. 
    Because the adjoint eikonal lines end at the field strength    in (\ref{expandprod3}), 
    corresponding to the color singlet  pair in the final state,
    the three- and four-point vertices have
    only two and three color indices, respectively.

As an example, corresponding to Fig.\ \ref{Fvertices}d, we have the expression
\bea
g\, P^{\mu'}\ U^\mu_{F,ac}(P,q,k)\; \left[ \; \frac{1}{-P\cdot k'+i\epsilon}\,  \frac{1}{(-P\cdot k + i\epsilon)^2}
+   \frac{1}{(-P\cdot k' + i\epsilon)^2}\, \frac{1}{-P\cdot k+i\epsilon}\;
\right]\, ,
\label{doublepoles}
\eea
where we have chosen the sign of the infinitesimal imaginary part
appropriate to the amplitude.  To avoid clutter in and proliferation of figures, we will
not introduce a graphical notation for squared propagators, but 
simply assume that the sum over terms is carried out in every diagram
with a field strength operator at  the largest time.

The three-point vertex, $U^\mu_F$ in Eq.\ (\ref{Udef}) is just the momentum representation
of the Maxwell term of the field strength.   It thus trivially decouples from
scalar-polarized gluons,
\bea
U^\nu_F(p,q,k) k_\nu = 0\, .
\label{Utransverse}
\eea
This result will lead to considerable simplification in our
calculations; in particular, it eliminates, on a diagram-by-diagram
basis, collinear poles associated with the octet eikonal
line in the $\ell$ direction.  This is because in any
covariant gauge, collinear divergences are associated
with gluons whose polarization is proportional to their
momenta in the collinear limit \cite{Ste78}.

In the next two sections, we apply these rules to study
the coupling of soft gluons to the heavy quark pair.  We begin
at NLO, and  then generalize to NNLO.

 \section{Next-to-leading Order}

\begin{figure}[h]
\begin{center}
\epsfig{figure=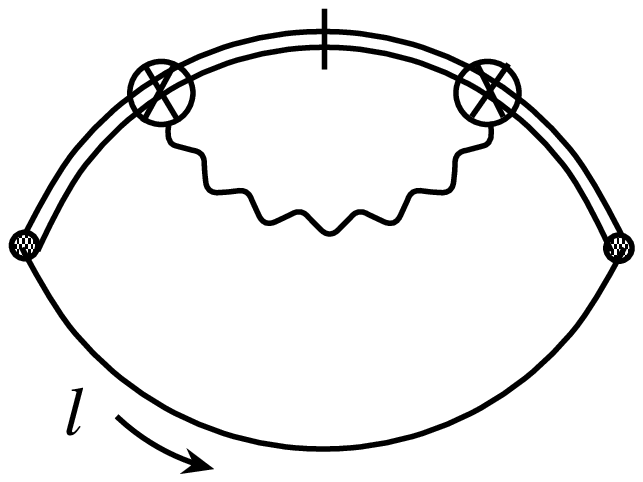,width=0.23\textwidth}
\hskip 0.1\textwidth
\epsfig{figure=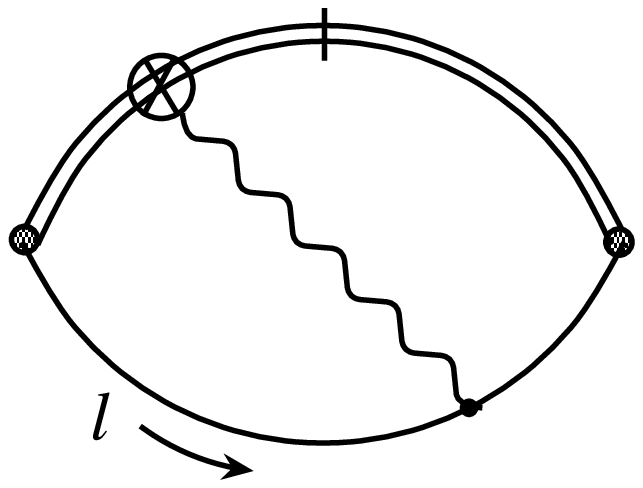,width=0.23\textwidth}

(a) \hskip 0.29\textwidth (b)
\caption{Representative NLO contributions to $g\to c\bar{c}$ fragmentation in 
eikonal approximation are found by all cuts of these diagrams.
In these figures, the parent gluon is contracted to a point, represented
by the dark circle, because it is off-shell by
order $m_c$. \label{nlofig} .}
\end{center}
\end{figure}

  Figs.\ \ref{nlofig}a and b illustrate
  the origin of infrared divergences
  in the fragmentation function at next-to-leading order in $\as$
  to order $v^2$.  This infrared structure is the same as the
  lowest order contribution to ${\cal I}_2$, Eq.\ (\ref{M2def}).
  As in Fig.\ \ref{velexp}, the sum over
gluon connections to quark and antiquark on each side of the cut 
has been replaced by 
a single field-strength vertex.
  Because the parent gluon is off-shell by order $m_c$,
  we may contract it to a point to study 
  soft gluon corrections.  For this purpose, it is
  then equivalent to study the matrix elements (\ref{sigma81}) 
  to next-to-leading order, and that is how we shall
  describe our calculation below.  We emphasize, however,
  that there is a trivial mapping from the matrix elements 
  to the fragmentation functions.

  The vertical lines in Fig.\ \ref{nlofig} represent the quark-antiquark
  pair in the final state, and a projection onto a color singlet 
  (implemented by a color trace) is understood, along
  with a sum over all connections of the gluon to the
  quark and antiquark.
The full set of diagrams is found by completing the cut, which
can be done in only one way for \ref{nlofig}a, where
the gluon must be in the final state.  For Fig.\ \ref{nlofig}b,
on the other hand, there are two
possibilities, one with a virtual gluon correction and
one with a real gluon.  In fact, of the two diagrams, only \ref{nlofig}a 
can contribute to ${\cal I}^{(8\to 1)}$, Eq.\ (\ref{sigma81}).
If we  require a color singlet pair in the final state
Fig.\ \ref{nlofig}b requires interference between 
octet and singlet in the hard scattering functions.
We consider this diagram because it follows a pattern
observed in the original arguments for NRQCD factorization,
given in Ref.\ \cite{bodwin94}, and because 
its square contributes to ${\cal I}^{(8 \to 1)}$ at NNLO.

Let us begin with Fig.\ \ref{nlofig}a, which has a 
 topologically-factorized form, in which
the soft gluon connects only to the heavy quarks, rather
than to other finite-energy final-state lines.
(In this case, the only such line is the eikonal line in direction $l$.)
From the perturbative rules described above, we
immediately write down the following integral, which is
readily evaluated in $D=4-2\varepsilon$ dimensions,
\begin{eqnarray}
\Sigma^{(8a)}(P,q)~
&=&~ 16\, g^2\mu^{2\varepsilon}\, \int \frac{d^D 
k}{(2\pi)^{D-1}}~\delta(k^2)\,
    [q_\nu (P\cdot k)-(q\cdot k) P_\nu] \nonumber\\
&& \hspace{30mm} \times
\ [q^\nu (P\cdot k) - (q\cdot k) P^\nu]
\frac{1}{[(P\cdot k)^2]^2} \nonumber\\
&=& {16\over 3}\, {\alpha_s\over \pi}\, {\vec{q}\, {}^2\over P^2}\, 
{1\over -\varepsilon} + \dots\, .
\label{nlopt}
\end{eqnarray}
Here we have suppressed color factors, including the factor
of 1/2 from the color trace mentioned above Eq.\ (\ref{expandprod3}).
The infrared pole in this result is familiar from 
NLO calculations of fragmentation \cite{ma94,braaten97,petrelli98},
in which it is matched to the relevant NRQCD matrix element.

We now turn to the cuts of
Fig.\ \ref{nlofig}b, which contribute only to
color interference terms.  These diagrams, in which the gluon connects the quark-antiquark
pair with the eikonal line, are not topologically factorized. Based
on the arguments of \cite{bodwin94}, we expect these to cancel,
and they do.  This was verified explicitly in Ref.\ \cite{ma05}
for the case of color octet pairs in the  final state.
 It will be instructive, however, to see how
this happens in our velocity-expanded form to
linear order in  $q$ with a color singlet final state, because the cancelation 
found here  will  be  relevant to NNLO.

\begin{figure}[h]
\begin{center}
\epsfig{figure=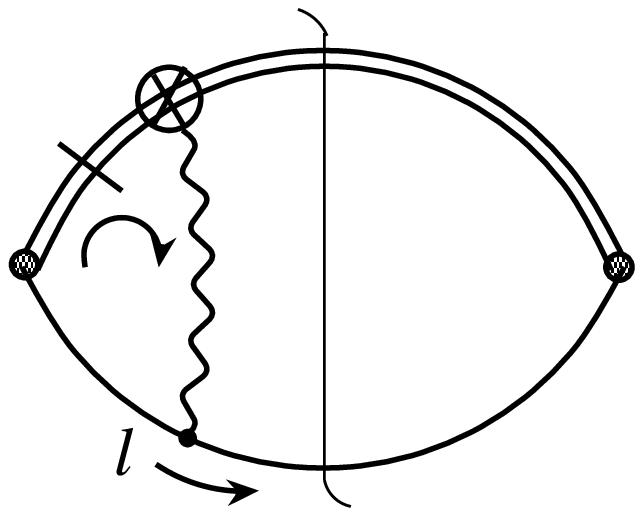,width=0.2\textwidth}
\hskip 0.05\textwidth 
\epsfig{figure=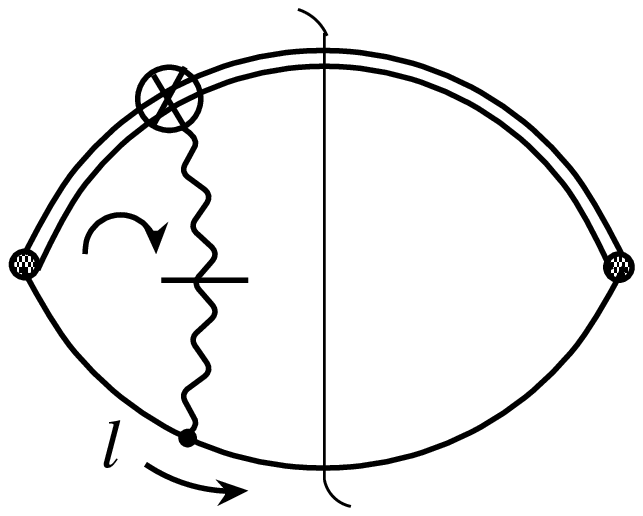,width=0.2\textwidth}
\hskip 0.05\textwidth 
\epsfig{figure=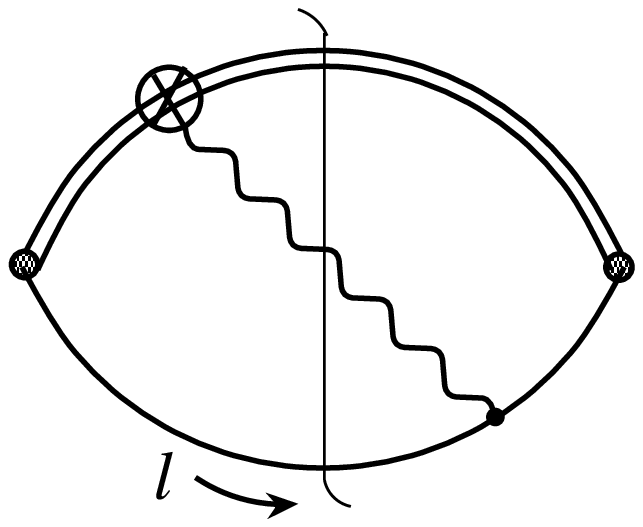,width=0.2\textwidth}
\hskip 0.2\textwidth

(a) \hskip 0.23\textwidth
(b) \hskip 0.23 \textwidth
(c)
\caption{a,b) Poles resulting from closing the minus loop
integral in cut diagrams corresponding to Fig.\ \ref{nlofig}b.  
c) Corresponding representation
of real gluon emission. \label{1looppolefig}}
\end{center}
\end{figure}

We consider first the cut diagram with a virtual
gluon loop in the amplitude.  For our purposes,
the overall normalization of the diagram is 
arbitrary, and we  write
\bea
\Sigma^{(8b)}_{\rm virtual}
=
g^2 \int \frac{d^Dk}{(2\pi)^D}\, N(P,k,q,\ell)\, {1 \over (P\cdot k+i\epsilon)^2}\;
{1 \over k^2+i\epsilon}\ {1 \over - \ell\cdot k + i\epsilon}\, ,
\label{sigma8bvirt}
\eea
with numerator factor 
\bea
N(P,k,q,\ell)
&=&
2\, \left[
q\cdot \ell\, (P\cdot k) - P\cdot \ell\, (q\cdot k)
\right] 
\nonumber\\
&=& 
\sqrt{2} \; P_0 \; \ell^- \,
\left[ \sqrt{2}k^+q_3 + q_\perp\cdot k_\perp\right]\, .
\eea
In this diagram, as in subsequent loop integrals, we will integrate 
first the minus loop momentum, by closing contours in the lower half-plane
and picking up the relevant poles.   Certain regularities and
cancellations are conveniently represented in this manner,
reducing the number of diagrams that must be computed explicitly.
The result is shown in Fig.\  \ref{1looppolefig}.
The double pole from  the quark pair octet eikonal denominator,
$(P\cdot k+i\epsilon)^2$, is always in the $k^-$
lower half-plane, while the pole of the exchanged gluon
is in the lower half-plane only when $k^+$ flows in
the direction indicated in the figure.   Closing in the lower half-plane, 
and neglecting the term odd in $k_\perp$ we find only a terms proportional to $q_3$,
\bea
\Sigma^{(8b)}_{\rm virtual,1}
&=&
 i\, 4\, \frac{q_3}{P_0}\ \frac{g^2}{(2\pi)^{D-1}}\, 
\int d^{D-2}k_\perp \int_{-\infty}^\infty dk^+\;
\Bigg \{\, {d \over dk^-}
\left({1 \over k^2+i\epsilon}\right) \Bigg |_{k^-=-k^+}  \nonumber\\
&\ & \hspace{55mm} 
+ \theta(k^+)\, \frac{1}{2k^+}\, 
{1\over (k^+{}+k_\perp^2/2k^+)^2}\, \Bigg \}\nonumber\\
&\ & \hspace{-10mm} 
=  
  i\, 4\, \frac{q_3}{P_0} \frac{g^2}{(2\pi)^{D-1}}\, \int d^{D-2}k_\perp
\left\{  \int_{-\infty}^\infty dk^+\;
{ -2 k^+ \over (2k^+{}^2 + k_\perp^2)^2} 
\; + \int_0^\infty {dk^+\over 2k^+}
{1\over (k^+ + k_\perp^2/2k^+)^2}\, \right \}\, .  
\nonumber\\
\label{nlopoleterms}
\eea
These two terms correspond to the diagrams in Fig.\ \ref{1looppolefig}a and b,
where the straight line through the eikonal or gluon line indicates 
the $k^-$  pole  chosen.
Both of these terms are logarithmically divergent by power-counting
in the soft limit.   As in the case above, however, they are 
collinear finite.   In fact, the first term on the right-hand side
is odd in $k^+$ and vanishes after symmetric integration.
The pole at $k^+=0$, which would correspond to an on-shell
intermediate state, has vanishing residue at order $q$.

In the second term of the right-hand side of Eq.\ (\ref{nlopoleterms}),
the exchanged gluon is on-shell with
positive plus momentum flowing from the heavy quark pair to the 
eikonal lines.  Its contribution to the cross section, as illustrated in
Fig.\ \ref{1looppolefig}b,
is real, and it is straightforward to verify that it cancels the
corresponding diagram for real gluon emission, illustrated by
Fig.\ \ref{1looppolefig}c, which has a
standard flow of gluon momentum ($k^+>0$) in the cut diagram.  We will use the
notation shown in these figures to help organize our NNLO
computations below.

At the level of NLO, we have found that one-loop corrections
indeed follow the expected pattern: they cancel except when
topologically-factorized, and are thus consistent with
matching to conventional NRQCD matrix elements.  The presence of
the octet Wilson line in our gauge-completed matrix elements
does not change this pattern at NLO, as observed in Ref.\ \cite{ma05}

Before going on to the details of the NNLO calculations, 
we make a comment on gauge independence. 
   As defined, the factorized fragmentation functions are gauge invariant,
   since the eikonal and the pair creation operators
   $\Phi^{(A)}_\ell\psi^\dagger {\cal K}_n\chi$ of
   Eq.\ (\ref{replace}) are contracted to form a color
   singlet vertex.  Also, as we have seen,
   because of Eq.\ (\ref{Utransverse}),  there are immediate
   cancellations of many gauge terms like $k^\mu k^\nu/(k^2)^2$
  for a gluon of momentum $k$, because of
  the field strength vertices that appear when we expand in the
   relative velocity of the quark pair.  
  We can, of course, decouple the eikonal
  gauge line entirely, by choosing an $\ell\cdot A=0$ gauge.  
  The gauge invariance of the matrix elements assure that
  the result would be the same.   For our
  purposes, however, Feynman gauge is most convenient.

   \section{The Fragmentation Function at NNLO}

In this section we study in detail the infrared behavior 
of the gauge-completed gluon fragmentation function at NNLO.
Specifically, we will study non-topologically
factorized diagrams at order $\as^2$ in Eq.\ (\ref{sigma81}).
We will find uncanceled infrared divergences for
this set of diagrams, corresponding to ${\cal O}(v^2)$ 
contributions to the fragmentation functions.
We emphasize that the same infrared poles (proportional
to $\as^2/(-\vep)$)
appear in cross sections, associated
with soft gluon exchanges between the quark
pair and a recoiling gluon.  For this reason,
the gauge-completion of matrix elements
is necessary for factorization.  At the same time,
we will observe that the infrared pole is independent
of the direction of the vector $\ell^\mu$.  
This shows that
the gauge-completed fragmentation function
is universal to NNLO.  The same fragmentation function
will match  infrared poles for 
the quark pair recoiling against a gluon
in any frame, or indeed (as we shall see),
for any set of recoiling jets at this order of soft gluon exchange.
We are not yet able to show, however, that this redefinition
is universal at all orders
in soft gluon exchange.

   \subsection{The diagrams}

\begin{figure}[h]
\begin{center}
\epsfig{figure=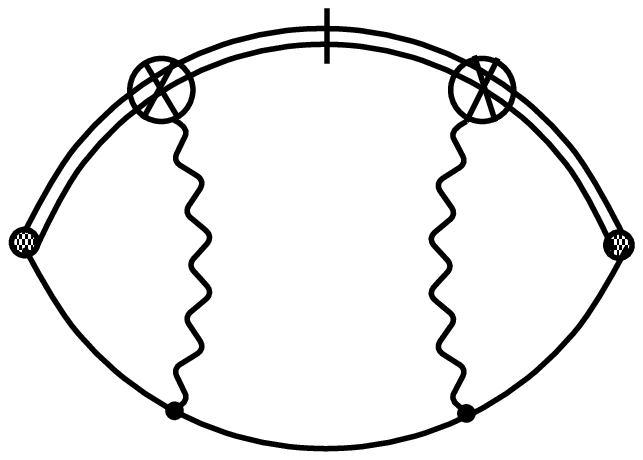,width=0.19\textwidth}
\hskip 0.14\textwidth
\epsfig{figure=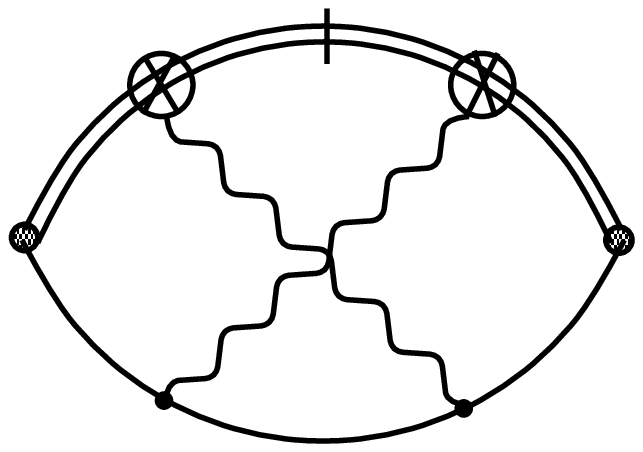,width=0.19\textwidth}
\hskip 0.14\textwidth
\epsfig{figure=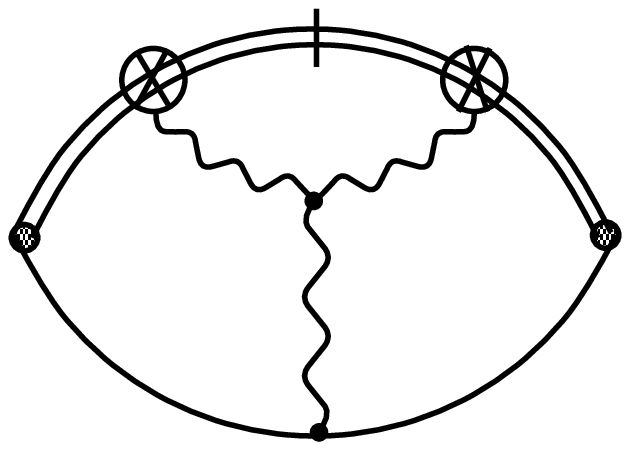,width=0.19\textwidth}

(I) \hskip 0.3\textwidth
(II) \hskip 0.3\textwidth
(III)

\epsfig{figure=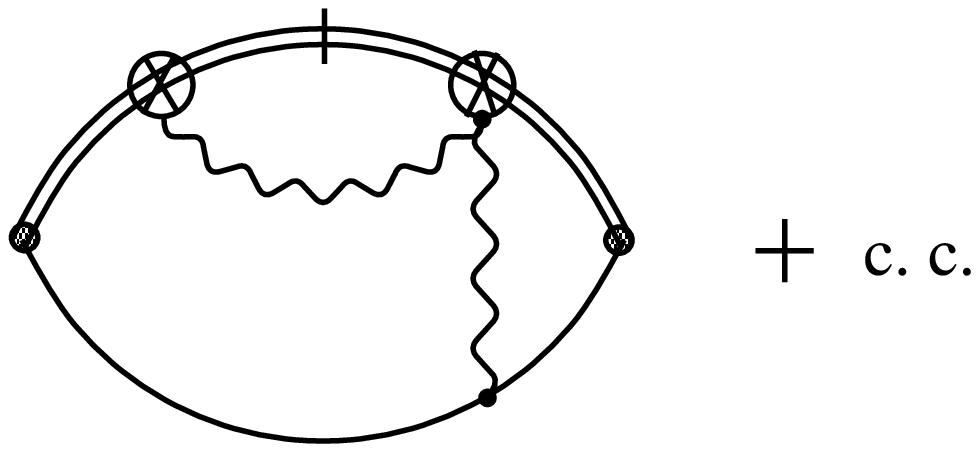,width=0.30\textwidth}
\hskip 0.03\textwidth
\epsfig{figure=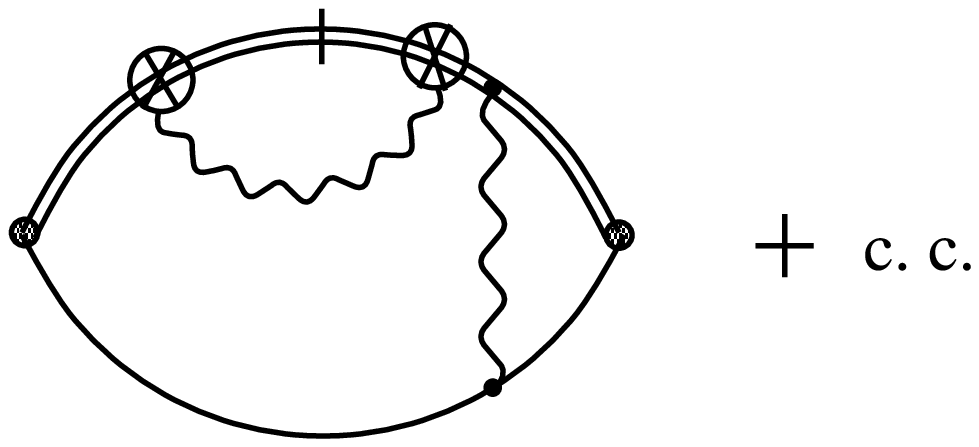,width=0.30\textwidth}
\hskip 0.03\textwidth
\epsfig{figure=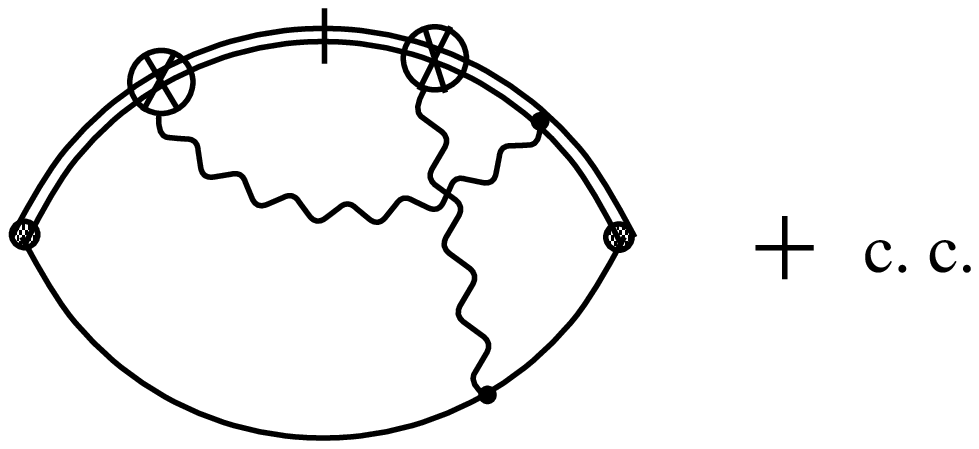,width=0.30\textwidth}

(IV) \hskip 0.3\textwidth
(V) \hskip 0.3\textwidth
(VI)

\caption{Diagrams I - VI discussed in the text. 
We sum over all cuts of these diagrams that can
produce a color singlet quark pair.   At the hard
vertex, the pair is created in an octet state.
 \label{softgluonfig}}
\end{center}
\end{figure}

   The soft-gluon diagrams that we will evaluate are shown in
   Fig.\ \ref{softgluonfig}.   
   Here, we compute only those contributions that
   correspond to the transition of a color octet pair to
   color singlet in both amplitude and complex conjugate.   
   For NRQCD factorization to hold,
   all infrared divergences should either cancel or
   factorize into octet matrix elements.

   We will discuss the diagrams of Fig.\ \ref{softgluonfig}
   one at a time.
   In evaluating each diagram, $k_1$ is defined as the momentum of
   the gluon that attaches to the quark pair at the left-most
   vertex, and it is always chosen to flow left to right in the
   diagrams.  We label the momentum of the remaining
   gluon line attached to the quark pair as $k_2$, choosing
   it to flow upward to the pair in each case.  As observed
   in Sec.\ 5, the structure of the field strength vertex 
   automatically eliminates collinear poles associated
   with soft gluons parallel  to the $\ell$ direction.

   For the purposes of this section, momenta associated
   with the quark eikonal lines, $P$, $p=P/2$ and $q$, will
   all be scaled by the quark mass, $m_c$.  In the quarkonium
   rest frame, then, we have $P = (2, \vec 0)$, $p=(1, \vec 0)$
   and $q = (0, \vec v/2)$.   

\subsection{Summary of results}

In the remainder of this section, we have given the
calculations that confirm our claims above in substantial
detail.   Since this discussion is of necessity rather detailed,
it may be useful to summarize our results at the outset.
Each of the diagrams in Fig.\ \ref{softgluonfig} contributes to the NNLO infrared
factor in three separate quantities: first, the inclusive cross section
for the production of a color-singlet charm pair of total momentum
$P$ to order $q^2$ in their relative momentum, through the fragmentation
of an off-shell gluon; second, the fragmentation function for a gluon
to the same color-singlet pair; and third, the 
gauge-completed production matrix element
Eq.\ (\ref{replace}) for the production of the color-singlet pair from a local
color-octet combination of quark and antiquark operators.  
The matching of the cross section with the fragmentation
function was shown in Sec.\ 2, and the matching of
the fragmentation function (\ref{fragfn1}) with the production
matrix element (\ref{replace}) in Sec.\ 3.  Notice, however,
that these diagrams do not appear in the conventional
production matrix element (\ref{Ondef1}).  Since all
other diagrams are held in common between 
the two matrix elements (\ref{replace}) and (\ref{Ondef1}), we
can confidently conclude that the modification of
the matrix element is necessary for matching at NNLO,
at least in Feynman gauge.
Indeed, because the two sets of diagrams
are actually the same in a light-cone $\ell\cdot A=0$ gauge,
this is a quick way to see that the matrix elements
without the $\ell$ eikonal lines are not gauge invariant.

In the following subsections, we identify the diagrams
by the numbers in Fig.\ \ref{softgluonfig}.  In view of the above,
we recognize that each of the quantities: cross section,
gluon fragmentation function, and  production matrix element
is proportional to the sum of these diagrams, multiplied by 
infrared-safe factors.  There is no question that
for individual final states, these diagrams are
infrared sensitive.  The question that we address
in this calculation is whether, when all the final
states of all the  diagrams are combined, 
the infrared poles remain.  As indicated above, the answer is yes.

Discussing the diagrams one-by-one, the actual results (in covariant 
gauge) are rather simple to summarize.  {\it The infrared poles
in dimensional regularization for diagrams I, II, IV, V and VI
all cancel.}   Only diagram III provides a noncancelling pole,
given by\footnote{The factor of two on the left is a convention in our calculation below.}
\bea
2{\rm Re} III = - \alpha_s^2~\frac{1}{3 \varepsilon}~\frac{{\vec{v}}{\, }^2}{4}\, ,
\label{IIIdivergence}
\eea
with $\vec v$ the relative velocity.
Multiplied by the appropriate color factors, 
this pole will appear in the calculation of all
of the three quantities just discussed.  Eq.\ (\ref{IIIdivergence}) 
is the basic result of our calculation.  Because it is nonzero,
the gauge-completion of NRQCD matrix elements appears
to be necessary to extend this formalism to production
processes at NNLO.  Conventional matrix elements
simply will not match the infrared poles that  are
encountered in cross sections and fragmentation
functions at this order.

Having said this much, the reader who wishes to avoid,
or delay, the details of the NNLO calulation may skip to
the final subsection of this section, where we discuss how
the applicability of this result to cross secctions involving
the production of the pair with arbitrary numbers of
hard jets, and to the conclusions, for a brief recapitulation.

   \subsection{Ladder-like diagrams}

  Diagrams I and II have a ladder and crossed-ladder
  structure.  We discuss the calculation of infrared poles
  in II in some detail; diagram I has a very similar structure.
  We will show that the single IR pole of diagram II has
  an imaginary residule.  
  
\begin{figure}
\begin{center}
\epsfig{figure=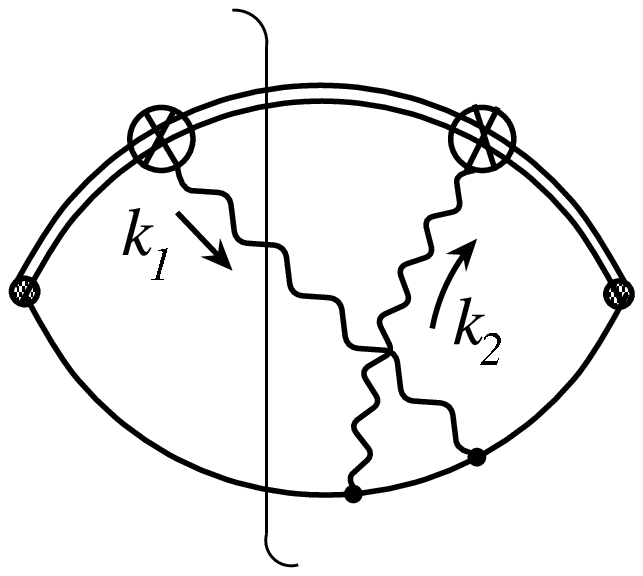,width=0.20\textwidth}
\hskip 0.1\textwidth
\epsfig{figure=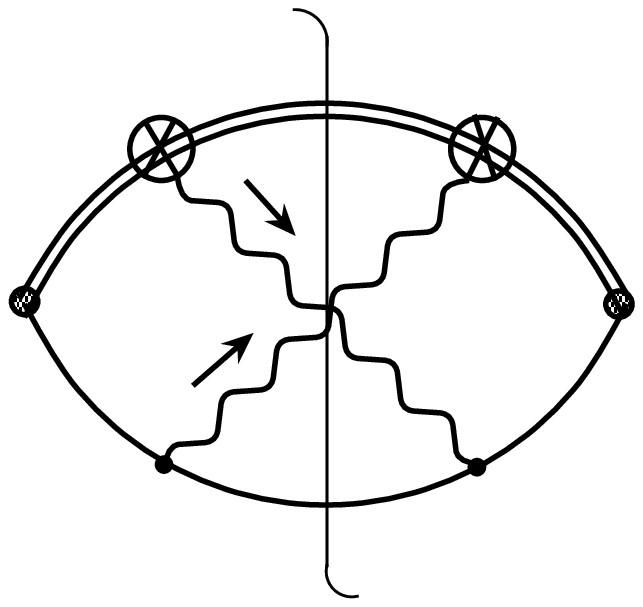,width=0.20\textwidth}
\hskip 0.1\textwidth
\epsfig{figure=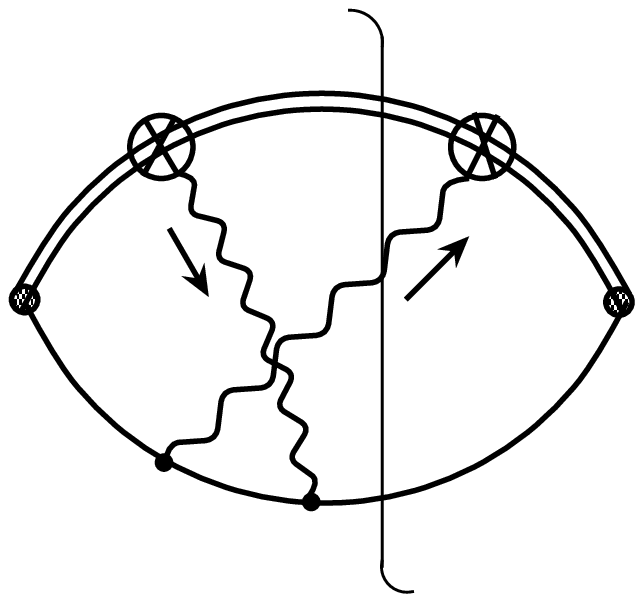,width=0.20\textwidth}

(IIA) \hskip 0.25\textwidth
(IIB) \hskip 0.25\textwidth
(IIC)

(a)

\epsfig{figure=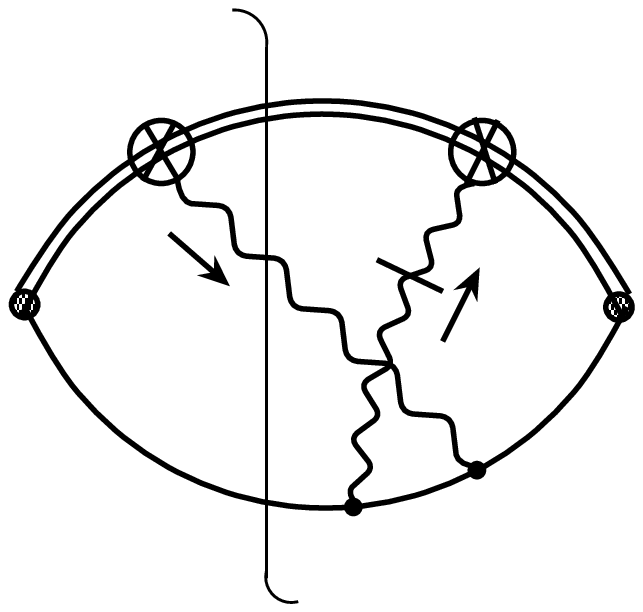,width=0.20\textwidth}
\hskip 0.1\textwidth
\epsfig{figure=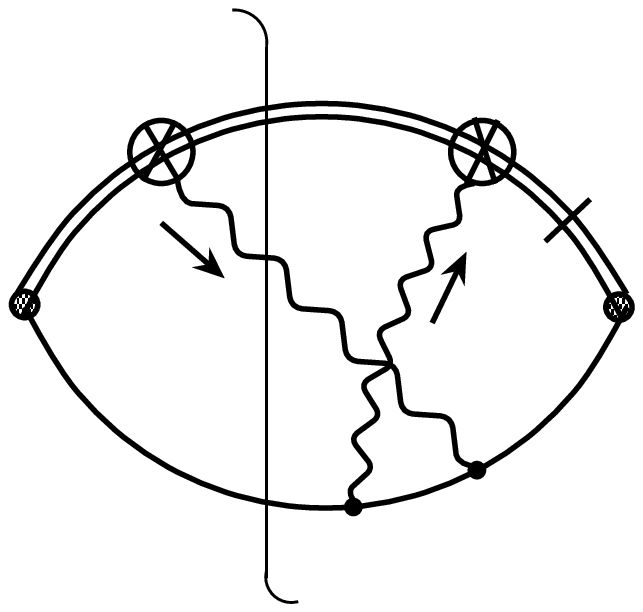,width=0.20\textwidth}

(b)

\caption{(a) Cuts of diagram II; (b) $k_2^-$ poles of Fig.\ 2a. \label{diaIIcuts}}
\end{center}
\end{figure}

  The cuts of diagram II, that is the 
  contributions from various final 
  states, are
  shown in Fig.\ \ref{diaIIcuts}a.   We begin with diagram IIA, in
  which a single gluon appears the final state.  IIA is the
  complex conjugate of IIC, while IIB is real.  Thus, at the
  order to which we work,
  we need consider only the real parts of each diagram.

 After dropping terms that are linear in $k_{i\perp}$, $i=1,2$,
 the integral becomes
 \bea
 IIA &=&  
\frac{- 16\, i\, g^4}{(2\pi)^{2D-1}}\; 
 \int^\Lambda d^Dk_1\int^\infty  d^D k_2\
 \delta_+(k_1^2)\; {1 \over k_2^2-i\epsilon}
 \nonumber\\
&\ & { [\, (P\cdot  \ell)(q\cdot k_1) - (q\cdot \ell)(P\cdot k_1)\, ]\;
[\, (P\cdot  \ell)(q\cdot k_2) - (q\cdot \ell)(P\cdot k_2)\, ]
\over
(P\cdot k_1)^2\, (P\cdot k_2 - i\ep)^2 (- \ell \cdot k_2 -i\ep)\,
(\ell \cdot (k_1 - k_2) -  i\ep)}
 \nonumber\\
 &=& 
 {16\, i\, g^4 q_3^2 \over (2\pi)^{2D-1}}
 \int^\Lambda d^Dk_1\int^\infty  d^D k_2\
 \delta_+(k_1^2)\; {1 \over k_2^2-i\epsilon}
 \nonumber\\
 &\ & \hspace{5mm} \times {k_1^+
 \over
 \left(k_1^- + k_1^+ \right)^2\, 
 \left(k_2^- + k_2^+-i\epsilon\right)^2\, \left(-k_2^++k_1^+  - i\epsilon\right)}\, .
 \eea
 The first expression gives IIA in terms of $P=(2,\vec 0)$,
 normalized  as in the perturbative expansion of Eq.\ (\ref{M2def}), but
 suppressing color factors.  In particular,  as in the NLO case,
 we  divide by $(1/2)^2$ to compensate for the traces in
 quark representation. 
 Here and below, $\Lambda \sim m_c$ is an ultraviolet cut-off for
 real soft gluon radiation.

 Again as in the one-loop examples, we do the minus integrals first,
closing the $k_2^-$ contour in the upper half-plane.  This gives
two terms, one from the quark-pair eikonal, another from the
gluon propagator,
\be
 IIA = IIA^{(k_2^0)} + IIA^{(k_2^2)}\, ,
 \ee
 represented in Fig. \ref{diaIIcuts}b.
Again as in the one-loop example, we readily verify that
the gluon pole term, combined with
the corresponding gluon
pole of $IIC$, cancels the entire contribution
of diagram $IIB$, where both gluons appear in the
final state.  For the remaining contribution we find 
after the minus integrations
\bea
IIA^{(k_2^0)}
&=& {16g^4q_3^2 \over (2\pi)^{2(D-1)}} 
\int d^{D-2} k_{1\perp} d^{D-2}k_{2\perp}\
\int dk_2^+  {k_2^+ \over (2k_2^+{}^2 + k_{2\perp}^2)^2} \nonumber \\
&\ & 
\hspace{20mm} \times 
\int_0^\Lambda dk_1^+\; {4\, k_1^+{}^2 \over (2k_1^+{}^2 
+ k_{1\perp}^2)^2(k_2^+-k_1^+ + i\epsilon)}
\, .
\eea
      In this expression it is clear that we may extend the 
      $k_{1\perp}$ upper limit to infinity without changing the infrared
      behavior of the integrtal.  Then, performing both transverse integrals,
      and changing variables to $x\equiv \sqrt{2}\, k_1^+$
and $y=k_2^+/k_1^+$, we find
 \bea
 IIA^{(k_2^0)}
&=&  4
\left(\frac{\as}{\pi}\right)^2\, (4\pi)^{2\varepsilon}\, 
\Gamma^2(1+\varepsilon)\, q_3^2
\nonumber\\
&\ & \hspace{5mm} \times  
\int_0^{\sqrt{2}\Lambda} dx\, {1\over x^{1+4\varepsilon}}\
\int_{-\infty}^\infty  dy\, {y \over (y{}^2 )^{1+\varepsilon}\, (y-1+i\epsilon)}\, .
\eea
This integral is infrared regularized for $\varepsilon<0$, that is in more
than four dimensions.  
The pole is found from the identity
\be
{1\over x^{1+N\varepsilon}} = {1\over -N\varepsilon} \delta(x) + \left[ {1\over x}\right]_+
+ {\cal O}(\varepsilon)\, ,
\label{Nepspole}
\ee
with a residue that is given by the $\varepsilon=0$ limit of the
remaining expression.   The $y$ integral has no pole,
because for $\vep<0$, the poles from $y \to 0^+$ and $y \to 0^-$
cancel.  The
$y$ integral at $\vep=0$  is then found to be
\be
{\rm lim}_{\vep \to 0} \ \int_{-\infty}^\infty dy\, {y \over (y{}^2 )^{1+\varepsilon}\, (y-1+i\epsilon)}
=
-i\pi\, .
\ee
We conclude that although $IIA^{(k_2^0)}$, and hence
the complete diagram $II$, is
infrared divergent, its divergence is imaginary,
and does not contribute to the fragmentation function, which is real.
Essentially identical considerations apply to the uncrossed
ladder diagram, $I$.

\subsection{Diagrams with three gluons on the quark pair eikonal}

The diagrams with three gluons connected to the
quark lines, are IV, V and VI of Fig.\ \ref{softgluonfig}.
We first consider diagram IV, which involves the
commutator term of the field strength.  Diagram V
and VI both have an additional eikonal vertex
at which a gluon couples to the quark pair in an octet color state.

\subsubsection{Diagram IV}

\begin{figure}[h]
\begin{center}
\epsfig{figure=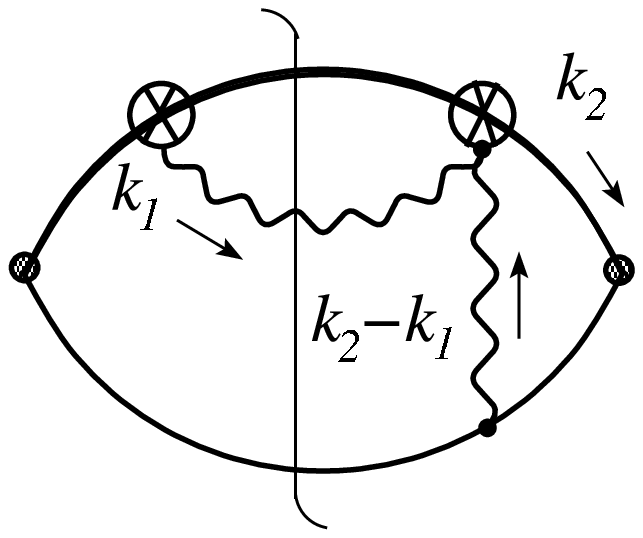,width=0.20\textwidth}
\hskip 0.1\textwidth
\epsfig{figure=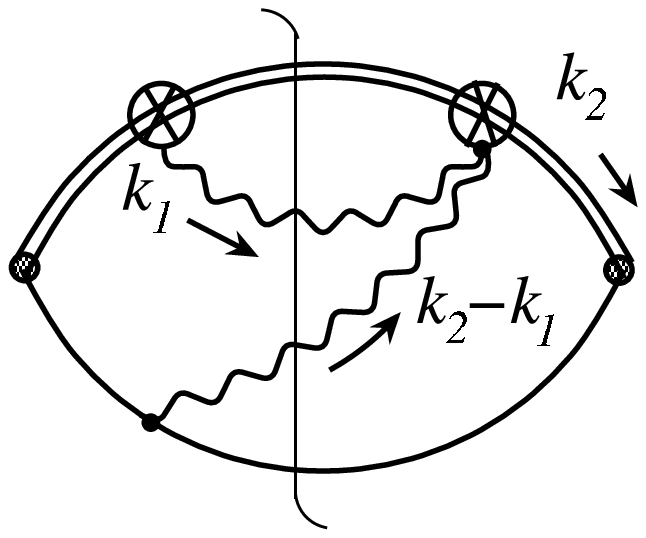,width=0.20\textwidth}

(IVA) \hskip 0.25\textwidth
(IVB)

\caption{Cuts of diagram IV. \label{IVabfig}}
\end{center}
\end{figure}

The relevant cuts of diagram IV are shown in Fig.\ \ref{IVabfig}.
As in the case of the ladder diagrams in the previous
subsection, 
diagram IVB, which is real, cancels against the $(k_1-k_2)^2=0$ pole
of IVA found by   closing $k^-_2$ in the upper half-plane.
We thus need only evaluate the real part of IVA
from the double pole at $P\cdot k_2=2k_2^0=0$.  The imaginary
part, of course, cancels against the complex conjugate diagram.

In the same normalization as above, the numerator momentum
factor for diagram IV is independent of $k_2$, and is given by
(after dropping terms linear in $k_{1\perp}$),
\bea
n_{IV} &=& \left( P\cdot k_1 q^\mu - q\cdot k_1 P^\mu\right)\,
\left(q_\mu P_\nu - P_\mu q_\nu\right)\, l^\nu
\nonumber \\
&=& 4[- q_3^2k_1^+ - (1/2) q_\perp^2\, \left( k_1^+ + k_1^-\right)]\, ,
\eea
and the full contribution of diagram IV is given by  the
real part of
\bea
IVA + IVB &=& IVA^{(k_2^0)}
\nonumber\\
&=& 
-32\left({\as \over \pi}\right)^2\, {1\over (4\pi^2)^{1-2\vep}}\,
\int_0^\Lambda {dk_1^+\over 2k_1^+}\, \int_{-\infty}^\infty dk_2^+
\int d^{D-2} k_{1\perp}\, \int d^{D-2}k_{2\perp}
\nonumber\\
&\ & \hspace{5mm} \times 
{ - q_3^2k_1^+ - (1/2) q_\perp^2\, \left( k_1^+ + k_{1\perp}^2/2k_1^+ \right)
\over 
\left(k_1^++k_{1\perp}^2/2k_1^+\right)^2\,
\left [  -2(k_2^ + -k_1^+)(k_2^++k_{1\perp}^2/(2k_1^+)) 
- (k_{2\perp}-k_{1\perp})^2 -i\ep  \right]^2}\, .
\nonumber\\
\eea
As in the previous diagrams, we change variables to $y=k^+_2/k_1^+$.  
In addition, we rescale the transverse integrals as $\kappa_i\equiv k_{i\perp}/(\sqrt{2}k_1^+)$.
The $\kappa_i$ integrals are finite, but we find that
an explicit infrared pole appears from the limit $k_1^+\rightarrow 0$.
After the $\kappa_2$ integral, the diagram becomes
\bea
IVA^{(k_2^0)} &=& 4
\left({\as \over \pi}\right)^2\, 2^{2\vep}\pi^{3\vep-1}\, \Gamma(1+\vep)\;
\int_0^\Lambda {dk_1^+\over k_1^+{}^{1+4\vep}}\; \int d^{2(1-\vep)} \kappa_1
\nonumber\\
&\ & \hspace{5mm} \times \left( {q_3^2\over [1+\kappa_1^2]^2 } + 
{q_T^2 \over 2[1+\kappa_1^2]} \right)\;
\int_{-\infty}^\infty {dy \over \left[- \kappa_1^2 - y(1-\kappa_1^2) + y^2 +i\ep\right]^{1+\vep}}\, .
\label{IVyintegral}
\eea
The $y$ integral is readily carried out (after changing variables
to $y'= y-(1/2)(1-\kappa_1^2)$), and we verify that 
the residue of the infrared single pole in $\vep$ is imaginary.
The real contribution of diagram IV to the fragmentation
function is then infrared finite.

\subsubsection{Diagram V}

\begin{figure}[h]
\begin{center}
\epsfig{figure=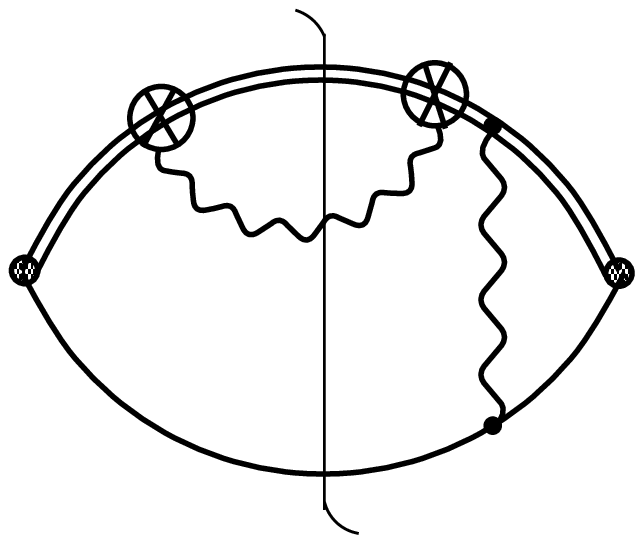,width=0.20\textwidth}
\hskip 0.1\textwidth
\epsfig{figure=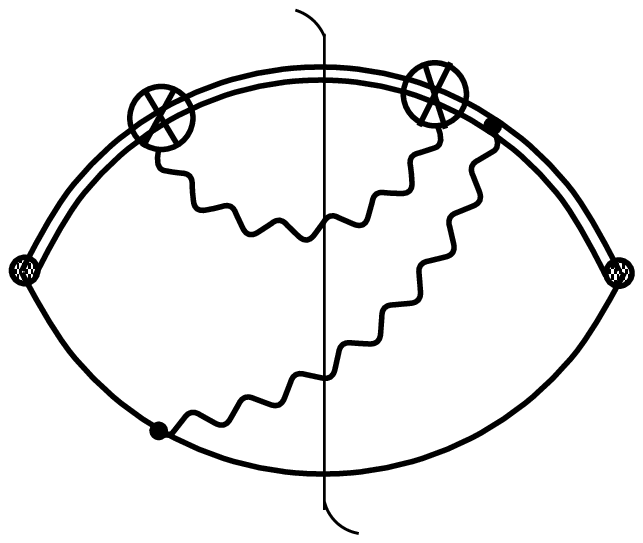,width=0.20\textwidth}

(VA) \hskip 0.25\textwidth
(VB)

\caption{Cuts of diagram V. \label{Vabfig}}
\end{center}
\end{figure}

As for diagram IV, we will find the
infrared pole of the real
part of VA and VB, given in Fig.\ \ref{Vabfig}, and once
again the latter, with two gluons in the
final state, cancels against the $(k_1-k_2)^2$
pole in the former, when the $k_2^-$ contour
is closed in the upper half-plane.

The Feynman rules for the 
field strength vertex lead to a sum of terms (in this
case two) in which each of the denominators on the octet
ordered exponential in Eq.\ (\ref{expandprod3}) is squared,
just as in Eq.\ (\ref{doublepoles}).
Thus, diagram VA has two terms.  
The momentum numerator factor, which depends only
on $k_1$ is the same for both.  
We choose to route the $k_1$ momentum across the gluon
to the $\ell$-eikonal line on the bottom of the diagram,
so that the right-most quark pair eikonal carries momentum $k_2$
to the right, and the exchanged gluon $k_2-k_1$ up.
The VA integral
is then given by
\bea
 VA &=&
 {i2^{5/2}g^4 \over (2\pi)^{2D-1}}
 \int^\Lambda d^Dk_1\int^\infty  d^D k_2\
 \delta_+(k_1^2)\; {1 \over (k_2-k_1)^2-i\epsilon}
 \nonumber\\
 &\ & \hspace{-15mm} \times {n_{VA}(k_1,k_2) \over k_1^+ - k_2^+-i\ep}
\; \left[\,
 {1\over \left(k_1^- + k_1^+ \right)^4 (k_2^++k_2^- - i\ep)}
 + 
 {1\over \left(k_1^- + k_1^+ \right)^3 (k_2^++k_2^- - i\ep)^2}\, \right]\, ,
 \label{VAfull}
 \eea
 which exhibits the squaring of poles to the right of the 
 field strength vertex in the diagram.
The  momentum numerator $n_{VA}$ is
\bea
n_{VA} &=& \left( P\cdot k_1 q^\mu - q\cdot k_1 P^\mu\right)\,
\left( P\cdot k_1 q_\mu - q\cdot k_1 P_\mu\right)\, (P\cdot \ell/2)
\nonumber \\
&=& \sqrt{2}\, 
\left(\, - 4q_3^2k_1^+ k_1^-  -  q_\perp^2\, \left( k_1^+ + k_1^-\right)^2 + 
2(q_\perp \cdot k_{1\perp})^2\, \right)\, .
\eea
After performing the $k_i^-$ integrals of  VA and VB, and noting the
cancellation of the exchange gluon pole, we are left with the contributions
of the $k_2^0$ pole
\bea
V^{(k_2^0)}&=&
\left({\as \over \pi}\right)^2\, {8 \over (4\pi^2)^{1-2\vep}}\,
\int_0 {dk_1^+\over 2k_1^+}\, \int_{-\infty}^\infty dk_2^+
\int d^{D-2} k_{1\perp}\, \int d^{D-2}k_{2\perp}\ 
{1 \over (k_1+ + k_{1\perp}^2/2k_1^+)^3}
\nonumber\\
&\ & \hspace{5mm} \times 
{\left[ - 2q_3^2k_\perp^2 - q_\perp^2\, \left( k_1^+ + k_{1\perp}^2/2k_1^+\right)^2
- 2(q_\perp\cdot k_{1\perp})^2\, \right]
\over
\left( 2(k_2^+  - k_1^+ - i\ep)(-k_2^+ - k_{1\perp}^2/2k_1^+)- (k_{2\perp}-k_{1\perp})^2 - i\ep\right)}
\nonumber \\
&\ & \hspace{-15mm} \times
\left[ {1\over (k_1^+ - k_2^+-i\ep)\, (k_1^+  + k_{1\perp}^2/2k_1^+)}
+{2 \over \left( 2(k_2^+-k_1^+)(-k_2^+ - k_{1\perp}^2/2k_1^+)- (k_{2\perp}-k_{1\perp})^2 - i\ep\right)}\, \right]\, .
\nonumber\\
\label{Vintegral}
\eea
We rescale $k_2^+$ and both of the transverse momenta as
$y=k_2^+/k_1^+$ and
$\kappa_i = k_{i\perp}/\sqrt{2}k_1^+$, which again isolates an overall
infrared divergence at the lower limit of the $k_1^+$ integration.  
The result can be expressed as
\bea
V^{(k_2^0)} &=& 8
 \left({\as \over \pi}\right)^2\, {2^{-2\vep} \over (4\pi)^{1-2\vep}}\,
\int_0 {dk_1^+\over k_1^+{}^{1+4\vep}}\, \int d^{2-2\vep}\kappa_1
\;
{ 4q_3^2\kappa_1^2 + q_\perp^2\, \left( 1+\kappa_1^2 \right) - 4(q_\perp\cdot \kappa_1)^2
\over
(1+\kappa_1^2)^3}\; J_V(\kappa_1)
\nonumber\\
\label{Vkoneplus}
\eea
where the function $J_V$ is defined by
\bea
J_V(\kappa_1)
&=&
-\, \int_{-\infty}^\infty dy \int d^{2-2\vep}\kappa_2\;
\bigg[ {1\over (1+\kappa_1^2)\, (1-y-i\ep)\,
 \left( (y-1)(y+\kappa_1^2) + (\kappa_2-\kappa_1)^2 + i\ep\right) }
\nonumber\\
&\ & \hspace{10mm}
-  {1\over \left[ (y-1)(y+\kappa_1^2) + (\kappa_2-\kappa_1)^2 + i\ep\right]^2}
\,
\bigg]\, .
\label{Vrescaled}
\eea
Performing the $\kappa_2$ integral we find
\bea
J_V(\kappa_1)
&=&
- \pi^{1-\vep}\Gamma(1+\vep)\, \int_{-\infty}^\infty dy \;
\bigg[ {1\over \vep}\, {1\over (1+\kappa_1^2)\, (1-y-i\ep)\,
 \left( y^2 + y(\kappa_1^2-1) - \kappa_1^2 + i\ep\right)^{\vep} }
\nonumber\\
&\ & \hspace{10mm}
-\, {1\over \left( y^2 + y(\kappa_1^2-1) - \kappa_1^2 + i\ep\right)^{1+\vep}}
\,
\bigg]\, .
\eea
The $1/\vep$ pole in the first term in square brackets comes from the
term in which the squared denominator on the quark
eikonal is outside the loop, and this pole is of ultraviolet origin.
The corresponding $k_2$ virtual loop integral, as in the one-loop example of Fig.\ \ref{nlofig}b,
Eq.\ (\ref{sigma8bvirt}),
is infrared finite, and hence may be absorbed into a
coefficient function in the NRQCD expansion.
At the same time, the remaining, $k_1$, integral of this term is analogous 
to the $k$ integral in Fig.\ \ref{nlofig}a, Eq.\ (\ref{nlopt}), 
and its infrared divergence is topologically factorized in the
NRQCD expansion.  
Next, comparing the second term in brackets to Eq.\ (\ref{IVyintegral}), we see that
it is the same as the $y$ integrand in that case, and
when combined with the $k_1^+$ integral in Eq.\ (\ref{Vkoneplus})
gives a purely imaginary infrared pole.  In summary,
the infrared sensitivity of diagram V is fully consistent with
the NRQCD expansion.

\subsubsection{Diagram VI}

\begin{figure}[h]
\begin{center}
\epsfig{figure=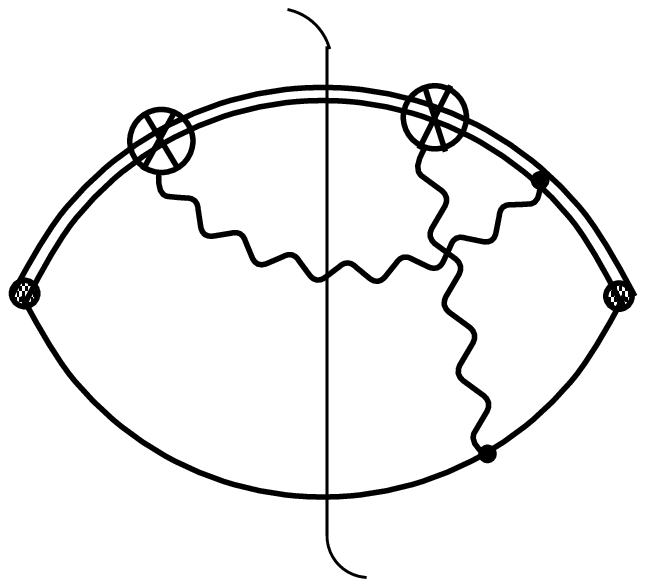,width=0.20\textwidth}
\hskip 0.1\textwidth
\epsfig{figure=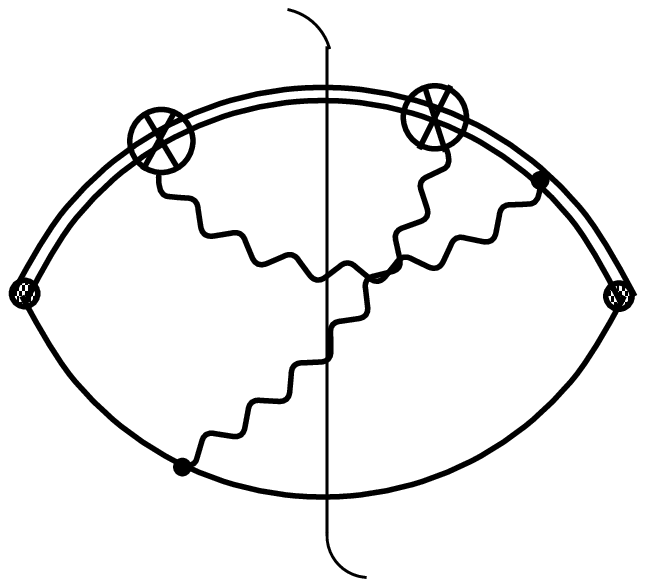,width=0.20\textwidth}

(VIA) \hskip 0.25\textwidth
(VIB)

\caption{Cuts of diagram VI. \label{VIabfig}}
\end{center}
\end{figure}

Diagram VI, with cuts shown in  Fig.\ \ref{VIabfig},
 is treated in a similar way to the previous two
diagrams with three gluons connected to the eikonal
quark pair line.     In this case the $k_1$ line is again
connected to the left-most (field strength) vertex,
while we route momentum $k_2-k_1$ from
the gluon eikonal ($\ell$) to the other field strength
vertex.  Once again the pole from the $k_2 - k_1$ line
of diagram VIA cancels the two-gluon final state,
diagram VIB.  

The corresponding integral for VIA is
\bea
 VIA &=&
 {i2^{5/2}g^4  \over (2\pi)^{2D-1}}
 \int^\Lambda d^D k_1\int^\infty  d^D k_2\
 \delta_+(k_1^2)\; {1 \over (k_2-k_1)^2-i\epsilon}\; {1\over (k_1^++k_1^-)^2}
 \nonumber\\
 &\ & \hspace{5mm} \times {n_{VIA}(k_1,k_2) \over k_1^+-k_2^+-i\ep}
\; \bigg[\,
 {1\over \left(k_2^+ - k_1^+ +k_2^- - k_1^- - i\ep \right)^2 (k_2^+ + k_2^- +i\ep)}
  \nonumber\\
 &\ & \hspace{35mm}
 + {1\over \left(k_2^+ - k_1^+ +k_2^- - k_1^- - i\ep \right) (k_2^++k_2^-+i\ep)^2}\, \bigg]
 \nonumber\\
 &=& 
-\, {i2^{5/2}g^4  \over (2\pi)^{2D-1}}
 \int^\Lambda d^D k_1\int^\infty  d^D k_2\
 \delta_+(k_1^2)\; {1 \over (k_2-k_1)^2-i\epsilon}\; {1\over (k_1^++k_1^-)^2}
 \nonumber\\
 &\ & \hspace{5mm} \times {n_{VIA}(k_1,k_2) \over k_1^+-k_2^+-i\ep}
\; \frac{d}{dk_2^+}\, \bigg[\,
 {1\over \left(k_2^+ - k_1^+ +k_2^- - k_1^- - i\ep \right) (k_2^+ + k_2^- +i\ep)}\, \bigg]
 \, .
  \label{VIAfull}
 \eea
 The first equality exhibits the squaring of poles to the right of the 
 field strength vertex in the diagram.   In the second, we note that the
 term in  square brackets is a derivative with  respect to $k_2^+$,
 and that the expression is simplified by an integration by parts
 in that variable.

The momentum numerator factor is
\bea
n_{VIA}
&=&
\left( P\cdot k_1 q^\mu - q\cdot k_1 P^\mu\right)\, (P_\mu/2)\,
\left( P\cdot (k_1-k_2) q^\nu - q\cdot (k_1-k_2) P^\nu\right)\, \ell_\nu
\nonumber\\
&=& 2\sqrt{2} q_3^2\, (k_1^+-k_1^-)\, (k_1^+-k_2^+) + \dots \, ,
\label{VInum}
\eea
where the terms linear in the $k_{i\perp}$ 
will not contribute, and are omitted in the second line.

In the second form of Eq.\ (\ref{VIAfull}), the $k_2^-$ integral
has three simple poles.
After the $k_2^-$ and $k_1^-$ integrals, using
the cancelation of the exchange gluon pole, we have
\bea
VI &=& VIA+VIB
\nonumber\\
&= & - 32 \left({\as \over \pi}\right)^2\, {1\over (4\pi^2 )^{1-2\vep}}\,
\int_0 {dk_1^+\over 2k_1^+}\, \int_{-\infty}^\infty dk_2^+
\int d^{D-2} k_{1\perp}\, \int d^{D-2}k_{2\perp}
\nonumber\\
&\ & \hspace{5mm} \times 
{ q_3^2\, ( k_1^+ - k_{1\perp}^2/2k_1^+)
\over
 (k_1^+  + k_{1\perp}^2/2k_1^+)^3}
\nonumber \\
&\ & \hspace{-15mm} \times
\left[  { k_1^+ - k_2^+ \over \left (- 2(k_2^+  -  k_1^+)^2 - (k_{2\perp}-k_{1\perp})^2\right)^2}
+  {k_2^+ + k_{1\perp}^2/2k_1^+ \over 
\left( - 2(k_2^+ - k_1^+)(k_2^+ + k_{1\perp}^2/2k_1^+)- (k_{2\perp}-k_{1\perp})^2 - i\ep\right)^2 }\, \right]
\, .
\nonumber\\
\label{VIintegral}
\eea
Carrying out our by-now standard
rescalings and performing the $\kappa_2$ transverse integration,
we  find
\bea
VI  &=& 
- 4
\left({\as \over \pi}\right)^2\, q_3^2\, (2\pi)^{2\vep}\pi^{\vep-1} \,
\int_0 {dk_1^+\over k_1^+{}^{1+4\vep}}\, 
\int d^{n-2} \kappa_{1}\, {1-\kappa_1^2 \over (1+\kappa_1^2)^3}
\nonumber\\
&\ & \hspace{5mm} \times  \int_{-\infty}^\infty dy\; (1-y)
\left[\, {1 \over (2(y-1)^2 -i\ep)^{1+\vep}}
- 
{ 1 \over (2(y-1)(y+\kappa_1^2) - i\ep)^{1+\vep}}\,
\right]\, .
\label{VIyintegral}
\eea
The two $y$ integrals both give finite and purely imaginary
contributions at $\vep=0$, so that once again the sum of contributions
to the fragmentation function from the cuts of diagram VI  is infrared finite.
We have now shown that
of the six classes of diagrams generated from Fig.\ \ref{softgluonfig},
all but diagram III are consistent with standard NRQCD
factorization.  We now turn to this diagram, which is
the most complex to compute.

\subsection{Three-gluon rescattering contribution}

Diagram III is distinguished by its three-gluon coupling.
It connects a subdiagram analogous to Fig.\ \ref{nlofig}a,
Eq.\ (\ref{nlopt}),
which was infrared divergent but topologically factorized,
with the eikonal line $\ell$.   It describes a process in
which the soft gluon that transforms the color octet
pair to a color singlet pair rescatters on the 
adjoint eikonal to lowest order
by exchanging a  gluon.  We recall that the gluon eikonal
represents the influence of the remainder of the
high-$p_T$ process.  We are thus testing the possible dynamical 
influence of this process on the soft hadronization itself.
We shall find that it is a nontrivial influence, with a
noncancelling infrared divergence.  
Nevertheless, the residue of the infrared poles will
be rotationally invariant, and hence consistent with
an NRQCD factorization in terms of our modified 
matrix elements.

Before doing any  integrals, diagram IIIA is of the form
\begin{eqnarray}
&&IIIA(q)~
=~ -16i\, g^4\mu^{4\varepsilon}\, \int \frac{d^D k_1}{(2\pi)^D} 
\frac{d^D
k_2}{(2\pi)^D}~2\pi~\delta(k_1^2) \nonumber\\
&& \hspace{40mm} \times
 \; n_{III}(k_1,k_2)\ \frac{1}{[P\cdot k_1 +i\epsilon]^2~
[P\cdot k_2 - i\epsilon ]^2} \nonumber\\
&& \hspace{40mm} \times \frac{1}{[k_2^2 - i \epsilon]~[(k_2- k_1)^2 - 
i\epsilon]~
  [l\cdot (k_1 - k_2) - i\epsilon]}
\, ,
\label{nnlopt4}
\end{eqnarray}
with a numerator factor $n_{III}$ that we shall
define below.
As usual, we choose the rest frame of heavy quarkonium,
$P^\mu~=~(2,0,0,0)$, and we will perform the $k_2^-$ integral by
closing the contour in the upper half-plane.  

\begin{figure}[h]
\begin{center}
\epsfig{figure=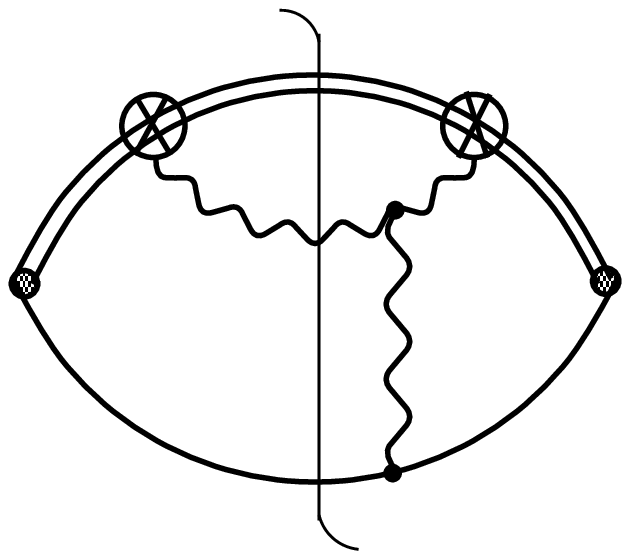,width=0.20\textwidth}
\hskip 0.1\textwidth
\epsfig{figure=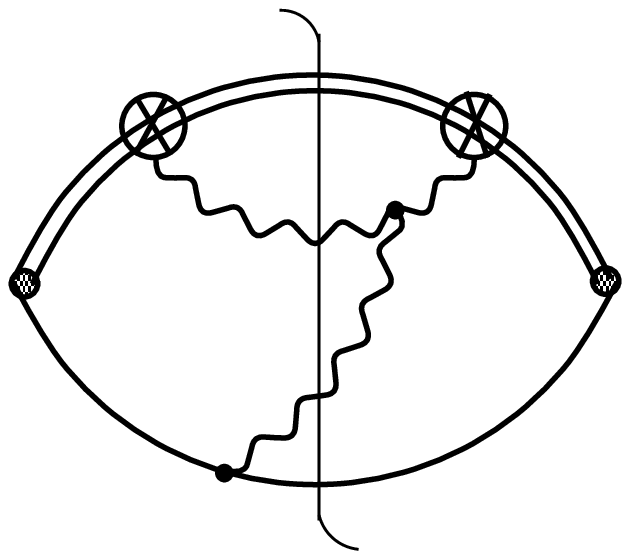,width=0.20\textwidth}

(IIIA) \hskip 0.25\textwidth
(IIIB)

\caption{IIIA and IIIB. \label{IIIAandIIIB}}
\end{center}
\end{figure}

\begin{figure}[h]
\begin{center}
\epsfig{figure=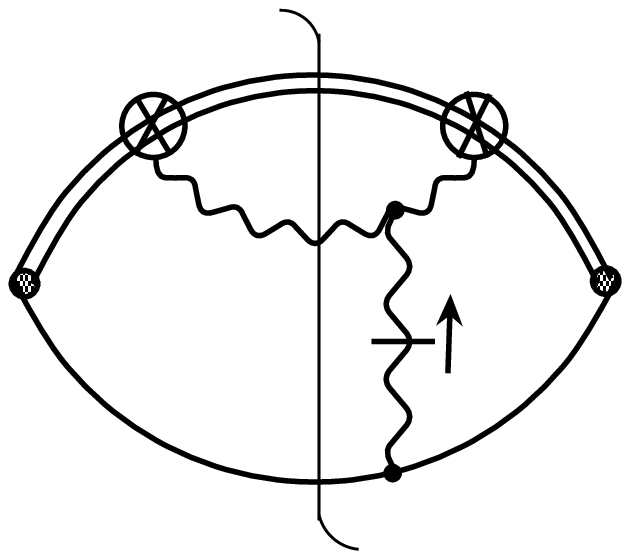,width=0.20\textwidth}
\hskip 0.1\textwidth
\epsfig{figure=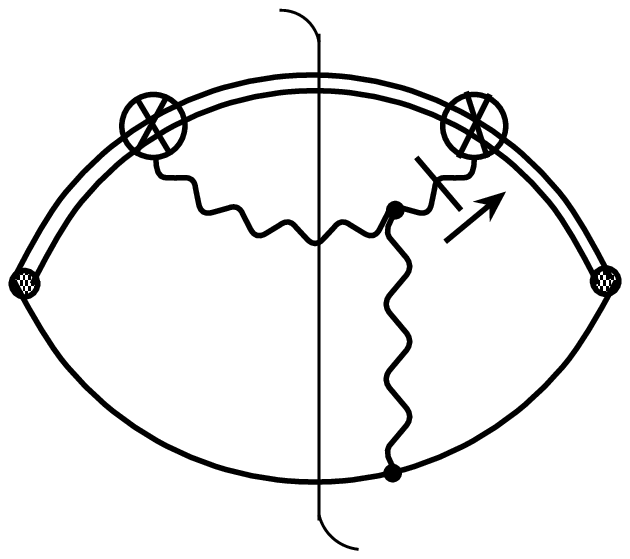,width=0.20\textwidth}
\hskip 0.1\textwidth
\epsfig{figure=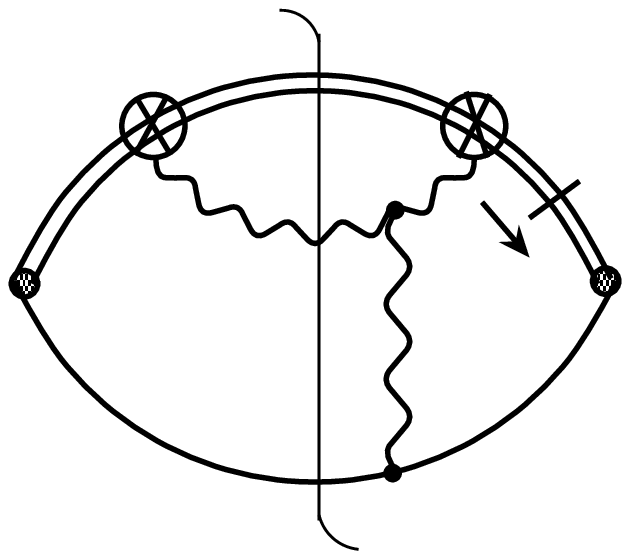,width=0.20\textwidth}

(a) \hskip 0.25\textwidth
(b) \hskip 0.25\textwidth
(c)

\caption{IIIA poles. \label{IIIApoles}}
\end{center}
\end{figure}

The basic pattern for diagram III in 
Fig.\ \ref{softgluonfig} is similar to those above:
the two-gluon cut in Fig.\ \ref{IIIAandIIIB}, IIIA, cancels the pole in
$k_2^-$ from the exchanged gluon in IIIB that is
 attached to the 
octet eikonal line $\ell$.  As for 
diagrams V and VI, we choose the momentum
of this gluon as $k_1-k_2$, flowing down.
There are two additional poles in diagram
IIIA when we close the $k_2^-$ integral
in the upper half-plane, as shown in Fig.\ \ref{IIIApoles}.
After the cancellation with IIIB, only 
the contributions from poles (b) and (c) remain.  

\subsubsection {The numerator and the $k_2^2$ pole}

The numerator factor $n_{III}$ is
\bea
n_{III} &=& \left( P\cdot k_1 q^\mu - q\cdot k_1 P^\mu\right)\, \left( P\cdot k_2 q^\nu - q\cdot k_2 P^\nu\right)\nonumber\\
&\ & \hspace{20mm} \times\; \ell^\sigma\, \left[\, g_{\mu\nu}(-k_1-k_2)_\sigma
+ g_{\sigma\mu}(2k_1-k_2)_\nu + g_{\nu\sigma}(2k_2 - k_1)_\mu\, \right]
\nonumber\\
&=& 2\, [-2(k_1^++k_2^+)(q^2k_{10}k_{20}-(q\cdot k_1 k_{20} + q\cdot k_2 k_{10})q_0
+q\cdot k_1 q\cdot k_2) \nonumber \\
&\ & \hspace{20mm} +4(l \cdot q k_{20} -q\cdot k_2 l^0)(q\cdot k_2 k_{10} -q \cdot k_1 k_{20})
\nonumber\\
&\ & \hspace{30mm}
+4(l \cdot q k_{10} -q\cdot k_1 l^0)(q\cdot k_1 k_{20} -q \cdot k_2 k_{10})] 
\nonumber\\
\label{nnlopt3}
\end{eqnarray}
This is a fairly complex expression, but is clearly symmetric
in $k_1$ and  $k_2$. 

When we take the contribution of the 
$k_2^2$ pole, Fig.\ \ref{IIIApoles}b, we find
\begin{eqnarray}
IIIA^{(k_2^2)}(q)~
&=&  \frac{4}{(4\pi^2)^{1-2\vep}}~ 
\left (\frac{\alpha_s}{\pi} \right)^2 \mu^{4\varepsilon}\, 
\int_0 \frac{dk_1^+}{2k_1^+}\, \int_0 \frac{dk_2^+}{2k_2^+}\, 
\int d^{D-2} k_{1\perp} \, \int d^{D-2} k_{2\perp}  \nonumber \\
&\ & \times { n_{III}(k_1,k_2)_{k_1^2=k_2^2=0}
\over 
(2(k_2^+ - k_1^+)\left( k_{2\perp}^2/2k_2^+ - k_{1\perp}^2/2k_1^+)^2 - (k_{1\perp} - k_{2\perp})^2
- i \ep \right)}\,
\nonumber\\
&\ & {1\over \left[ k_2^+ + k_{2\perp}^2/2k_2^+\right]^2\, \left[ k_1^+ + k_{1\perp}^2/2k_1^+\right]^2}
\, {1\over k_1^+ -k_2^+ - i\ep}\, .
\end{eqnarray}
This is an antisymmetric expression in $k_1$ and $k_2$, 
except for the imaginary contribution at $k_1^+=k_2^+$.  
As a result, once again all infrared poles in $IIIA^{(k_2^2)}$
are imaginary and do not contribute to the
fragmentation function.

\subsubsection{The $k_2^0=0$ double pole}

 We are  left with the evaluation of the pole of diagram \ref{IIIApoles}c,
 the double eikonal pole at $k_2^0=0$ as the only 
 potential source of infrared singularities in the fragmentation
 function that are not topologically factorized in the usual
 sense.  As we have anticipated, we will find an infrared
 pole in dimensional regularization.  Since the calculation
 is a substantial one, we will give most of the details.
 To make it a bit more manageable, we first set
 $q_\perp$ =0 in the numerator (\ref{nnlopt3}), and extend the
 result to nonzero transverse momentum in the appendix.

At zero $q_\perp$, the momentum numerator factor (\ref{nnlopt3}) simplifies to
\bea
n_{III}(q_3,q_\perp=0)  = 2q_3^2[2(k_1^++k_2^+) (k_1^+k_2^-+k_2^+k_1^-)~+~4(k_2^+-k_1^+)
(k_1^+k_2^- - k_1^- k_2^+)]\, .
\eea
We are now ready to pick up the pole in $k_2^-$
corresponding to the diagram of Fig.\ \ref{IIIApoles}c, with the result
\begin{eqnarray}
&&
IIIA^{(k_2^0)}(q)~
=~ \frac{4\, q_3^2}{(4\pi^2)^{1-2\vep}}~ 
\left (\frac{\alpha_s}{\pi} \right)^2 \mu^{4\varepsilon}\, \int d^{D-2} k_{1\perp} ~\int_0^\Lambda  \frac{dk_1^+}{k_1^+}
\int d^{D-2} k_{2\perp} ~\int_{-\infty}^\infty dk_2^+ \nonumber \\
&&
 \hspace{20mm} \times\;
 \frac{1}{k_1^+ - k_2^+-i\epsilon}~\frac{1}{[k_1^++ \frac{{k_{1\perp}}^2}{2k_1^+} ]^2} \nonumber \\
&&
 \times \; \frac{d}{dk_2^-}\,
 \left [ \frac{2(k_1^++k_2^+) (k_1^+k_2^-+k_2^+ \frac{{k_{1\perp}}^2}{2k_1^+} 
)~+~4(k_2^+-k_1^+)(k_1^+k_2^- - \frac{{k_{1\perp}}^2}{2k_1^+} k_2^+)]}
{[2 (k_2^+-k_1^+)(k_2^--\frac{{k_{1\perp}}^2}{2k_1^+} 
)-(k_{1\perp}-k_{2\perp})^2 -i\epsilon]~[2k_2^+k_2^--{k_{2\perp}}^2-i\epsilon]} \right]
\Bigg |_{k_2^-=-k_2^+}\, .
\nonumber\\
\end{eqnarray}
As above we work in $D=4-2\varepsilon$
dimensions, and we rescale the transverse and $k_2^+$ momenta as
\bea
 \kappa_1 = \frac{k_{1\perp}}{\sqrt 2 k_1^+}\, ,
 \qquad
 \kappa_2= \frac{k_{2\perp}}{\sqrt 2 k_2^+}\, ,
 \qquad
 y = \frac{k_2^+}{k_1^+}\, ,
 \eea
which again isolates the infrared pole in the $k_1^+$ integral,
\begin{eqnarray}
&&IIIA^{(k_2^0)}(q_3)
= 
 \frac{2^{4-2\vep} q_3^2}{(4\pi^2)^{1-2\vep}}~ 
\left (\frac{\alpha_s}{\pi} \right)^2 \mu^{4\varepsilon}\, 
 \int_0 \frac{dk_1^+}{{k_1^+}^{1+4\varepsilon}}
 \nonumber \\
&& \hspace{30mm}
\times \int_{-\infty}^\infty dy~ \int d^{2-2\varepsilon} \kappa_1 ~ 
\frac{1}{(1+\kappa_1^2)^2}~
\int d^{2-2\varepsilon} \kappa_2 ~ \frac{1}{1-y-i\epsilon} \nonumber \\
&& \hspace{50mm} \times \Bigg[\,
\frac{1-3y}{2[y^2+\kappa_2^2 ]~[(1-y)(y+\kappa_1^2)-(\kappa_1-\kappa_2)^2 -i\epsilon]} \nonumber \\
&& \hspace{50mm} -\;
\frac{y(y-1)\, [\, y(3+\kappa_1^2)-(1+3\kappa_1^2)\, ]}{2[y^2+\kappa_2^2 ]~[(1-y)(y+\kappa_1^2)-(\kappa_1-\kappa_2)^2 -i\epsilon]^2} \nonumber \\
&& \hspace{50mm} +\; \frac{y^2\, [\, y(3+\kappa_1^2)-(1+3\kappa_1^2)\, ]}{2[y^2+\kappa_2^2 ]^2 
 \, [(1-y)(y+\kappa_1^2)-(\kappa_1-\kappa_2)^2 -i\epsilon]}\; \Bigg] \, .
 \label{IIIAscaled}
\end{eqnarray}
To do the $\kappa_2$ integration we introduce a Feynman parametrization, 
\begin{eqnarray}
&\ & \int d^{2-2\varepsilon} \kappa_2 ~  
~\frac{1}{[y^2+\kappa_2^2 ]^a~[(1-y)(y+\kappa_1^2)-(\kappa_1-\kappa_2)^2 -i\epsilon]^b} \nonumber \\
&\ & \hspace{2mm}
= \frac{\Gamma(a+b)}{\Gamma(a)~\Gamma(b)} \
\int_0^1~dx~x^{b-1}~(1-x)^{a-1} ~(-1)^b~\nonumber \\
&\ & \hspace{10mm} 
\times \int d^{2-2\varepsilon} \kappa_2 ~ \frac{1}{[\kappa_2^2-2x\kappa_2 \cdot \kappa_1 +x\kappa_1^2+x(y^2+y(\kappa_1^2-1)-\kappa_1^2)
+(1-x)y^2+i\epsilon]^{a+b}} \nonumber \\
&\ & \hspace{2mm} = 
(-1)^b \pi^{1-\varepsilon}~\frac{\Gamma(a+b-1+\varepsilon)}{\Gamma(a)~\Gamma(b)}
\nonumber\\
&\ & \hspace{10mm} \times
 \int_0^1~dx~x^{b-1}~(1-x)^{a-1} ~
\frac{1}{[y^2+xy(\kappa_1^2-1)-x^2\kappa_1^2 +i\epsilon]^{a+b-1+\varepsilon}} \, .
\nonumber\\
\label{xparameter}
\end{eqnarray}
After the $\kappa_2$ integration we get
\begin{eqnarray}
&&IIIA^{(k_2^0)}(q_3)~
=~ 2^{2+2\vep} \pi^{3\vep-1}\, q_3^2\, \left (\frac{\alpha_s}{\pi} \right)^2 \mu^{4\varepsilon}\, 
\int_0 \frac{dk_1^+}{{k_1^+}^{1+4\varepsilon}}~
\int_{-\infty}^\infty dy~ \int d^{2-2\varepsilon} \kappa_1 ~ 
\frac{1}{(1+\kappa_1^2)^2}~
\frac{1}{1-y-i\epsilon}  \nonumber \\
&& \hspace{40mm}
\times\; \int_0^1 dx~
\Bigg [\, -\, \frac{1}{2}~ \Gamma(1+\varepsilon)~\frac{1-3y}{[y^2+xy(\kappa_1^2-1)-x^2\kappa_1^2 +i\epsilon]^{1+\varepsilon}} \nonumber \\
&& \hspace{55mm}
- \frac{1}{2}~ \Gamma(2+\varepsilon)~x~\frac{y(y-1)\, [\, (y-1)(\kappa_1^2+3)+2(1- \kappa_1^2)\, ]}
{[y^2+xy(\kappa_1^2-1)-x^2\kappa_1^2 +i\epsilon]^{2+\varepsilon}} \nonumber \\
&& \hspace{55mm}
-\frac{1}{2}~ \Gamma(2+\varepsilon)~(1-x)~\frac{y^2\, [\, (y-1)(\kappa_1^2+3)+2(1- \kappa_1^2)\, ]}
{[y^2+xy(\kappa_1^2-1)-x^2\kappa_1^2 +i\epsilon]^{2+\varepsilon}} 
\, \Bigg]\, .
\nonumber\\
\label{IIIAk20}
\end{eqnarray}
We represent the above equation as 
\bea
IIIA^{(k_2^0)}(q_3)~
&=&~ 2^{2+2\vep} \pi^{3\vep-1}\, q_3^2~ \left (\frac{\alpha_s}{\pi} \right)^2 \mu^{4\varepsilon}\, \int \frac{dk_1^+}{{k_1^+}^{1+4\varepsilon}}
\nonumber\\
&\ & \hspace{20mm} \times\; 
\int d^{2-2\varepsilon} \kappa_1 ~ \frac{1}{[1+\kappa_1^2]^2} ~[I^{(1)}(\kappa_1)~+~I^{(2)}(\kappa_1)]
\, ,
\label{i1i2}
\eea
where $I^{(1)}(\kappa_1)$ organizes a set of terms terms in which the pole at $y=1$ 
has been cancelled by the numerator, and 
$I^{(2)}(\kappa_1)$ summarizes a set  in which the factor $\frac{1}{1-y}$
remains.   To effect this separation,
we rewrite $1-3y = 3(1-y) -2$ in the first term in brackets of Eq.\ (\ref{IIIAk20})
and to combine the second and third terms we use $-x(1-y)y +(1-x)y^2 = y(y-x)$.
After these manipulations, we have
\begin{eqnarray}
&&I^{(1)}(\kappa_1) =~ \int_{-\infty}^{\infty} dy~ \int_0^1 dx ~\Bigg[\,
-\, \frac{3}{2}~ \Gamma(1+\varepsilon)~
\frac{1}{[\, (y+x(\kappa_1^2-1)/2)^2-x^2(\kappa_1^2+1)^2/4 +i\epsilon\, ]^{1+\varepsilon}}
\nonumber \\
&&~+\frac{1}{2}~ \Gamma(2+\varepsilon)~
\frac{y(y-x)(\kappa_1^2+3)}{[\, (y+x(\kappa_1^2-1)/2)^2-x^2(\kappa_1^2+1)^2/4+i\epsilon\, ]^{2+\varepsilon}} \, \Bigg]\, ,
\label{i1}
\end{eqnarray}
where we have completed the squares in the denominators.
For the $1/(1-y)$ terms we have
\begin{eqnarray}
&&I^{(2)}(\kappa_1) =~ \int_{-\infty}^{\infty} \frac{dy}{1-y-i\epsilon}~ \int_0^1 dx~\Bigg[\,
\Gamma(1+\varepsilon)~\frac{1}{[\, (y+x(\kappa_1^2-1)/2)^2-x^2(\kappa_1^2+1)^2/4 +i\epsilon\, ]^{1+\varepsilon}} 
\nonumber \\
&&~-~ \Gamma(2+\varepsilon)~\frac{y(y-x)(1-\kappa_1^2)}
{[\, (y+x(\kappa_1^2-1)/2)^2-x^2(\kappa_1^2+1)^2/4 +i\epsilon\, ]^{2+\varepsilon}}\, \Bigg]
\, . 
\label{i2}
\end{eqnarray}

The $y$ integral for $I^{(1)}$, Eq. (\ref{i1}),  is straightforward.
We change variables to $y'=y+x(\kappa_1^2-1)/2$ and
note that in the numerator $y(y-x)= y'^2 + x^2(\kappa_1^4-1)- y'x\kappa_1^2$,
where the last term vanishes because it is odd in $y'$.
In this way, we find
\begin{eqnarray}
I^{(1)}(\kappa_1) &=& \frac{\sqrt{\pi}}{2} ~ \int_0^1 dx~\ \Bigg \{\,
~[-3~ +(1+\varepsilon)~(\kappa_1^2+3)]~\Gamma(1/2+\varepsilon)~
\frac{1}{[-x^2(\kappa_1^2+1)^2/4 + i\ep ]^{1/2+\varepsilon}} \nonumber \\
&\ &~+2\kappa_1^2~(\kappa_1^2+3)~\Gamma(3/2+\varepsilon)~ (x^2(\kappa_1^2+1)/4)\
\frac{1}{[-x^2(\kappa_1^2+1)^2/4 + i\ep]^{3/2+\varepsilon}}\, \Bigg \} \, .
\end{eqnarray}
The $x$ integration is now trivial, and using the expansion 
$(-1+i\ep)^{-\varepsilon}~=~e^{-i\pi \varepsilon}~\sim (1-i\pi \varepsilon)$, we isolate the 
imaginary pole in $I^{(1)}$, and a corresponding finite real part,
\begin{eqnarray}
I^{(1)}(\kappa_1) =~-i\pi~\frac{1}{\varepsilon}~(1-i\pi \varepsilon)~ \left[\, \frac{\kappa_1^2}{(1+\kappa_1^2)^2}\, \right ]~+~\dots\, .
\label{i1i}
\end{eqnarray}
The real term in this expression,  when substituted into Eq.\ (\ref{i1i2}), gives
a real, single pole, contribution to the fragmentation function from
the $k_1^+$ integral.  This
is the generic mechanism we are after.

The complete result at $q_\perp=0$, of course, requires $I^{(2)}(\kappa_1)$,
which is a bit more complicated, because of the extra denominator
$1/(1-y - i\epsilon)$.
We give the detailed calculation of 
$I^{(2)}$ in Sec.\ A.1 of the appendix, where we show that
\begin{eqnarray}
I^{(2)}(\kappa_1) =~i\pi~\frac{1}{\varepsilon}~(1-i\pi \varepsilon)~ \left [\, \frac{2\kappa_1^2}{(1+\kappa_1^2)^2}\, \right ]~+~ \dots \, .
\label{I2i}
\eea
Substituting  Eqs.\ (\ref{i1i}) and (\ref{I2i}) 
in Eq.\ (\ref{i1i2}) we find that the remaining, $\kappa_1$, integration 
is convergent because $\kappa_1$ of order unity corresponds
to  $k_{1\perp}$ of order $k_1^+$.  Thus, the transverse momentum
integration of the real gluon converges at a scale  far below the
fixed quark mass, and is effectively independent of  the
phase space cut-off.
 Performing the $\kappa_1$ integral, and adding the contribution
   from the complex conjugate of the diagram, we find
for the leading, real $\frac{1}{\varepsilon}$ divergent term,
\begin{equation}
2\, {\rm Re}\, IIIA^{(k_2^0\, {\rm pole})}(q_3)~=~ -\, 
\alpha_s^2~\frac{1}{3 \varepsilon}~q_3^2
\qquad (q_\perp=0)\, .
\label{sq3}
\end{equation}
The leading imaginary double pole, of course,  cancels in  
the full fragmentation function. 

We evaluate the corresponding $q_\perp$-dependent pole
in  the appendix.  We note that all $q_3\, \times\, q_\perp$ 
interference terms vanish because they are linear in
the $k_{i\perp}$ integrations.
Combining Eq.\ (\ref{sq3}) for $q_\perp=0$ 
with the result Eq.\ (\ref{sqT})
from the appendix for the $q_\perp^2$ term, we obtain
a rotationally invariant result
\begin{equation}
2\, {\rm Re}\,  IIIA^{(k_2^0\, {\rm pole})}(q)~=~ 
- \alpha_s^2~\frac{1}{3 \varepsilon}~{\vec{q}}^2
~=~ - \alpha_s^2~\frac{1}{3 \varepsilon}~\frac{{\vec{v}}{\, }^2}{4}\, .
\label{svq}
\end{equation}
This is the full result for diagram III and hence, as 
discussed in Sec.\ 6.2 above, for the entire
NNLO infrared pole term in the cross section
and fragmentation function, matched by
the gauge-completed production matrix
elements at the same order.

\subsection{Rotational invariance and universality}

The significance of rotational invariance is that the infrared pole is 
independent of the 
relative orientation of the pair's relative velocity
$\vec v$ and  the gluon eikonal direction $l$. 
The complete result shows first, that the gauge invariant
redefinition of the NRQCD matrix element is necessary,
but also shows that once this is done, the factorized form is consistent
with universality of the factorization.   

As we have emphasized above,
the same reasoning applies to cross sections in which 
the pair recoils against a gluon jet.  In fact, 
the matching of cross sections with our matrix elements 
is even more general, as a result of
the rotational invariance of Eq.\ (\ref{svq}).  This 
follows from the nature of the gluon rescattering diagrams
that give this result, in which
two soft gluons attach to the pair, leaving only a single soft gluon
to attach to the other jet.  At the same time, the
 exchange of soft gluons at NNLO
between the heavy quarkonium palr and each hard jet in
the final state will give the same pole factor, given by Eq.\ (\ref{svq}),
up to the effect of color.   

Because the momentum factors are the same
for the pole found by coupling the soft gluons
to each final-state jet, we easily show that the complete color factor
turns out to be independent of the number and directions
 of the jets, and of the color representations of their parent partons.
 We outline the proof as it applies
 to leptonic annihilation cross sections, where all
 jets are in the final state.
 The result follows from gauge invariance.

Recalling the discussion of Sec.\ 2, we suppose
that we are at a leading region of
phase space where there is
an arbitrary number of jets, of momenta $p_j$,  $j=1\dots n$.
At any leading region, in the absence of soft gluon
exchange, the cross  section factorizes into 
a product of jet subdiagrams $J_j(p_j){}_{b_j,a_j}$,
contracted in color indices $a_j$ and $b_j$ with a
hard scattering function $h_{a_0,a_j}$ in the amplitude,
and a corresponding function in the complex conjugate
amplitude.  In addition to the final-state jets,  
the short-distance function $h$ is also contracted
with the  parent parton (gluon above)  of the heavy
quark  pair, though color index $a_0$.
Concentrating just on the amplitude,
the fragmentation function is thus proportional to the combination
\bea
{\cal M}_{b_n \dots b_1,a_0}
=
 \sum_{a_1\dots a_n}\, 
\left[\; \prod_{j=1}^n\, J_j(p_j){}_{b_j,a_j}\; \right]\; h_{a_0,a_1  \dots  a_n}\,  .
 \label{calJ}
 \eea
 We suppress the function associated with the
 jet in which the pair appears.

We now consider  the effect of adding soft gluons at NNLO
in this leading region,
and we again discuss  the case when the
quark-antiquark pair is an octet at short distances and a  singlet
in the final state.  This requires that two gluons 
attach to the pair.
Recalling the factorization property of jet-soft interactions derived in
Sec.\ 2, the infrared behavior of each set of diagrams
where soft gluons couple to jet $j$
can be replaced by diagrams in which the soft gluons
attach  to an eikonal line in the direction of $p_j$.
Once again only soft-gluon diagrams like III in Fig.\ \ref{softgluonfig},
with a three-gluon coupling, can give rise to 
a real infrared pole in the cross section.
In the set of such diagrams, the single exchanged
gluon attaches to the $n$ jets one at a time.

We denote the color index of the exchanged gluon by $e$,
and the flavor of the
parent parton of jet $j$ by $f_j$.   At fixed
values of the pair relative velocity  $\vec v=2\vec q$,  the effect of this insertion
is to  multiply $\cal  M$ of Eq.\ (\ref{calJ}) by the same pole
term $IIIA^{(k_0,pole)}(q)$, Eq.\ (\ref{svq}), that we have determined above
 for the fragmentation function
at NNLO.   This factor is independent of the jet to which the exchanged
gluon attaches.

The effect of the exchanged gluon's color, of course, differs from jet to jet,
but is still quite simple  after it has been factorized.
The short-distance color tensor is multiplied by
the  matrix through which the soft gluon couples
to the eikonal  line in the $p_j$  direction, that is the color
generator $T^{(f_j)}$.  In summary, the structure of the NNLO
pole term in the cross section is
\bea
{\cal M}^{(NNLO)}_{b_n \dots b_1,a_0}(q)
&=&
IIIA^{(k_0,pole)}(q)\ \times\
 \sum_{l=1}^n\,  \sum_{a_1\dots  \dots a_n}\, 
\left[\; \prod_{j=1}^n\, J_j(p_j){}_{b_j,a_j}\; \right]   \nonumber\\
&\ & \hspace{35mm} \times\;
\sum_{a'_l}\, 
 h_{a_0,a_1 \dots a'_l \dots  a_n}\, [T^{(f_l)}_e]_{a'_l,a_l} \, .
 \eea
 We now observe that multiplication by the color generator for 
 a given external line of the short-distance function
 $h$ is equivalent to an infinitesimal color rotation
 of the corresponding  external line.  The sum of
color rotations on all its external lines vanishes by the
 gauge invariance of the theory.
 The sum of color rotations on all the final-state jets, therefore,
 is  the negative of a color rotation on
 the parent gluon of the pair,
 \bea
\sum_{l=1}^n\, 
\sum_{a'_l}\, h_{a_0,a_1 \dots a'_l \dots  a_n}\, [T^{(f_l)}_e]_{a'_l,a_l} 
=
- \sum_{a'_0}\, h_{a'_0,a_1 \dots   a_n}\, [T^{(f_0)}_e]_{a'_0,a_0}\, .
\eea
 The sum of the color
factors associated with attaching a  single soft gluon
to all recoiling jets is therefore independent of
the number and/or flavor of the final-state jets.
The same argument can be applied to the color
factors of the fragmentation function,  with the
same result.    The gauge-completed
matrix element is therefore universal up to  NNLO,
for arbitrary numbers of hard jets in the final state.

\section{Conclusions}

We have investigated the proposal of NRQCD factorization in
production processes at large transverse
momentumm ($p_T$), and have demonstrated that factorization 
holds to NNLO in production from an octet pair, after a
redefinition of the nonperturbative matrix elements in the effective
theory.  We have seen, in fact, that this matrix element is
universal at NNLO for high-$p_T$ quarkonium production with
arbitrary final states.
Many questions remain, however, and it is unclear
to us whether the pattern we have found, uncanceled
infrared divergences that can be absorbed into universal
gauge-completed matrix elements, will survive at higher
orders.  On the other hand, the very nontrivial organization
of the  NNLO infrared divergences
into a single power of ${\vec q}{\, }^2$ is encouraging.

So far our analysis has involved infrared
structure associated only with electric dipole couplings,
at momentum scales that are characteristically of order
$mv$. Such a higher-order analysis will also require
study of the lower momentum scale characteristic of
binding energies $mv^2$, which, as we have observed
above do not enter into
our NNLO octet-to-singlet calculations \cite{potential}.  
Finally, our study of fragmentation at
large $p_T$ strongly suggests that the low-$p_T$
cross sections for quarkonium production
deserves a fresh look \cite{kho04}.

In summary, the calculations and reasoning presented in
this paper have, we believe, demonstrated
that further investigation is crucial to
provide a theoretical grounding for  the analysis of the
production of heavy quarkonia.

\subsection*{Acknowledgements}

   This work was supported in part
by the National Science Foundation, grants PHY-0071027, PHY-0098527,
PHY-0354776 and PHY-0345822,
and by the Department of Energy, grant DE-FG02-87ER40371.
We thank Geoff Bodwin for useful exchanges and for
emphasizing the importance of this topic, and 
Yu-Qi Chen for helpful discussions.

\begin{appendix}

\section{Appendix}

The appendix provides more details on the evaluation
of the infrared pole in diagram IIIA.  The integrals
presented here are all reasonably straightforward,
and are complex only because of the rather large numbers
of terms.
Nevertheless, because they can be performed ``by hand",
we feel an interested reader who wishes to
reconstruct the calculation in detail may find
the following relatively extensive presentation 
useful.

\subsection{The integral $I^{(2)}$ of diagram $IIIA$ with $q_\perp=0$}

We continue here with the detailed evaluation of  the real
$\as^2\, \varepsilon^0$ contribution from the function
 $I^{(2)}(\kappa_1)$ of Eq. (\ref{i2}).   Compared to
 the case considered above, $I^{(1)}(\kappa_1)$, $I^{(2)}(\kappa_1)$
 differs primarily by having an extra $1/(1-y)$ denominator,
 which requires an additional Feynman parameterization.

 To simplify the $y$ integral, we eliminate the explicit $y^2$
 numerator factor by using the identity
 $y^2 - yx = [\, y'^2 -x^2(\kappa_1^2 + 1)^2/4\, ] - x(y-x)\kappa_1^2$,
 where the term in brackets cancels a power
 in the denominator and where, as above, $y' = y +  x(\kappa_1^2-1)/2$.
 This gives the slightly simpler form
 \begin{eqnarray}
&&I^{(2)}(\kappa_1) = \int_0^1 dx\; \int_{-\infty}^{\infty} \frac{dy}{1-y-i\epsilon} \nonumber \\
&& \hspace{5mm} \times\; \Bigg[\,
(\,\Gamma(1+\varepsilon) - (1-\kappa_1^2)\Gamma(2+\vep)\, )
\frac{1}{[\, (y+x(\kappa_1^2-1)/2)^2-x^2(\kappa_1^2+1)^2/4 +i\epsilon\, ]^{1+\varepsilon}} 
\nonumber \\
&& \hspace{15 mm} + \Gamma(2+\varepsilon)~\frac{x(y-x)\kappa_1^2(1-\kappa_1^2)}
{[\, (y+x(\kappa_1^2-1)/2)^2-x^2(\kappa_1^2+1)^2/4 +i\epsilon\, ]^{2+\varepsilon}}\, \Bigg]
\nonumber\\
&& \hspace{10mm} =
(\,\Gamma(1+\varepsilon) - (1-\kappa_1^2)\Gamma(2+\vep)\, )
l^{(2)}(\kappa_1)
+ \Gamma(2+\varepsilon)~\kappa_1^2(1-\kappa_1^2)\; j^{(2)}(\kappa_1)\, ,
\label{i22dform}
\end{eqnarray}
where the second equality serves to define $l^{(2)}(\kappa_1)$ and $j^{(2)}(\kappa_1)$.
Consider the first $y$ integral of Eq. (\ref{i22dform}),
\bea
&&l^{(2)}~ =~ \int_{-\infty}^{\infty} \frac{dy}{1-y-i\epsilon}~ \int_0^1 dx~ 
\frac{1}{[(y+x(\kappa_1^2-1)/2)^2-x^2(\kappa_1^2+1)^2/4 +i\epsilon]^{1+\varepsilon}}. 
\eea
Introducing an additional Feynman parameter,  $x^\prime$, 
and expanding the square of the second denominator we get 
\bea
&&l^{(2)}~ =~ -\frac{\Gamma(2+\varepsilon)}{\Gamma(1+\varepsilon)}~
\int_0^1 dx~ \int_0^1 dx^\prime\, x'{}^\ep
\nonumber\\
&& \qquad \times 
\int_{-\infty}^{\infty} ~dy~ 
\frac{1}{[\, x^\prime[y^2+xy(\kappa_1^2-1)-x^2\kappa_1^2 ]~+~(1-x^\prime)(y-1)~+~i\epsilon\, ]^{2+\varepsilon}}. 
\eea
The $y$ integration is now easily performed, and gives 
\bea
&&l^{(2)}~ =~ -\frac{\Gamma(3/2+\varepsilon)}{\Gamma(1+\varepsilon)}~\sqrt{\pi}\, (-1+i\ep)^{-3/2-\vep}
 \int_0^1 dx~ \int_0^1 \frac{dx^\prime}{{{x^\prime}}^{1/2-\vep}}~\nonumber \\
&& \hspace{20mm} \times \frac{1}{[x^\prime x^2 (\kappa_1^2+1)^2/4+(1-x^\prime)[x(\kappa_1^2-1)/2+1]+\frac{(1-x^\prime)^2}{4x^\prime}]^{3/2+\varepsilon}}
\, .
\label{I2Ddef}
\eea

To isolate the infrared pole of this expression, it is useful to 
change variables to $u=x^2$ and $v^\prime=(1-x^\prime)/u$.
Also using  $(-1+i\ep)={\rm e}^{i\pi}$, we have
\bea
&&l^{(2)}~ =~ -\frac{\Gamma(3/2+\varepsilon)}{2\Gamma(1+\varepsilon)}~
\sqrt{\pi}~e^{-i\pi(3/2+\varepsilon)}~
\int_0^1 \frac{du}{u^{1+\varepsilon}}~ \int_0^{1/u} \frac{dv^\prime}{{{(1-uv^\prime)}}^{1/2-\vep}}
\nonumber \\
&& \hspace{15mm}
\times\, \frac{1}{[(1-uv^\prime) (\kappa_1^2+1)^2/4+v^\prime[\sqrt{u}(\kappa_1^2-1)/2+1]+\frac{u{v^\prime}^{2}}{4(1-uv^\prime)}]^{3/2+\varepsilon}}
\, .
\label{I2final}
\eea

The $u$ and $v'$ integrals in (\ref{I2final})
are finite for $\vep<0$, characteristic of an infrared pole.
The $1/\varepsilon$ pole comes from $u\to 0$, and is isolated using
(\ref{Nepspole}).  Its residue is purely imaginary.  There is a corresponding real contribution
to $I^{(2)}$ at $\vep=0$, however, found from the expansion of the exponential.
The $v'$ integral is trivial at $\vep=0$ and $u=0$, and we find
\bea
&&l^{(2)}~ =~ i\pi~\frac{1}{\varepsilon}~(1-i\pi \varepsilon)~\frac{1}{(1+\kappa_1^2)}
+{\cal O}(i\ep^0) \, .
\eea
This term will contribute at the level  of $\as^2/\vep$ in the fragmentation function
after the integrals over $\kappa_1$ and $k_1^+$.

An identical procedure can be used to evaluate 
the second term, $j^{(2)}(\kappa_1)$ in Eq.\ (\ref{i22dform}), 
\bea
j^{(2)} &=&   \frac{\Gamma(5/2+\varepsilon)}{2\Gamma(2+\varepsilon)}\,
\sqrt{\pi}~e^{-i\pi(5/2+\varepsilon)}~
 \int_0^1 \frac{du}{u^{1+\ep}}\; \int_0^{1/u} dv^\prime\, {(1-uv^\prime)}^{1/2-\vep}
 \nonumber \\
&\ &\hspace{10mm} \times
~\frac{ (\kappa_1^2+1)/2+ \sqrt{u}\, v^\prime/[2(1-uv^\prime)]}
{[(1-uv^\prime) (\kappa_1^2+1)^2/4+v^\prime[\sqrt{u}(\kappa_1^2-1)/2+1]+\frac{u{v^\prime}^{2}}{4(1-uv^\prime)}]^{5/2+\varepsilon}}\, ,
\nonumber\\
\eea
with the same denominator as in
Eq.\ (\ref{I2final}).
The relevant singular behavior of this expression is
\bea
&&j^{(2)}~ =~ \frac{\Gamma(5/2+\varepsilon)}{4\Gamma(2+\varepsilon)}\,
\sqrt{\pi}~e^{-i\pi(5/2+\varepsilon)}~
 \int_0^1 \frac{du}{u^{1+\varepsilon}}~ \int_0^{\infty} dv^\prime 
 \frac{ 1+\kappa_1^2 }{[(\kappa_1^2+1)^2/4+v^\prime]^{5/2+\vep}}\, .
\nonumber\\
\eea
Isolating the imaginary pole and the accompanying real finite part in the
same way as for $l^{(2)}$, we get
\bea
&&j^{(2)}~ =~ i\pi \frac{1}{\varepsilon}~(1-i\pi \varepsilon)~\frac{1}{(1+\kappa_1^2)^2}\
+\ {\mathcal O}(i\ep^0)\, .
\eea
Finally, substituting the results for $l^{(2)}$ and $j^{(2)}$ into Eq. (\ref{i22dform}) we find
\begin{eqnarray}
I^{(2)}(\kappa_1) =~i\pi~\frac{1}{\varepsilon}~(1-i\pi \varepsilon)~
\frac{2 \kappa_1^2}{(1+\kappa_1^2)^2} + \dots \, ,
\label{i2i}
\end{eqnarray}
which is the result quoted in Eq.\ (\ref{I2i}).

\subsection{Transverse momenta for $q$ in diagram $IIIA$}

When $q_3$ =0  the numerator for diagram III is
\bea
n_{III}(q_3=0,q_\perp) &=& 
2\left[q_\perp^2(k_1^++k_2^+) (k_1^++k_1^-) (k_2^++k_2^-) 
       -2(q_\perp \cdot k_{1\perp})^2 (k_2^++k_2^-)
\right.
\nonumber\\
&\ & \hspace{10mm}
\left.
-2(q_\perp \cdot k_{2\perp})^2 (k_1^++k_1^-) 
+2(q_\perp \cdot k_{1\perp}) (q_\perp \cdot k_{2\perp}) (k_1^-+k_2^-)
\right]\, .
\label{n3qperp}
\eea
Diagram $IIIA^{(k_2^0)}$ with $q_3=0$ is given by
\begin{eqnarray}
&&IIIA^{(k_2^0)}(q_\perp)~
=~ \frac{4}{(4\pi)^{1-2\vep}}~ \left (\frac{\alpha_s}{\pi} \right)^2 \mu^{4\varepsilon}\; 
\int_0^\Lambda \frac{dk_1^+}{2k_1^+}\;  \int  d^{D-2} k_{1\perp}\; \int dk_2^+\; 
\int d^{D-2} k_{2\perp}
\nonumber \\
&& \hspace{20mm} \times \frac{1}{k_1^+-k_2^+-i\epsilon}\;
\frac{1}{[\, k_1^+ +  \frac{{k_{1\perp}}^2}{2k_1^+}\, ]^2} \nonumber \\
&&~ \times\, \left[\frac{d}{dk_2^-}\;
\left[ \frac{ n_{III}(q_3=0,q_\perp)}{[\, 2 (k_2^+-k_1^+)(k_2^--  \frac{{k_{1\perp}}^2}{2k_1^+} )
      - (k_{2\perp}-k_{1\perp})^2 -i\epsilon\, ]\; [\, 2k_2^+k_2^--{k_{2\perp}}^2-i\epsilon\, ]}\right] 
\right]_{k_2^-=-k_2^+}\, .
\nonumber\\
\label{IIIAqperp}
\end{eqnarray}
Defining as above the scaled variables,
$\kappa_1=\frac{k_{1\perp}}{\sqrt 2 k_1^+}$, $\kappa_2=\frac{k_{2\perp}}{\sqrt 2 
k_2^+}$ and $y=\frac{k_2^+}{k_1^+}$,
this integral becomes
\begin{eqnarray}
IIIA^{(k_2^0)}(q_\perp)
&=&  -\; 4\frac{(2\mu^2)^{2\ep}}{(\pi^2)^{1-2\vep}}~ 
\left (\frac{\alpha_s}{\pi} \right)^2 \, \int 
\frac{dk_1^+}{{k_1^+}^{1+4\epsilon}}~
 \int d^{2-2\varepsilon} \kappa_1 ~ \frac{1}{(1+\kappa_1^2)^2}~
 ~\int dy~  \frac{1}{1-y-i\epsilon} 
 \nonumber \\
&\ & \hspace{0mm} \times \int d^{2-2\varepsilon} \kappa_2\;
[\frac{1}{4}~\frac{(1+y)(1+\kappa_1^2)q_\perp^2-4(q_\perp \cdot \kappa_1)^2+4(q_\perp 
\cdot \kappa_1)(q_\perp \cdot \kappa_2)}{[y^2+\kappa_2^2 
+i\epsilon]~[(1-y)(y+\kappa_1^2)-(\kappa_1-\kappa_2)^2 -i\epsilon]} 
\nonumber \\
&\ & \hspace{5mm}+~y~
\frac{-(1+\kappa_1^2)(q_\perp \cdot \kappa_2)^2
      +(q_\perp \cdot \kappa_1)(q_\perp \cdot \kappa_2)~(\kappa_1^2-y)}
     {[y^2+\kappa_2^2 +i\epsilon]^2~
      [(1-y)(y+\kappa_1^2)-(\kappa_1-\kappa_2)^2 -i\epsilon]} 
\nonumber \\
&\ & \hspace{5mm} + ~(1-y)~
\frac{-(1+\kappa_1^2)(q_\perp \cdot \kappa_2)^2
      +(q_\perp \cdot \kappa_1)(q_\perp \cdot
                                        \kappa_2)~(\kappa_1^2-y)}
     {[y^2+\kappa_2^2 +i\epsilon]
     ~[(1-y)(y+\kappa_1^2)-(\kappa_1-\kappa_2)^2 -i\epsilon]^2}] \, ,
\nonumber\\
\end{eqnarray}
which is  analogous to Eq.\ (\ref{IIIAscaled}) for the $q_3^2$ terms in the numerator.

To perform $\kappa_2$ integration we can
again introduce a Feynman parameter $x$ as 
in (\ref{xparameter}).  The change of variables
$\kappa_2' = \kappa_2 -x\kappa_1$
completes the square in the denominator,
and the integral results in the rather lengthy expression
\begin{eqnarray}
IIIA^{(k_2^0)}(q_\perp)~
&=& - \;4\, ~ 
\left (\frac{\alpha_s}{\pi} \right)^2 \frac{2^{2\ep}\mu^{4\varepsilon}}{\pi^{1-3\ep}}\, \int 
\frac{dk_1^+}{{k_1^+}^{1+4\varepsilon}}~
 \int d^{2-2\varepsilon} \kappa_1 \frac{1}{(1+\kappa_1^2)^2}
 \int_0^1 dx~ \nonumber \\
 &\ & \hspace{-25mm}
\times \int_{-\infty}^\infty dy\; 
\Bigg \{\, x\, \Bigg [\, 
- \frac{\Gamma(1+\varepsilon)}{2}\; 
\frac{(1+\kappa_1^2)q_\perp^2}{[y^2+xy(\kappa_1^2-1)-x^2\kappa_1^2 +i\epsilon]^{1+\varepsilon}}\,
 \nonumber \\
 \, \nonumber \\
 &\ &  \hspace{25mm} +\; \Gamma(2+\varepsilon)~
\frac{-(q_\perp \cdot 
\kappa_1)^2[x^2(1+\kappa_1^2) - x(\kappa_1^2-y)]}{[y^2+xy(\kappa_1^2-1)-x^2\kappa_1^2 +i\epsilon]^{2+\varepsilon}}\ \Bigg ] 
\nonumber \\
\, \nonumber \\
&\ & \hspace{-15mm}
+\, \frac{1}{1-y-i\epsilon}\,
 \Bigg [\, \frac{\Gamma(1+\varepsilon)}{4}~\frac{(1+y)(1+\kappa_1^2)q_\perp^2-4(1-x)(q_\perp 
\cdot \kappa_1)^2}{[y^2+xy(\kappa_1^2-1)-x^2\kappa_1^2 
+i\epsilon]^{1+\varepsilon}} \nonumber \\
&\ &  +y(1-x) \bigg ( \, -\, \frac{\Gamma(1+\varepsilon)}{2}\;
\frac{(1+\kappa_1^2)q_\perp^2}{[y^2+xy(\kappa_1^2-1)-x^2\kappa_1^2 +i\ep\, ]^{1+\vep}}
\nonumber\\
\, \nonumber \\
&\ &  \hspace{10mm}+\; \Gamma(2+\varepsilon)~\frac{-(q_\perp \cdot 
\kappa_1)^2[x^2(1+\kappa_1^2)-x(\kappa_1^2-y)]}{[y^2+xy(\kappa_1^2-1)-x^2\kappa_1^2 +i\epsilon]^{2+\varepsilon}}\ \bigg)\; \Bigg ]\ \Bigg\}
\nonumber\\
\, \nonumber \\
&\ & \hspace{-20mm}  \equiv 
4\, \frac{1}{\pi^{1-3\vep}}~ 
\left (\frac{\alpha_s}{\pi} \right)^2 2^{2\ep}\mu^{4\varepsilon}\, \int 
\frac{dk_1^+}{{k_1^+}^{1+4\varepsilon}}~ 
\int d^{2-2\varepsilon} \kappa_1 \frac{1}{(1+\kappa_1^2)^2}~ 
[I^{(3)}(\kappa_1)~+~I^{(4)}(\kappa_1)]\, ,
\nonumber\\
\label{si3i4}
\end{eqnarray}
where in the second relation, we define $I^{(3)}(\kappa_1)$ to include the terms without
the $1/(1-y)$
denominator, and $I^{(4)}$ to include the remaining terms, all with this denominator.
The infrared poles of $I^{(3)}(\kappa_1)$ are identified in the same way as
those of the corresponding $q_3^2$ integral, $I^{(1)}(\kappa_1)$, Eq.\ (\ref{i1}),
while those of $I^{(4)}(\kappa_1)$ are found in the same way as for $I^{(2)}(\kappa_1)$, Eq.\ (\ref{i2}).

For $I^{(3)}$ the $y$ integration is elementary, and we find
\begin{eqnarray}
 I^{(3)}(\kappa_1) &=& \sqrt{\pi}~\int_0^1 dx~ x 
 \Big[\, \frac{\Gamma(1/2+\varepsilon)}{2}\;
 \frac{(1+\kappa_1^2)q_\perp^2}{[-x^2(1+\kappa_1^2)^2 +i\epsilon]^{1/2+\varepsilon}} 
 \nonumber \\
\nonumber \\
&\ & \hspace{10mm}
-\; \Gamma(3/2+\varepsilon)~\frac{-(q_\perp 
\cdot 
\kappa_1)^2~[x^2(1+\kappa_1^2)-x\kappa_1^2 - x^2(\kappa_1^2-1)/2]}{[-x^2(1+\kappa_1^2)^2 +i\epsilon]^{3/2+\varepsilon}}\, \Big]\, . \nonumber \\
\end{eqnarray}
The overall infrared pole in this expression
is easily identified as arising from the limit $x\to 0$.  It comes entirely
from the the middle term in the numerator of the second fraction,
\bea
&& I^{(3)}(\kappa_1)~ =~2\pi i\ \frac{(q_\perp \cdot 
\kappa_1)^2\, \kappa_1^2}{(1+\kappa_1^2)^3}
~\frac{(1-i\pi\varepsilon)}{\varepsilon} \, .
\label{i3k1}
\eea
Once again the pole is purely imaginary with, however,
an associated finite real part.  Finite corrections are all imaginary.

To evaluate $I^{(4)}(\kappa_1)$ we again introduce a
Feynman parameter
$x^\prime$, which enables us to do the $\kappa_2$ integral
just as for $I^{(2)}$ in Sec.\ A.1, giving
\begin{eqnarray}
&& I^{(4)}(K_1)~ = - \int_0^1 
dx~\int_0^1~dx^\prime {x^\prime}^\varepsilon\;
\int_{-\infty}^{\infty}~dy   \nonumber \\
&& \times
\Bigg[\,
\frac{\Gamma(2+\varepsilon)}{4}~\frac{(1+\kappa_1^2)q_\perp^2-4(1-x)(q_\perp \cdot 
\kappa_1)^2}
{[\, x^\prime[y^2+xy(\kappa_1^2-1)-x^2\kappa_1^2 ]~+~(1-x^\prime)(y-1)~+~i\epsilon\, ]^{2+\varepsilon}} \nonumber \\
&&
~+\frac{1}{4}\, y(2x-1)~\Gamma(2+\varepsilon)
\frac{(1+\kappa_1^2)q_\perp^2}
{[\, x^\prime[y^2+xy(\kappa_1^2-1)-x^2\kappa_1^2 ]~+~(1-x^\prime)(y-1)~+~i\epsilon\, ]^{2+\varepsilon}} \nonumber \\
&&~-y(1-x)xx'\, \Gamma(3+\varepsilon)~\frac{(q_\perp \cdot 
\kappa_1)^2[x(1+\kappa_1^2)-\kappa_1^2]}
{[\, x^\prime[y^2+xy(\kappa_1^2-1)-x^2\kappa_1^2 ]~+~(1-x^\prime)(y-1)~+~i\epsilon\, ]^{3+\varepsilon}} \nonumber \\
&&~-y^2(1-x)x x^\prime \Gamma(3+\vep)~\frac{(q_\perp \cdot \kappa_1)^2}
{[\, x^\prime[y^2+xy(\kappa_1^2-1)-x^2\kappa_1^2 ]~+~(1-x^\prime)(y-1)~+~i\epsilon\, ]^{3+\varepsilon}}\, \Bigg]\, .
\nonumber\\
\label{i4}
\end{eqnarray}
These integrals are
precisely of the form of those in $I^{(2)}$, but
to limit the rather  large number of terms, we introduce the $y$-independent
quantities
\bea
&& P~=~\frac{1}{2}\, \left[ x(\kappa_1^2-1)~+~(1-x^\prime)/x^\prime\, \right]   \nonumber \\
&& M^2~=~x^2\kappa_1^2~+~(1-x^\prime)/x^\prime\, .
\label{PMdef}
\eea
In this notation, the denominators of Eq.\ (\ref{i4})
are
\bea
x^\prime[y^2+xy(\kappa_1^2-1)-x^2\kappa_1^2 ]~+~(1-x^\prime)(y-1) = x'(y^2 +2Py - M^2)\, .
\eea
After the $y$ integral, $I^{(4)}(\kappa_1)$ can be written as a sum of five terms,
\begin{eqnarray}
&& I^{(4)}(\kappa_1)~ = 
- \sqrt{\pi}\, ie^{-i\pi \varepsilon}\, \int_0^1 dx ~\int_0^1 dx^\prime~   \nonumber \\
&&
\times \Bigg[\,  \frac{\Gamma(3/2+\varepsilon)}{4}\;
\frac{(1+\kappa_1^2)q_\perp^2}{{x^\prime}^2[P^2+M^2]^{3/2+\varepsilon}} \nonumber \\
&& \hspace{10mm}- \frac{\Gamma(3/2+\varepsilon)}{4}\, (2x-1)\,
P~\frac{(1+\kappa_1^2)q_\perp^2}{{x^\prime}^2[P^2+M^2]^{3/2+\varepsilon}} \nonumber \\
&& \hspace{10mm} -  \Gamma(3/2+\varepsilon)\, (1-x)\, \frac{(q_\perp \cdot \kappa_1)^2}{{x^\prime}^2[P^2+M^2]^{3/2+\varepsilon}} \nonumber \\
&& \hspace{10mm} - \Gamma(5/2+\varepsilon)\, x(1-x)\, P~\frac{(q_\perp \cdot 
\kappa_1)^2(x-(1-x)\kappa_1^2)}{{x^\prime}^2[P^2+M^2]^{5/2+\varepsilon}} \nonumber \\
&&\hspace{10mm}  - \frac{\Gamma(3/2+\varepsilon)}{2}\, x(1-x)\,
\frac{(q_\perp \cdot \kappa_1)^2(M^2-2P^2+\varepsilon  M^2)}
{{x^\prime}^2[P^2+M^2]^{5/2+\varepsilon}}\, \Bigg]
\nonumber\\
&& \equiv
- \sqrt{\pi}\, ie^{-i\pi \varepsilon}\, \Gamma(3/2+\varepsilon)\, 
[\, i^{(1)}(\kappa_1)~+~ i^{(2)}(\kappa_1)~+~ i^{(3)}(\kappa_1)~+~ i^{(4)}(\kappa_1)~+~ i^{(5)}(\kappa_1)\, ]\, ,
\nonumber\\
\label{ii4}
\eea
where the final line is the notation we will use for the five terms,
taken in order, with $i^{(1)}$ the first, and $i^{(5)}$ the last.

The infrared pole of each of the $i^{(i)}$ can be found by
the straightforward, if slightly tedious,  application of the following steps:
1) re-express $P$ and $M^2$ in terms of $x$, $x'$
and $\kappa_1^2$ using (\ref{PMdef}), 2) 
change variables as above to $u~=~x^2$ and $v^\prime~=~(1-x^\prime)/u$,
3) identify  the residue of the singular  $u^{-1-\vep}$ behavior,
where it is present.  In fact, of the five terms, only $i^{(1)}$, $i^{(3)}$
and $i^{(4)}$ are singular at  $\vep=0$.  Their poles are determined from
\bea
~i^{(1)}(\kappa_1)~=~ \frac{(1+\kappa_1^2)q_\perp^2}{8}~\int_0^1~
\frac{du}{u^{1+\varepsilon}}~\int_0^\infty~dv^\prime~
~\frac{1}{[v^\prime ~+~(1+\kappa_1^2)^2/4]^{3/2}}\, ,
\label{i1k1}
\eea
\bea
~i^{(3)}(\kappa_1)~=~-\, \frac{(q_\perp \cdot \kappa_1)^2}{2}~
\int_0^1~ \frac{du}{u^{1+\varepsilon}}~\int_0^\infty~dv^\prime~
~\frac{1}{[v^\prime ~+~(1+\kappa_1^2)^2/4]^{3/2}}. 
\label{i3k1a}
\eea
\bea
~i^{(4)}(\kappa_1)~=~ \frac{3(q_\perp \cdot \kappa_1)^2}{8}~\kappa_1^2~(\kappa_1^2-1)~
\int_0^1~ \frac{du}{u^{1+\varepsilon}}~\int_0^\infty~dv^\prime~
~\frac{1}{[v^\prime ~+~(1+\kappa_1^2)^2/4]^{5/2}}\, .
\label{i4k1}
\eea
The remaining two terms
behave as $u^{-1/2}$ for  $u\to  0$ and are hence of order $\vep^0$
and imaginary.

The contributions of the poles in Eqs.\ (\ref{i1k1})-(\ref{i4k1})
are also purely imaginary because of the
overall factor of $-i\pi$ in  (\ref{ii4}), and will
enter the fragmentation function as an imaginary
double pole, which therefore cancel.   Correspondingly, all real terms of
order $\vep^0$  from the $u$, $v'$ and $\kappa_1$ integrals
contribute only at the level of an imaginary  single pole.
A real single pole in the final result can only result from
a relative factor $-i\vep$, which is found  as above
from the expansion of ${\rm e}^{-i\pi\vep}$.

The final  result for $IIIA^{(k_0)}(q_\perp)$, 
defined by Eq.\ (\ref{IIIAqperp}) and 
(\ref{n3qperp}),  is
 therefore found from: 4) isolating the
finite real part  from the expansion  of the overall factor of $e^{-i\pi\vep}$ 
in (\ref{ii4}), 5) performing the remaining $\kappa_1$ integration  at $\vep=0$,
and finally 6) replacing the final $k_1^+$ integral by $1/(-4\vep)$,
according to Eq.\ (\ref{Nepspole}).
In this way, we  obtain 
\bea
2\, {\rm Re}\, IIIA^{(k_2^0\, {\rm pole})}(q_\perp)
~=~ -\, \alpha_s^2~\frac{1}{3 \varepsilon}~q_\perp^2
\qquad (q_3=0)\, ,
\label{sqT}
\eea
matching Eq.\ (\ref{sq3}) for the $q_3^2$ term.   There are no
terms linear in $q_3$ and $q_\perp$, and the complete
result  is thus rotationally invariant.

\end{appendix}

     \end{document}